# Experimental and Computational Study of Isomerically Pure Soluble Aza-phthalocyanines and Azasubphthalocyanines of Varying Number of Aza Units

Martin Liebold[a], Eugen Sharikow[a], Elisabeth Seikel[a], Lukas Trombach[b], Klaus Harms[b], Petr Zimčík[c], Veronika Nováková[c], Ralf Tonner[b,*], and Jörg Sundermeyer[a,*]

[a] *Fachbereich Chemie and Wissenschaftliches Zentrum für Materialwissenschaften (WZMW), Philipps-Universität Marburg, Hans-Meerwein-Straße 4, 35032 Marburg, Germany. e-mail: jsu@chemie.uni-marburg.de*

[b] *Fachbereich Chemie, Philipps-Universität Marburg, Hans-Meerwein-Straße 4, 35032 Marburg, Germany. e-mail: tonner@chemie.uni-marburg.de*

[c] *Department of Pharmaceutical Chemistry and Drug Control, Faculty of Pharmacy in Hradec Kralove, Charles University in Prague, Heyrovskeho 1203, Hradec Kralove, 500 05 Czech Republic e-mail: petr.zimcik@faf.cuni.cz*

**Abstract:** Herein, we present a series of isomerically pure, peripherically alkyl substituted, soluble and low aggregating azaphthalocyanines as well as their new, smaller hybrid homologues, the azasubphthalocyanines. The focus lies on the effect of the systematically increasing number of aza building blocks [-N=] replacing the non peripheral [-CH=] units and its influence on physical and photophysical properties of these chromophores. The absolute and relative HOMO-LUMO energies of azaphthalocyanines was analyzed using UV-Vis, CV and compared to density functional theory calculations (B3LYP, TD-DFT). Crystals of substituted subphthalocyanine, $N_2$-Pc*$H_2$ and $N_4$-[Pc*Zn·$H_2$O] were obtained out of DCM. For the synthesis of the valuable tetramethyltetraline phthalocyanine building block a new highly efficient synthesis involving a nearly quantitative Co$^{II}$ catalyzed aerobic autoxidation step is introduced replacing inefficient $KMnO_4$/pyridine as oxidant. The lowering of the HOMO level is revealed as the determining factor for the trend in the adsorption energies by electronic structure analysis.

## Introduction

Phthalocyanines (Pc) and structurally related isosteric octaazaphthalocyanines, in particular pyrazinoporphyrazines (Ppz), probably became one of the technologically most prominent classes of organic dyes. There are numerous applications, including absorbers in dye sensitized solar cells (DSCs),[1,2,3,4] semiconductors in organic field effects transistors (OFETs),[5] or photoconductors in laser printers.[6] Traditionally, they serve as dye ingredient in lacquers, paints,[7] and plastics.[8] More recently, they are used as colour filters in liquid crystal display,[8] as fluorescent markers,[9] as sensors or as sensitizers in photodynamic therapy (PDT).[10] On one side, the characteristic aggregation and $\pi$-stacking of these 42 $\pi$ electron extended HÜCKEL aromatic systems causes their optoelectronic materials properties. On the other side, aggregation typically leads to molecules completely insoluble in organic solvents.[8] By introducing bulky axial metal ligands,[11,12] and/or peripheral or non peripheral ring substituents,[13,14] the solubility in common organic solvents is increased and aggregation decreased. In current research, the investigation of [–CH=], [-N=] units is a hot topic. Recently, symmetrical $A_4$-type phthalocyanines and naphthalocyanines as well as their aza-analogues were investigated with respect to their singlet oxygen generation and photostability.[15] A selective access to a class of *meso* substituted Pc derivatives, namely tetrabenzotriazaporphyrins (TBTAP), has been reported as well.[16] Unfortunately, one *tert*-butyl group at the peripheral position commonly used to gain solubility of a Pc always leads to mixtures of up to four stereoisomers as a consequence of the cyclotetramerisation.[14] Commonly used two peripheral long n-alkyl or -OR, -SR substituents per isoindoline unit lead to one stereoisomer, however, is becomes difficult to get highly ordered crystalline Pc phases. Linking two peripheral *tert*-butyl groups via a joint C-C bond together leads to chromophores soluble enough for chromatography and $^1$H NMR spectroscopy, at the same time sterically demanding enough to avoid intermolecular deactivation paths of their excited state.[17] Such a best compromise phthalocyanine builing block PDN* **1** was first reported by MIKHALENKO *et al*. (Fig. 1).[18] Related symmetrical $A_4$-type Pc*MX complexes have a higher tendency to form well-ordered and even single-crystalline phases.[11] However, the 8-step synthesis of Pc*$H_2$ might not have looked appealing to followers, probably because a side chain oxidation of the 1,1,4,4,5,6-hexamethyltetraline intermediate to the corresponding *o*-dicarboxylic acid (Scheme 1) by $KMnO_4$ in pyridine was drastically limiting the overall yield of the ligand.[18] Here, we report a much more attractive method and selective catalytic aerobic oxidation as replacement of this synthesis step. It is suggested to call this sterically demanding soluble and compact phthalocyanine derivative with 16 methyl groups Pc*H$_2$, following the common terminology of prominent organometallic ligands Cp (C$_5$H$_5$) and Cp* (C$_5$Me$_5$). The corresponding soluble pyrazinoporphyrazine Ppz*H$_2$ has been introduced by us only recently.[12] The aim of this study is to apply the precursors PDN* (**1**) and PyzDN*(**2**) to synthesize and fully characterize a complete series of diaza-, tetraaza-, and hexaaza-N$_x$-Pc*M hybrid compounds as well as their subphthalocyanine homologues Spc*BCl, diaza-, tetraaza-, and hexaaza-N$_x$-[Spc*BCl] (or [Sppz*BCl] for n=6), respectively (Fig. 2 and 3). Despite of the fact, that a large number of lower-symmetry phthalocyanines of A$_{4-n}$B$_n$-type and subphthalocyanines of an A$_{3-n}$B$_n$-type is described in literature, only few of such aza derivatives are systematically studied.[19,20] We were interested in a systematic study of photophysical properties and optoelectronic response upon stepwise substitution of [–CH=] units by [–N=] units in these hybrid chromophores. In the following we correlate UV-Vis data with electro-analytically gained oxidation and reduction potentials measured by CV, and compare these results with results of DFT calculations. Density functional theory has been found to perform well for the discussion of trends in excitation energies along a series,[21] although the derivation of accurate absolute excitation energies needs more sophisticated methods.[22] It was found before, that the investigation of the HOMO-LUMO energy gap can be a good indicator for the excitation energy observed.[21] The reduced symmetry in protonated H$_2$Pc* or A$_{4-n}$B$_n$-type aza-analogues N$_x$-Pc*M should lead to splitting of the Q-band.[19]

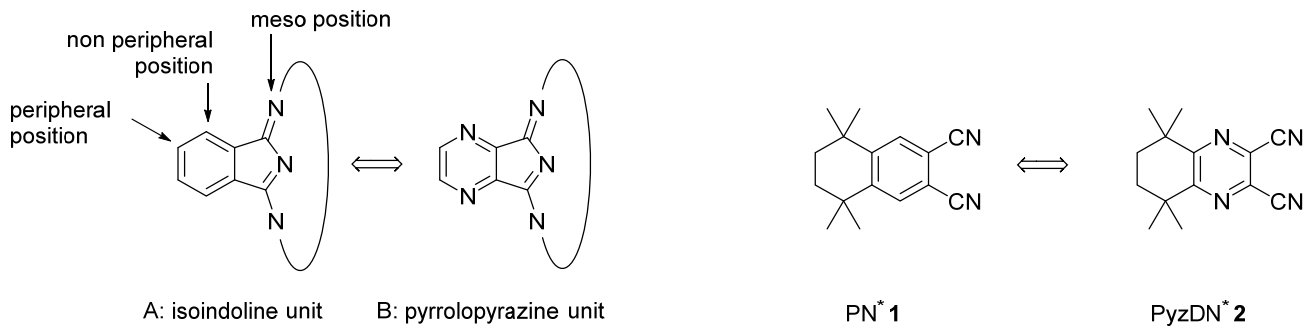

**Figure 1.** Discussed isoindoline and pyrrolopyrazine units derived from phthalodinitile **1** and pyrazinedinitril **2**.

Knowing and predicting the trend in the HOMO-LUMO gap, designing the absolute HOMO-LUMO energies, as well as adding a permanent dipole moment to these chromophores and control their aggregation are fundamentals needed to control level alignment at internal interface to wide band gap semiconductors such as TiO$_2$ or ZnO.[1,3,4] Understanding exciton dynamics and exciton dissociation into mobile charges at ideal model heterojunctions is of fundamental importance, e.g. for optimizing DSSCs.

## Results and Discussion

**Synthesis.** The respective dinitrils PDN* **1**[18,23,24] and PyzDN* **2**[12] have been described in literature. The overall yield limiting key step in the 8-step synthesis of PDN* **1** is the oxidation of 1,1,4,4,5,6-hexamethyltetraline to the 1,1,4,4-tetramethyl-tetraline-6,7-dicarboxylic acid building block. So far this stoichiometric oxidation was accomplished by treatment with large excess of potassium permanganate in hot pyridine. This procedure is hampered with unreliable yields of 15-60% and large amount of poisonous waste. Here we report a green catalytic access to this key intermediate, which will promote drastically its future use in the synthesis of soluble, non-aggregating, isomerically pure phthalocyanines (Scheme 1).

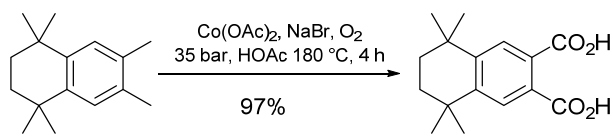

**Scheme 1**. New catalytic oxidation of 1,1,4,4,5,6-hexamethyltetraline to phthalic acid key building block.

Our protocol is similar to conditions used in the industrial autoxidation of *para*-xylene to terephthalic acid with dioxygen in the presence of cobalt acetate catalyst and a bromide promotor.[23] The yield increases to 97%, the waste is negligible, water is formed as by-product. The product precipitated in crystalline state, solvent glacial acid and even catalyst can be fully recycled. With large excess of glacial acid or in the presence of acetic anhydride the isolated product is corresponding phthalic anhydride, which saves even one reaction step in the synthesis of PDN* **1**. The radical chain mechanism induced by intermediate Co(OAc)$_3$ is discussed in the electronic supplement. H atom abstraction occurs exclusively at the benzylic sp$^3$ CH bond. This type of side-chain oxidation can also be carried out in lower yield without bromide promotor, but higher amounts of [Co(OAc)$_2$]·4H$_2$O and MEK are necessary.[24]

The azaphthalocyanines N$_x$-Pc*H$_2$, shown in Fig. 2, were synthesized in a cyclotetramerization reaction following the reliable protocol of reacting PDN* **1** and PyzDN* **2** in a suspension of lithium in *n*-octanol at 180 °C.[19] After workup with phosphoric acid and washing with methanol the mixture of the resulting four N$_x$-Pc*H$_2$ **4a**, **5a**, **6a**, and **7a**, Pc*H$_2$ **3a**, and Ppz*H$_2$ **8a** was separated via gradient column chromatography PE/EE 10:1 → 1:1. Ppz*H$_2$ **8a** could be finally eluted using DCM. Using a 1:1 ratio of PDN* **1** and PyzDN* **2** in this random co-cyclisation an overall yield of 39% of mixed (aza)phthalocyanines was obtained and separated by MPLC into six fractions: Main product is the A$_2$B$_2$ isomer **5a** (17%) tetraazaphthalocyanine N$_4$-Pc*H$_2$, followed by AB$_3$ N$_6$-Pc*H$_2$ **7a** (9%) and Ppz*H$_2$ **8a** (9%). Minor product was the ABAB isomer of N$_4$-Pc*H$_2$ **6a** (<1%) as well as A$_3$B N$_2$-Pc*H$_2$ **4a** (3%) and Pc*H$_2$ **3a** (1%). Even by increasing the amount of PDN* **1**, or by increasing the temperature of the reaction, **5a** was obtained as main product. KOBAYASHI et al. reported about this phenomenon, when different electron deficient phthalonitriles with different sterical demand were used.[19] Due to the similar steric bulk of both dinitriles, we assume a strong electronic influence in the reaction leading to a precyclisation of electron poor PyzDNs. A$_2$B$_2$ N$_4$-Pc*H$_2$ isomer **5a** was differentiated from ABAB isomer **6a** by means of $^1$H NMR and UV/Vis spectroscopy. After chromatographic isolation of the pure N$_x$-Pc*H$_2$, all zinc complexes Pcs, N$_x$-[Pc*Zn] **3b**-**8b** were obtained using [Zn(hmds)$_2$] or [Zn(OAc)$_2$]·2H$_2$O as metallating agents. Attempts to purify mixtures of zinc complexes by chromatography were of limited success.

In addition, we attempted to synthesize compounds **5a** and **6a** via KOBAYASHI ring expansion.[25] Following this synthetic strategy, we synthesized a series of azasubphthalocyanines N$_x$-[Spc*BCl] **9**-**12** shown in Fig. 3. [Spc*BCl] **9** and [Sppz*BCl] **12** was synthesized according to literature known procedures, using boron trichloride in a solvent such as toluene or o-xylene.[20] Compared to [Spc*BCl] **9** with 13% yield, [Sppz*BCl] **12** could be obtained in surprisingly high yields of 35%. N$_2$-[Spc*BCl] **10** and N$_4$-[Spc*BCl] **11** were obtained by using both dinitriles PN* **1** and PyzDN* **2** in a ratio 1:1. The ring expansion was attempted with [Spc*BCl] **9**, [Sppz*BCl] **12**, and N$_4$-[Spc*BCl] **11**. Therefore, the isoindoline compound of PN* **1**, or PyzDN* **2** in case of [Spc*BCl] **9**, was generated in situ by introducing ammonia in a NaOMe/MeOH solution.

To [Sppz*BCl] **12**, the isoindoline of **1** was added in DMSO and the solution stirred at 120 °C for 8 h, yielding AB$_3$ N$_6$-[Pc*M] **7**. In the case of [Spc*BCl] **9**, reacting with the isoindoline of PyzDN* **2**, no ring expansion can be observed. A ring expansion is favoured, the more electron deficient the Spc is, the more electron rich the isoindoline compound is, respectively. In the case of the insertion of isoindoline of **1** in N$_4$-[Spc*BCl] **9**, A$_2$B$_2$ N$_4$-Pc*H$_2$ **5a** was obtained. This verifies the electronic influence of the pyrazine units upon cyclotetramerisation.

**UV/Vis spectroscopy and CV**. The synthesized N$_x$-Pc*H$_2$ have been characterized by using UV/Vis spectroscopy and cyclic voltammetry (CV). By increasing the number of pyrazine units, a gradual hypsochromic shift of the Q-band can be observed, the higher the number of inserted [-N=] units is: from 707 nm for Pc*H$_2$ **3a** to 652 nm for Ppz*H$_2$ **8a** in THF, and from 575 nm for

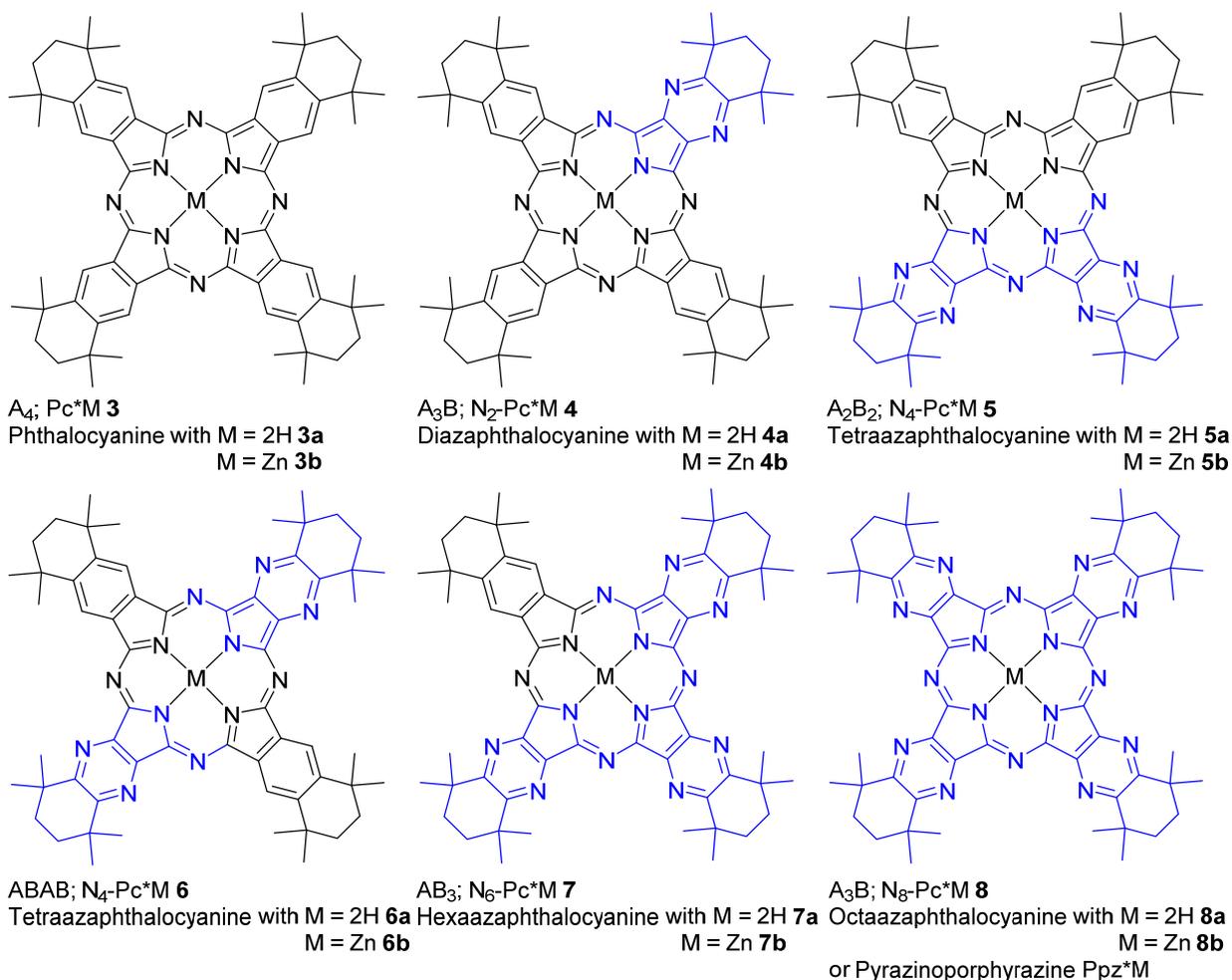

A$_4$; Pc*M **3**
Phthalocyanine with M = 2H **3a**
M = Zn **3b**

A$_3$B; N$_2$-Pc*M **4**
Diazaphthalocyanine with M = 2H **4a**
M = Zn **4b**

A$_2$B$_2$; N$_4$-Pc*M **5**
Tetraazaphthalocyanine with M = 2H **5a**
M = Zn **5b**

ABAB; N$_4$-Pc*M **6**
Tetraazaphthalocyanine with M = 2H **6a**
M = Zn **6b**

AB$_3$; N$_6$-Pc*M **7**
Hexaazaphthalocyanine with M = 2H **7a**
M = Zn **7b**

A$_3$B; N$_8$-Pc*M **8**
Octaazaphthalocyanine with M = 2H **8a**
M = Zn **8b**
or Pyrazinoporphyrazine Ppz*M

**Figure 2.** Discussed azaphthalocyanines N$_x$-[Pc*M] (M = 2H, Zn, n= 0,2,4,6,8).

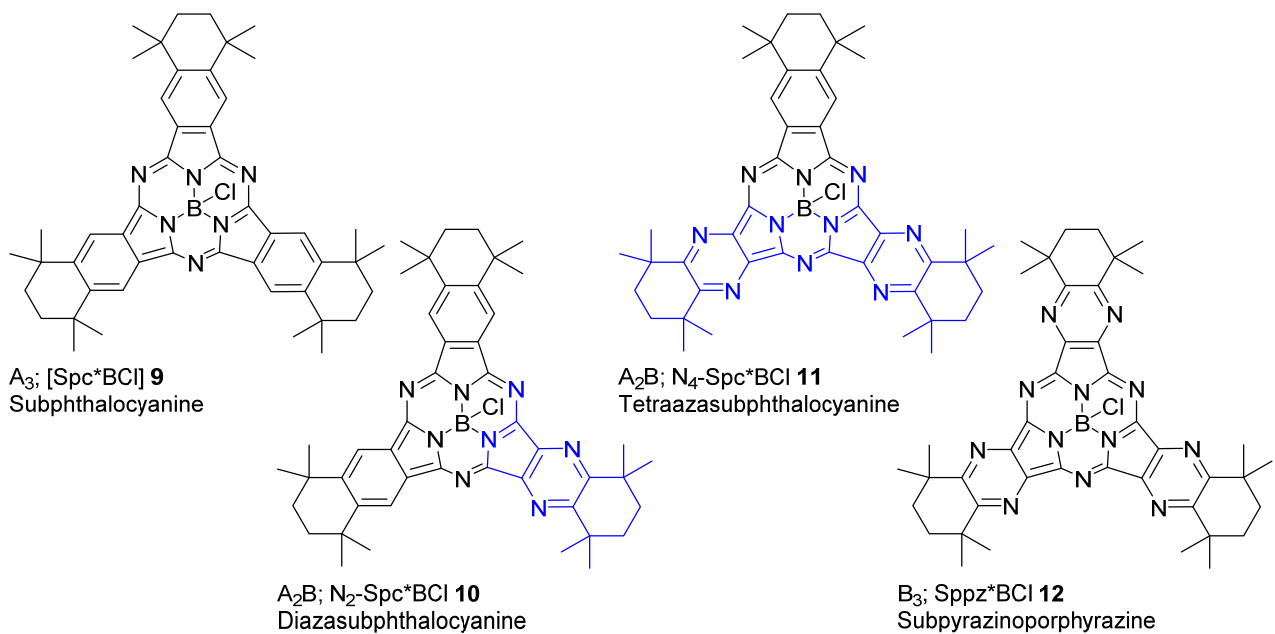

**Figure 3.** Discussed azasubphthalocyanines $N_x$-[Spc*BCl] (n= 0, 2, 4, 6).

[Spc*BCl] **9** to 533 nm for [Sppz*BCl] **12** in MeOH. The studied Pc macrocycles were of lower symmetry either due to the presence of the different number of pyrazine units or due to presence of central hydrogens in $N_x$-Pc*$H_2$ that was reflected in the splitting of the Q-bands. In the metal-free series $N_x$-Pc*$H_2$, the splitting occurred for all the macrocycles although, e.g. in case of **4a** and **7a** was only barely detectable and in particular for the former one manifested only by asymmetric broadening of the Q-band (Fig. 4). The strongest splitting was expected to be visible for ABAB type macrocycle **6a**. In this case, however, seemingly one main Q-band was detected. Closer inspection, however, revealed a shoulder at 701 nm as an indication of the split band. This small-intensity band at 701 nm was also detected in fluorescence excitation spectra (see electronic supplement, Fig. SA).

Considering this small band, the splitting of the Q-band is then the strongest in the series of all studied Pcs in accordance with literature data for lower symmetrical Pcs.[26] After insertion of a zinc atom, the lower symmetrical $A_3B$ and $AB_3$ $N_x$-[Pc*Zn] **4b** and **7b** showed a weak splitting of their Q-bands. This is caused by the $1e_g$ orbitals, which are not degenerate anymore. Single unperturbed Q-band was expectedly observed for both symmetrical **3b** and **8b** but also for $A_2B_2$ isomer **5b** although being of lower $C_{2v}$ symmetry. This observation is, however, typical for adjacent isomers of $A_2B_2$ type and has been reported several times in literature.[26,27]

The UV/Vis spectra correlate with the measured CVs. In Table 1, the measured and computed excitation energies are listed. The optical values, calculated by using UV/Vis spectroscopy, are in agreement with the determined redox excitation energies observed in CV measurements.

**Table 1.** Obtained oxidation and reduction potentials of $N_x$-Pc*$H_2$ in DCM.[a] potentials $E_{red}^1$, $E_{red}^2$, and $E_{ox}$ are expressed as $E_{1/2}$ (in V vs. SCE) with Fc/Fc$^+$ as the internal standard. [b] potential difference as the difference between the first oxidation and reduction potential. [c] UV/Vis data. [d] calculated HOMO-LUMO gap, from determined optical UV/Vis values.

|  | Ppz*$H_2$ (**8a**) | $N_6$-Pc*$H_2$ (**7a**) | $N_4$-Pc*$H_2$ (**6a**) | $N_4$-Pc*$H_2$ (**5a**) | $N_2$-Pc*$H_2$ (**4a**) | Pc*$H_2$ (**3a**) |
|---|---|---|---|---|---|---|
| $E_{ox}$ / V | 0.87 | 0.35 | 0.28 | 0.36 | 0.43 | 0.67 |
| $E_{red}^1$ / V | -1.05 | -1.47 | -1.47 | -1.37 | -1.23 | -0.97 |
| $E_{red}^2$ / V | -1.37 | -1.82 | -1.79 | -1.77 | -1.65 | -1.36 |
| $E_g^{CV}$ / V [b] | 1.92 | 1.82 | 1.75 | 1.73 | 1.67 | 1.58 |
| $\lambda_{max}$ / nm [c] | 655 | 660 | 674 | 680 | 687 | 711 |
| HOMO-LUMO$^{optical}$ / eV [d] | 1.89 | 1.87 | 1.85 | 1.82 | 1.80 | 1.74 |
| HOMO-LUMO$^{calc.}$ / eV | 1.52 | 1.44 | 1.36 | 1.41 | 1.35 | 1.35 |
| HOMO-(LUMO+1)$^{calc.}$ / eV | 1.59 | 1.56 | 1.52 | 1.46 | 1.45 | 1.37 |

The computed Q-band positions give the right trend in comparison to experiments, although the absolute numbers are around 100-130 nm smaller. The same conclusion holds for the HOMO-LUMO gap. It is found that the correlation is much better with the LUMO+1 orbital ($D_{HL+1}$) which is depicted in Figure SI-1 in the supporting information. A detailed discussion is given in the computational section. CV measurements were carried out with an rhd-instruments setup, using a 2000μL microcell in a glovebox. As working electrode a Pt electrode was used, a gold or Pt counter electrode as well as an Pt pseudo reference electrode. The compounds were measured in a 0.5 $N$ [TBA]PF$_6$ in DCM solution using a 5 $m$mol solution of the N$_x$-Pc*H$_2$. All measurements have been carried out using a 0.1 mmol Fc in DCM as reference. All measured N$_x$-Pc*H$_2$ show one reversible oxidation process and two reversible reduction potentials, with exception for the Ppz*H$_2$ **8a**, as shown in Figure 4. In this case, no reversible redox process was observed. In agreement with the Q-band position, as measured by UV/Vis spectroscopy, the energy gap was increasing with the number of pyrazine units in the N$_x$-Pc*H$_2$. The increasing electron deficiency of the aromatic system correlating with the increasing number of [-N=] is also seen in $^1$H NMR spectroscopy. An expected low field shift of the aromatic protons, the methylene protons, as well as the methyl protons by increasing the number of [-N=] can be observed. In summary, an almost linear trend could be observed in UV/Vis, CV and computational calculations. The origin of this trend is analysed in the computational discussion part below.

**Photophysical properties.** Electronic absorption spectra of the compounds are displayed in Fig. SI-1-3 of the supporting information. The Q-band can be shifted from 450 nm to 710 nm. The following trend is observed: the higher the number of inserted [-N=] atoms in the subphthalocyanine series (**9-12**), the more blue-shifted the Q-band within one type of π-system. At the high energy end we find the hexaaza-subphthalocyanines (533 nm) followed by tetraaza- (555 nm) and diaza-subphthalocyanines (566 nm) and finally the parent subphthalocyanine (575 nm). Same trend is observed for the phthalocyanine series in the region [Ppz*Zn] (640 nm) – Pc*H$_2$ (710 nm): The most blue-shifted Q-band is observed for octaaza-phthalocyanine, followed by hexaaza-, followed by the A$_2$B$_2$ and ABAB isomers of tetraaza-, of diaza- and finally by the parent phthalocyanine. Fluorescence emission spectra of all derivatives were measured in THF or MeOH (Fig. 4). They follow the same trends as the absorption spectra. The higher the number of [-N=] units replacing [-CH=], the more blue-shifted the maxima of the FL bands. Their shape and order mirrored the absorption Q-band with only small Stokes shifts (6 – 26 nm) between absorption and emission maximum. Position of the emission maximum can be finely tuned using the proper composition of the macrocycle (Fig. 4) that can be advantageously used in design of new fluorophores or photosensitizers with selected excitation and emission wavelength. Excitation spectra perfectly matched the absorption ones confirming that the compounds were in monomeric form in the solution during measurements and that the photophysical data were not affected by aggregation (Figs. 4, and SI-1-3). Exception from this observation was detected for the weakest fluorescent of **8a**. In this case, the excitation spectrum was substantially different from the absorption one indicating that a different species (e.g. partially deprotonated on the central NH)[28] is responsible for the fluorescence emission. It might be connected with very weak fluorescent properties of **8a** as suggested below that do not allow detection of its signal. Consequently, the emission from the new species prevails even when present as very minor component.

In the zinc series N$_x$-[Pc*Zn], the sum of the fluorescence quantum yield ($\Phi_F$ = 0.18-0.35) and singlet oxygen quantum yield ($\Phi_\Delta$ = 0.56-0.73) for a particular compound, were relatively constant (Table 2, Fig. 5) reaching values typically close to 0.9. The only difference between the compounds in the series was in relative contribution of the two processes. Different situation was observed in the metal-free series N$_x$-Pc*H$_2$. The sum of the quantum yields was significantly decreasing with the increasing number of aza-substitution with compounds **7a** and **8a** being almost photophysically inactive (from the point of view of $\Phi_F$ and $\Phi_\Delta$).

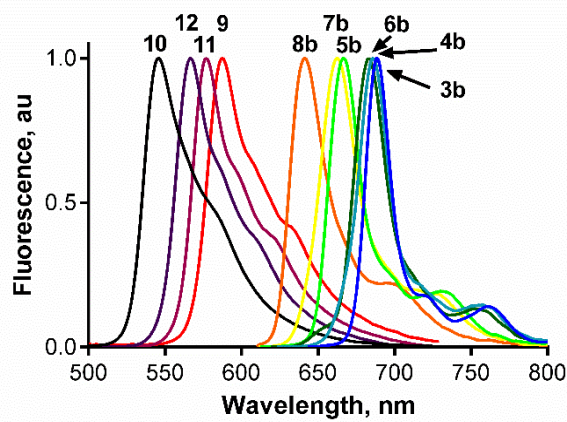

**Figure 4**. Normalized emission spectra of N$_x$-Pc*Zn in THF ($\lambda_{exc}$ = 598 nm) and N$_x$-[Spc*BCl] in MeOH ($\lambda_{exc}$ = 490 nm).

This interesting phenomenon does not have a clear explanation at this moment but is fully consistent with recently reported photophysical data for some metal-free Ppzs.[28,29] The presented series **3a**-**8a** is, however, unique as the photophysical data for metal-free derivatives are reported only from the members lying on the edges of the N$_x$-Pc*H$_2$ series (i.e. only for x = 0 and x = 8).

The trend observed on the Fig. 5 clearly indicated that the introduction of the nitrogens into the Pc ring increases the feasibility of this unknown non-radiative relaxation pathway in a stepwise manner. In the N$_x$-[Spc*BCl] series, the photophysical data resembled those obtained for zinc derivatives ($\Phi_F$ ~ 0.20,

**Table 2.** Electronic and photophysical data of N$_x$-Pc*M in THF and N$_x$-Spc*BCl in MeOH. [a] absorption maximum ($\lambda_{max}$), fluorescence emission maximum ($\lambda_F$), fluorescence lifetime ($\tau_F$), fluorescence quantum yield ($\Phi_F$), singlet oxygen quantum yield ($\Phi_\Delta$). [b] $\lambda_{exc}$ = 371 nm or 634 nm. [c] with [PcZn] ($\Phi_{F(THF)}$ = 0.32, $\lambda_{exc}$ = 598 nm) and rhodamin G6 ($\Phi_{F(EtOH)}$ = 0.94, $\lambda_{exc}$ = 490 nm) as reference for N$_x$-[Pc*M] and N$_x$-[Spc*BCl], respectively. [d] with ZnPc ($\Phi_{\Delta(THF)}$ = 0.53) and bengal rose ($\Phi_{\Delta(MeOH)}$ = 0.76) as reference for N$_x$-[Pc*M] and N$_x$-[Spc*BCl], respectively.

| | $\lambda_{max}$ (B-band) / nm | $\lambda_{max}$ (Q-band) / nm | $\lambda_F$ / nm | $\tau_F$ / ns[b] | $\Phi_F$[c] | $\Phi_\Delta$[d] | $\Phi_F + \Phi_\Delta$ |
|---|---|---|---|---|---|---|---|
| Pc*H$_2$ **3a** | 343 | 707, 673 | 711 | 5.52 | 0.42 | 0.16 | 0.58 |
| N$_2$-Pc*H$_2$ **4a** | 343 | 686 | 692 | 3.45 | 0.33 | 0.14 | 0.47 |
| N$_4$-Pc*H$_2$ **5a** | 358 | 678, 653 | 684 | 1.62 | 0.14 | 0.07 | 0.21 |
| N$_4$-Pc*H$_2$ **6a** | 353 | 668, (701) | 674 | 1.79 | 0.16 | 0.10 | 0.26 |
| N$_6$-Pc*H$_2$ **7a** | 349 | 658, 643 | 664 | 3.92 (26%), 0.63 (74%) | 0.06 | 0.04 | 0.10 |
| N$_8$-*H$_2$ **8a** | 345 | 652, 623 | 658 | 2.15 (29 %), 0.36 (71%) | 0.03 | 0.05 | 0.08 |
| [Pc*Zn] **3b** | 350 | 683 | 689 | 3.31 | 0.31 | 0.59 | 0.90 |
| N$_2$-[Pc*Zn] **4b** | 353 | 679, 668 | 686 | 2.69 | 0.25 | 0.63 | 0.88 |
| N$_4$-[Pc*Zn] **5b** | 356 | 658 | 667 | 3.08 | 0.35 | 0.56 | 0.91 |
| N$_4$-[Pc*Zn] **6b** | 355 | 679, 651 | 687 | 1.98 | 0.18 | 0.73 | 0.91 |
| N$_6$-[Pc*Zn] **7b** | 353 | 648 | 663 | 2.69 | 0.27 | 0.62 | 0.89 |
| N$_8$-[Pc*Zn] **8b** | 347 | 633 | 642 | 2.98 | 0.30 | 0.57 | 0.87 |
| [Spc*BCl] **9** | 266 | 575 | 587 | 2.04 | 0.18 | 0.65 | 0.82 |
| N$_2$-[Spc*BCl] **11** | 272 | 566 | 577 | 2.19 | 0.21 | 0.62 | 0.82 |
| N$_4$-[Spc*BCl] **12** | 288 | 555 | 567 | 2.85 | 0.23 | 0.64 | 0.87 |
| [Sppz*BCl] **10** | 301 | 533 | 546 | 2.98 | 0.19 | 0.69 | 0.87 |

$\Phi_\Delta$ ~ 0.65) with sum of $\Phi_F$ and $\Phi_\Delta$ over 0.80 indicating that both these processes are predominant in the relaxation of the excited states also in Spc derivatives.

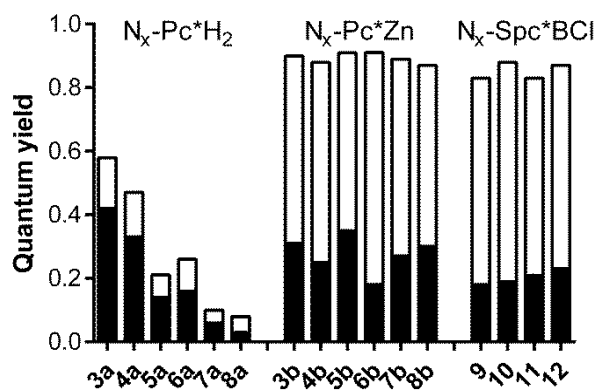

**Figure 5**. Fluorescence quantum yields ($\Phi_F$, full columns) and singlet oxygen quantum yields ($\Phi_\Delta$, blank columns) for N$_x$-[Pc*M] in THF and N$_x$-[Spc*BCl] in MeOH.

The data are well comparable with the reported values of Spcs with different peripheral substitution.[30] No significant differences between fluorescence and singlet oxygen production were observed in dependence on the number of nitrogen atoms. The fluorescence emission properties were complemented by the fluorescence lifetimes (Table 2).
The fluorescence decays were characterized by monoexponential curves for most of the studied compounds and $\tau_F$ values correlated well with the $\Phi_F$ values (Fig. SI-4) further validating the fluorescence data. Only in case of metal-free **7a** and **8a**, the decay curves were biexponential with increased contribution of the fast component. It supports the fact that another relaxation pathway occurs by the metal-free compounds as proposed above.

**Crystal structure of [Spc*BCl], N$_2$-Pc*H$_2$, and A$_2$B$_2$ N$_4$-[Pc*Zn·H$_2$O].** Suitable crystals for X-Ray diffraction of [Spc*BCl] **9**, N$_2$-Pc*H$_2$ **4a** and A$_2$B$_2$ N$_4$-[Pc*Zn·H$_2$O] **5b** were obtained from a saturated DCM-$d_2$ solution. Of all compounds, selected crystallographic datas are collected in the supporting information. [Spc*BCl] **9** crystallizes in the orthorombic crystal system Cmc2$_1$. The molecular structure of [Spc*BCl] **9** is shown in Fig. 6. Similar, to other literature known subphthalcyanines the boron(III)-atom in the [Spc*BCl] **9** is almost perfect tetrahedrally surrounded by one chlorine atom and three nitrogen atom. The boron is out of plane to the $\pi$-system. These are the reasons why only three isoindoline units are around the smaller boron atom, and causes the $\pi$-system to contract. In the crystal structure, the annulated methyl substituted cyclohexene ring shows the typical twisted chair conformation, which explains the multiplett of the methylene group observed in $^1$H NMR spectroscopy (Fig. SI-6). The characteristic bond lengths of the B-Cl and B-N$^{isoindoline}$ are listed in Table 3, compared to the first literature known [SpcBCl] by KIETAIBL.[31]

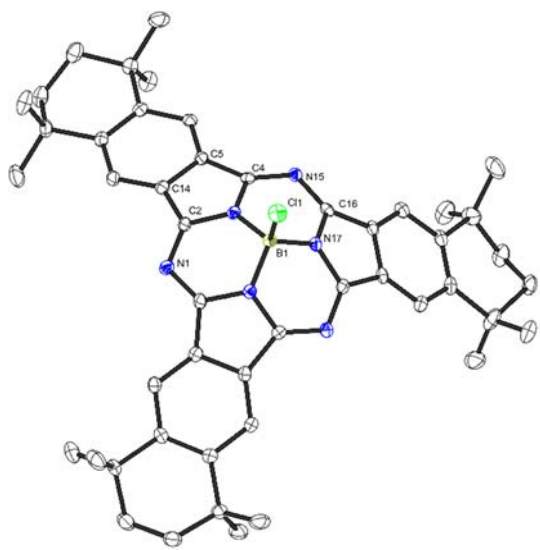

**Figure 6**. Molecular structure of [Spc*BCl] **9** in crystal. Bond distances and angles and the lattice structure are presented in the SI.

With a bond length of B-Cl of 1.85 Å, it is within the expected range, as well as all values found for C-C bonds in the aromatic system and C-N bonds of pyrrole units. The bowl depth of 0.56 Å, as a value for the position boron atom above the inclined [-BN$_3$-] plane is also comparable to [SpcBCl].[31] In addition to the [Spc*BCl] **9**, [Sppz*BCl] was crystallised, but because the low quality of the crystal an appropiate discussion of the bond length is not possible, but in comparison to [Spc*BCl] information of the molecular arrangement in solid state can be obtained. The structure of [Sppz*BCl] is attached to the SI.

**Table 3**: Selected bond length (Å) and angles (°) of [Spc*BCl] 3 and [SpcBCl].[a]

|  | [Spc*BCl] (**3**) | [SpcBCl] |
|---|---|---|
| B1-Cl1 | 1.852(4) | 1.863(7) |
| B1-N1 | 1.480(3) | 1.466(8) |
| B1-N3 | 1.479(4) | 1.468(5) |
| B1-N5 | 1.480(3) | 1.466(8) |
| N1-B1-Cl1, N3-B1-Cl1, N5-B1-Cl1 | 112.2(2), 115.7(2), 112.2(2) | 113.8(19), 112.8(1), 113.8(1) |
| N1-B1-N3, N3-B1-N5, N1-B1-N5 | 105.4(2), 105.0(3), 105.4(2) | 105.1(1), 105.3(0), 105.3(0) |
| ∀$_{Isoindoline}$ | 44.2 | 43.1-44.3 |

The free ligand of N$_2$-[Pc*H$_2$] **4a** was crystallized and a monoclinic system with a P2$_1$/c space group was determined. Here, also the twisted-chair conformation of the methylene ring is visible. The crystal structure is disorded, so in every molecule the non-peripheral position is partly occupied by a quarter [-N=] units and three quarters [-CH=], for the same reason, each NH position is occupied with a factor of 0.5 as shown in Fig. 7 in light blue. The molecular arrangement differs from unsubstituted Pc, where a *heringbone* pattern is typical.

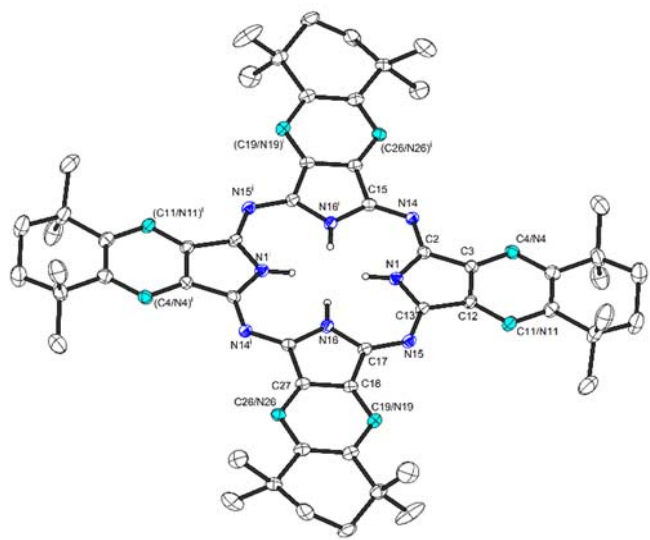

**Figure 7**. Molecular structure of N$_2$-Pc*H$_2$ **4a** in the crystal. Corresponding C and N positions of pyrazine and benzene rings are statistically 75:25 occupied. Bond distances and angles and the lattice structure are presented in the SI. Similar to the previously reported strucutre of Pc*H$_2$, in one unit cell the molecules are orientated perpendicular to each other.

For the first time, a lower symmetrical, alkyl substituted N$_4$-[Pc*Zn·H$_2$O] was structurally characterized. In comparison to known zinc phthalocyanines bearing water as axial ligand, such as [PcZn·H$_2$O]·2dmf,[32] in the molecular packing the described perpendicular orientation of the molecules is visible. Furthermore, the axial water pulls the zinc out of the ligand cavity [-MN$_4$-].

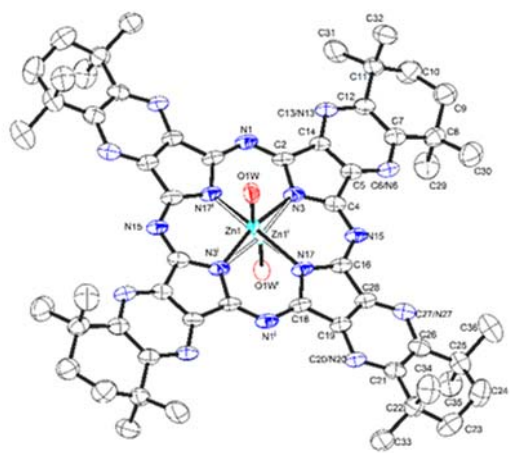

**Figure 8**. Molecular structure of $A_2B_2$ $N_4$-[Pc*Zn] **5b** in crystal. Corresponding C and N positions of pyrazine and benzene rings are statistically 75:25 occupied. Bond distances and angles and the lattice structure are presented in the SI.

## Computational Studies

The electronic structure of the compounds presented here are further analysed with density-functional theory using the PBE and B3LYP exchange-correlation functionals and including dispersion corrections (DFT-D3) for the structural optimization. With the aim to derive trends with increasing nitrogen substitution, the series from [Pc*M] to [Ppz*M] with M = $H_2$, Zn was investigated. All structures including regioisomers and tautomers are shown in Fig. SI-5 in the supporting information. For **6a**, two regioisomers $A_2B_2$ and ABAB, were found which are separated by only 5.6 kJ mol$^{-1}$. Furthermore, for compounds **4a**, **5a** and **7a**, two tautomers are observed. Only for the more favored regioisomer of **6a** (A2B2), a significantly lower stability of one tautomer is found ($\Delta E$ = 56.4 kJ mol$^{-1}$). For all other compounds, an equilibrium of tautomers under experimental conditions can be concluded ($\Delta E$ = 4.0 kJ mol$^{-1}$). For the Zn-substituted phthalocyanines, the two regioisomers **5b** and **6b** show an energy difference of 1.6 kJ mol$^{-1}$. Due to the small energy differences, the analysis of the electronic structure was therefore carried out for all isomers described.

The absorption characteristics for all compounds as derived from time-dependent density functional theory (TD-DFT) computations are summarized in Table SI-1 in the supporting information. The density functional used (B3LYP) has previously been found to provide good agreement with experiment.[21] The values for Pc*$H_2$ **3a** are comparable although a smaller basis set was used before.[21]

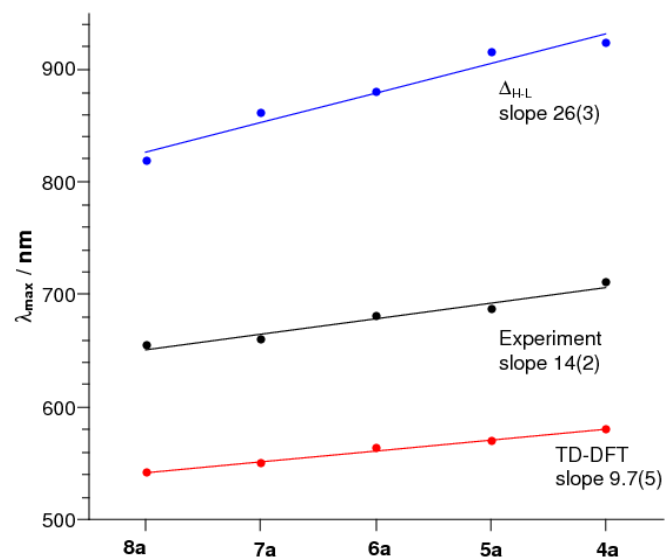

**Figure 9**: Orbital energies for HOMO (black), LUMO (red), and LUMO+1 (blue) for varying N-content of Pc*$H_2$. Orbitals for Pc*$H_2$ are depicted.

We find – with few exceptions – two absorption energies for the Q-bands as was found earlier for unsymmetric Pc derivatives.[19] The splitting is nevertheless very small and is not relevant for the interpretation of experimental UV/Vis spectra since the peaks observed are broader than the energy differences computed here (Fig. 4). The absorption maxima upon increase of the nitrogen content in going from Pc*$H_2$ **3a** to Ppz*$H_2$ **8a** are shifting to higher absorption energies, independent of hydrogen or Zn being in the center. An exception is tautomer **5a$_T$** which has a higher excitation energy than **4a**.

Since this tautomer is much less stable than **5a**, it will not be observed in the experimental spectrum. The shift to higher excitation energies here is stronger than was found for the substitution of the bridging carbon atom by nitrogen.[19] Oscillator strengths are rather large and very similar for both Q-bands. They do not change along the series of compounds investigated.

The character of the Q-bands is in all cases dominated by the HOMO→LUMO or the HOMO→LUMO+1 transition. The small energy differences in the absorption maxima of the Q-bands can therefore be understood by the close energetic proximity of the LUMO and LUMO+1 orbital (Fig. 9). But even for the most symmetric molecules carrying hydrogen, the orbitals do not have the same energy and therefore lead to different Q-band energies. This is a consequence of the necessity to choose one tautomer in the computations. The resulting structure exhibits inequivalent nitrogen atoms in the ring, leading to the different orbital energies of LUMO and LUMO+1 which are located either on the hydrogen-carrying or the hydrogen-free nitrogen atoms (orbital insets in Fig. 9).

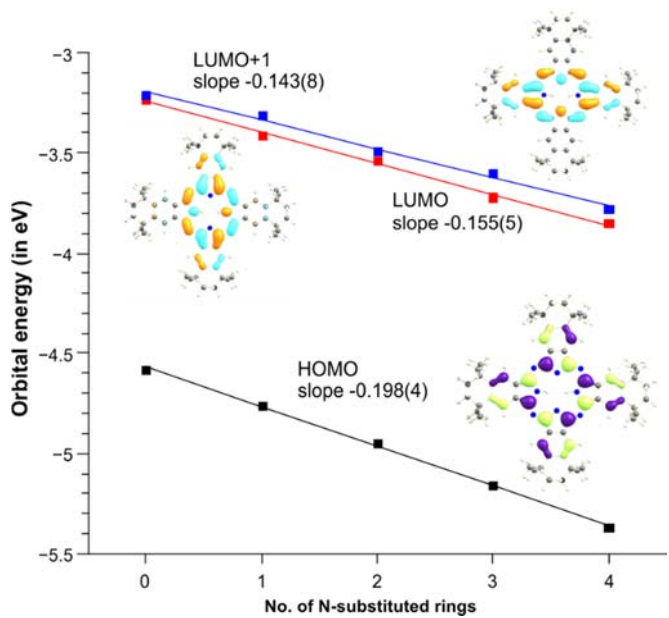

**Figure 10**: Trend of experimental (black), TD-DFT (red), and $D_{H-L}$ (blue) absorption maxima with linear regression curves for $Pc^*H_2$.

In experiment, the tautomeric equilibrium will result in an averaging of this difference and thus lead to only one, broader excitation in addition to the usual broadening effects in UV/Vis spectra.

The difference between frontier orbital energies (ΔHOMO/LUMO and ΔHOMO/LUMO+1 in Table SI-1) gives much higher values than the computed excitation energies due to the missing orbital relaxation in this estimate. Nevertheless, the trend is the same compared to the TD-DFT results: increase of the gap with increased nitrogen content. The comparison of experimental excitation energies and computed ones from TD-DFT and the ΔH-L approach is shown in Fig. 10. The experimental trend of increasing excitation energies with increasing nitrogen content is reproduced by both computational approaches. But both approaches fail to quantitatively reproduce the experimental slope and the absolute excitation energies, while the more accurate TD-DFT approach gives results in better agreement with experiment.

This indicates that both computational approaches are appropriate for deriving trends in excitation energies for substituted phthalocyanines but further improvement is needed to achieve a quantitative agreement.

Since the frontier orbital gap is a suitable description for the excitation energies, the question arises if the excitation energy trend observed in the aza-substituted phthalocyanines is due to a change in the energy levels of occupied or unoccupied orbitals. The orbital energies for the lowest-energy structures of the series of compounds investigated is shown in Fig. 10. While LUMO and LUMO+1 show a similar slope, the occupied HOMO shows a quicker decrease in energy upon nitrogen substitution (larger slope). Thus, lowering the HOMO energy is the determining factor for the experimentally observed trend in the excitation energies.

The dipole moment μ and polarizability α are crucial quantities for the application of phthalocyanines. Both values are listed in Table SI-1 for the series of compounds investigated. The observed dipole moment of up to 1.25 a.u. can easily be rationalized by looking at the structure of the compounds investigated. The N-substituted rings show higher electron density and are thus always at the negative end of the dipole moment vector. In case of more than one ring being N-substituted, the direction of the dipole moment can be derived from vector addition. The polarizability of the compounds is also included and shows a slight increase upon substitution of C-H groups by the more polarizable N atom. Every aza-substituted ring adds approx. 23 a.u. to the isotropic polarizability and the trend is irrespective of hydrogen or Zn being in the center of the molecule.

## Conclusions

We presented the first systematic study on the synthesis, the spectroscopic, electroanalytical, photophysical, structural, and computational analysis of three complete series of azaphthalocyanines and azasubphthalocyanines: $N_x$-$Pc^*H_2$ or $N_x$-$[Pc^*Zn]$ (n = 0, 2, 4, 6, 8), and $N_x$-$[Spc^*BCl]$ (n = 0, 2, 4, 6), respectively.

Whereas non-aggregating soluble $A_4$-type phthalocyanines and $B_4$-type pyrazinoporphyrazines (octaazaphthalocyanines) are easily isolated in pure form and well studied in literature, this paper is dedicated to the unsymmetrical hybrid chromophores of $A_{4-n}B_n$ and $A_{3-n}B_n$-type. The effect of increasing number of [–CH=] units being stepwise substituted by [–N=] units on photophysical (UV-Vis, fluorescence) and electrochemical (CV) properties were correlated with the results of computational calculations (TD-DFT). This study offers insight to predict trends of the absolute HOMO and LUMO levels and of the HOMO-LUMO gap, of the absolute chromophore dipole moment and polarizability, and maybe even of their singlet-oxygen versus fluorescence quantum yield by chemical design. These are fundamental properties relevant to applications of such chromophores as sensitizers in dye sensitized solar cells (DSSCs)[33], their injection of electrons into a wide band gap inorganic semiconductor CB[34] on the one hand, or as singlet oxygen generators in PDT[35] by energy transfer on the other hand. All chromophores described herein are suitable for organic molecular beam epitaxy, they can be sublimed in HV without decomposition while parent azaphthalocyanines, e.g. $PpzH_2$ or PpzZn, do not sublime without decomposition.

The synthesis of this class of weakly aggregating, soluble azaphthalocyanines, $N_x$-$Pc^*M$ and $N_x$-$[Spc^*BCl]$, has been greatly improved by introducing a highly selective catalytic aerobic oxidation of the 1,1,4,4,5,6-hexamethyltetraline to corresponding phthalic acid key building block. For the first time, insight into the molecular and lattice structures of symmetrical 2,2,5,5-tetramethylcyclohexane substituted [Spc*BCl] and Sppz [Sppz*BCl], as well as of lower symmetry N$_2$-Pc*H$_2$ and A$_2$B$_2$-type N$_4$-[Pc*Zn] was gained via single crystal XRD.

Upon increasing number of [–N=] units replacing [–CH=] a linear increase of the HOMO-LUMO gap and the excitation energies was observed. The optical HOMO-LUMO levels were correlating with the potentials determined by CV. Computational analysis gave the correct trend regarding the increase of excitation energies upon aza-substitution with ground state (DFT) and excited state (TD-DFT) methods although quantitative agreement needs refined methods. The trend can be traced back to the HOMO energy rising quicker than the LUMO and LUMO+1 energies. Transitions from the HOMO into these two orbitals determines the Q-bands. The trends in computed dipole moments and polarizabilities are in line with the substitution patterns. Strong dipole moments are observed for asymmetrically substituted N$_x$-[Pc*M] molecules while polarizability increases with the increasing nitrogen content. Both methods – frontier orbital gap and time-dependent density functional theory – provide the right trends but fails to reproduce the absolute values observed in experiment.

The photophysical determination of fluorescence versus singlet oxygen quantum yield evidenced that a fast non-radiative relaxation pathway occurs for the metal-free derivatives N$_x$-Pc*H$_2$ and that the feasibility of this process clearly increases with number of nitrogen atoms in the macrocycle core. For the N$_x$-[Spc*BCl] and N$_x$-[Pc*Zn] series, the photophysical data revealed essentially 100% quantum yield as sum of fluorescence quantum yields ($\Phi_F$) and singlet oxygen quantum yields ($\Phi_\Delta$), non-radiative decay paths are suppressed. Thus fine tuning of absorption and emission maxima was demonstrated in these two series of dyes.

## Experimental Section

6,7-Dicyano-1,1,4,4-tetramethyltetraline (PN* **1**),[18] 5,5,8,8-tetramethyl-5,6,7,8-tetrahydro-chinoxaline-2,3-carbodinitrile (PyzDN* **2**),[12,36] and precursors were prepared according to literature known procedures. The final dinitrils **1** and **2** were purified by using column chromatography or, if necessary, sublimed before use. nOctanol was obtained from Merck, dried according to standard methods and all reactions were handled using standard SCHLENK techniques.

The electronic UV/Vis spectra were recorded on an Avantes AvaSpec-2048 spectrometer in a glovebox. IR spectra were recorded on a BrukerAlpha FT-IR spectrometer with an ATR measurement setup (diamondcell) in a glovebox using neat samples. APCI-HR mass spectra were measured with a Finnigan LTQ-FT spectrometer using dichloromethane/acetonitrile as solvent. $^1$H and $^{13}$C NMR spectra were recorded on a Bruker AVII 300 MHz. X-ray diffraction (XRD) analyses was performed on a QUEST area detector system using Mo Kα radiation (λ= 71.073 pm) at 100 K. Stoe Quest software was used for integration and data reduction; structure solution and refinement was done with the WinGX program suite using BRUKER SAINT, XT V2014/1 and SHELX-2014/7. Molecular graphics were produced with Diamond 4.0. THE SQUEEZE routine of the PLATON program package was used in these cases to remove the corresponding delocalized electron density from the data sets.

Cyclovoltametric measurements were carried out with an rhd-instruments setup, using a closed microcell in glovebox. As working electrode a Pt electrode was used, a gold or Pt counter electrode as well as an Ag reference electrode. The compounds have been measured in a 0.5 M TBAF in DCM solution using 5 mmol N$_x$-Pc*H$_2$. All measurements have been carried out using a 0.1 mmol Fc in DCM as reference.

**Synthesis of 1,1,4,4-tetramethyltetraline-6,7-dicarboxylic acid**
0.05 g NaBr (486 μmol), 125 mg [Co(OAc)$_2$]·4H$_2$O (502 μmol), 3.00 g 1,1,4,4,6,7-hexamethyltetraline (13.9 mmol, 1 eq) were added in an autoclave. The autoclave was charged with 20 bar O$_2$ heated to 180 °C, 90 min. The pressure increased to 30 bar, dropped down to 20 bar. Autoclave was charged as often as necessary, until the pressure stayed the same. Carboxylic acid was precipitated out of the solution and washed with water. A second/third portion was yielded overnight, filtered, washed with water and finally dried in vacuum.
**Yield**: 97%. - **$^1$H NMR** (300 MHz, [D$_6$]DMSO, 25 °C): $\delta$ = 13.07 (s, 2 H, COO*H*), 7.57 (s, 2 H, Ar-*H*), 1.66 (s, 4 H, C*H*$_2$), 1.25 (s, 12 H, C*H*$_2$) ppm. - **MS** (ESI(+), MeOH): m/z = 277.1436, cal. for C$_{16}$H$_{20}$O$_4$+H$_1$: 277.1434. **Additional information**: Depending on the amount of glacial acid, the formation of the anhydride can be observed, which is the product of the following step in the synthesis of PN*. An up-scaling of the reaction is possible.

**General procedure for N$_x$-[Spc*BCl]:** 1 eq PN* **1** (or PyzDN* **2**, or as a mixture) were dissolved in 2 mL toluene per 5 mmol dinitril. Fresh boron trichloride solution (1 N in heptane or in o-xylene, 1 eq) was added at once and then heated to 150 °C in a closed Schlenk tube. The yellowish solution turned into a deep pink. After heating for 2-5 h the solution was added on a short pluck of silica to remove insoluble byproducts. Spcs were purified by a column chromatography with silica gel 60 Merck and eluted with PE/EE 10:1.

**[Spc*BCl]: Yield:** 13%. - **R$_f$** (Tol/THF 30:1) = 0.27. - **$^1$H NMR** (300 MHz, C$_6$D$_6$, 25 °C): $\delta$ = 8.91 (s, 6 H, Ar-C*H*), 1.65-1.52 (m, 12 H, C*H*$_2$), 1.42 (s, 18 H, C*H*$_3$), 1.10 (s, 18 H, C*H*$_3$) ppm. - **$^1$H NMR** (300 MHz, [D$_2$]DCM, 25 °C): $\delta$ = 8.84 (s, 6 H, Ar-C*H*), 1.94-1.80 (m, 12 H, C*H*$_2$), 1.66 (s, 18 H, C*H*$_3$), 1.44 (s, 18 H, C*H*$_3$) ppm. - **$^{13}$C NMR** (75 MHz, CDCl$_3$, 25 °C): $\delta$ = 150.0, 148.3, 130.0, 120.6, 35.5, 35.3, 32.6, 32.1, 30.1 ppm. - **IR:** $\tilde{v}$ = 2959 (m), 2925 (m), 2861 (m), 1729 (w), 1623 (w), 1466 (m), 1259 (m), 1088 (s), 796 (s), 662 (m) cm$^{-1}$. - **UV/Vis** (DCM): $\lambda$ = 582 (s), 528 (sh), 318 (s), 267 (s) nm. - **Fluorescence** (DCM, $\lambda_{ex}$ = 350 nm): $\lambda$ = 592 nm. - **MS** (APCI-HRMS(+)): m/z = 761.4247 [M+H]$^+$, cal. for C$_{48}$H$_{54}$B$_1$Cl$_1$N$_6$+H: 761.4272.

**[Sppz*BCl]: Yield:** 35%. - **R$_f$** (Tol/THF 30:1) = 0.35. - **$^1$H NMR** (300 MHz, CDCl$_3$, 25 °C): $\delta$ = 1.94-2.07 (m, 12 H, C*H*$_2$), 1.75 (s, 18 H, C*H*$_3$), 1.52 (s, 18 H, C*H*$_3$) ppm. - **$^{13}$C NMR** (75 MHz, CDCl$_3$, 25 °C): $\delta$ = 162.9, 146.2, 140.7, 39.3, 34.0, 30.2, 30.8 ppm. - **IR:** $\tilde{v}$ = 2956 (m), 2922 (m), 2859 (m), 1708 (w), 1557 (m), 1454 (s), 1371 (s), 1356 (m), 1243 (vs), 1107 (s), 1045 (s), 798 (s), 733 (m) cm$^{-1}$. - **UV/Vis** (DCM): $\lambda$ = 534 (s), 489 (sh), 334 (m), 303 (s) nm. -

**Fluorescence** (DCM, $\lambda_{ex}$ = 350 nm): $\lambda$ = 546 nm. - **MS** (APCI-HRMS(+)): m/z = 767.3964 [M+H]$^+$, cal. for $C_{42}H_{49}B_1Cl_1N_{12}$+H: 767.3980.

**N$_4$-[Spc*BCl]: Yield:** 2%. - **$^1$H NMR** (300 MHz, CDCl$_3$, 25 °C): $\delta$ = 8.76 (s, 2 H, ArCH), 1.76-1.48 (m, 12 H, CH$_2$), 1.67 (s, 6 H, CH$_3$), 1.65 (s, 6 H, CH$_3$), 1.64 (s, 6 H, CH$_3$), 1.36 (s, 6 H, CH$_3$), 1.36 (s, 6 H, CH$_3$), 1.34 (s, 6 H, CH$_3$) ppm. - **IR** (ATR, 400-4000 cm$^{-1}$): $\tilde{v}$ = 2957 (m), 2869 (w), 1455 (m), 1253 (m), 1088 (s), 1020 (s), 797 (s), 701 (m), 516 (m) cm$^{-1}$. - **UV/Vis** (DCM): $\lambda$ = 557 (s), 506 (sh), 330 (m), 296 (s) nm. - **MS** (APCI-HRMS(+)): m/z = 765.4075 [M+H]$^+$, cal. for $C_{44}H_{51}B_1Cl_1N_{10}$+H: 765.4082.

**N$_2$-[Spc*BCl]: Yield:** 3%. - **$^1$H NMR** (300 MHz, CDCl$_3$, 25 °C): $\delta$ = 8.87 (s, 2 H, ArCH), 8.78 (s, 2 H, ArCH), 1.73-1.50 (m, 12 H, CH$_2$), 1.67 (s, 6 H, CH$_3$), 1.42 (s, 6 H, CH$_3$), 1.38 (s, 6 H, CH$_3$), 1.36 (s, 6 H, CH$_3$), 1.05 (s, 6 H, CH$_3$), 1.03 (s, 6 H, CH$_3$) ppm. - **IR** (ATR, 400-4000 cm$^{-1}$): $\tilde{v}$ = 2960 (m), 2925 (w), 2861 (w), 1459 (w), 1259 (s), 1090 (s), 1016 (s), 791 (s), 695 (m) cm$^{-1}$. - **UV/Vis** (DCM): $\lambda$ = 572 (s), 519 (sh), 327 (m), 303 (s) nm. - **MS** (APCI-HRMS(+)): m/z = 763.4173 [M+H]$^+$, cal. for $C_{46}H_{53}B_1Cl_1N_8$+H: 763.4177.

The N$_x$-[Spc*BCl] are unstable towards light in solution and should be kept in the dark under inert gas. In the $^1$H NMR N$_x$-[Spc*BCl] free dinitril of degredated N$_x$-[Spc*BCl] was observed which could not be separated.

**Synthesis of N$_x$-Pc*H$_2$** 240 mg lithium was dissolved in 10 mL *n*octanol at 120 °C within 20 min. After cooling down to rt, 1.0 g PN* **1** (4.19 mmol, 2 eq) and 1.0 g PyzDN* **2** (4.16 mmol, 2 eq) were added and stirred for 10 h at 180 °C. The solutions had a change in color to deep green. After cooling down to rt, 100 mL MeOH and 3 mL H$_3$PO$_4$ were added. The solution was further washed 3 x 100 mL with MeOH. The product mixture was separated using gradient column chromatography with PE/EE 10:1 → 1:1. First, Pc*H$_2$ **3a** and N$_2$-Pc*H$_2$ **4a** was eluted, then the ABAB and A$_2$B$_2$ N$_4$-Pc*H$_2$ **5a**, **6a** was eluted. After changing the solvent to PE/EE 1:1 N$_6$-Pc*H$_2$ **7a** was eluted. Finally Ppz*H$_2$ **8a** was eluted with DCM.

**$^{(1)}$Pc*H$_2$: Yield:** 20 mg, 20.9 mmol, 1%. - **$^1$H NMR** (300 MHz, CDCl$_3$, 25 °C): $\delta$ = 9.91 (s, 8 H, Ar-H), 1.79 (s, 16 H, CH$_2$), 1.54 (s, 48 H, CH$_3$), -0.01 (s, 2 H, NH) ppm. - **UV/Vis** (DCM): $\lambda$ = 710 (s), 677 (s), 647 (sh), 615 (sh), 343 (s), 295 (sh), 231 (sh) nm. - **MS** (APCI-HRMS(+)): m/z = 955.6096 [M+H]+, cal. for $C_{64}H_{75}N_8$+H: 955.6109.

**N$_2$-Pc*H$_2$: Yield:** 48 mg, 50.1 mmol, 3%. - **$^1$H NMR** (300 MHz, CDCl$_3$, 25 °C): $\delta$ = 9.69 (s, 2 H, Ar-H), 9.65 (s, 2 H, Ar-H), 9.42 (s, 2 H, Ar-H), 2.19 (s, 4 H, CH$_2$), 2.07 (s, 8 H, CH$_2$), 2.04 (s, 4 H, CH$_2$), 1.90 (s, 12 H, CH$_3$), 1.83 (s, 12 H, CH$_3$), 1.81 (s, 12 H, CH$_3$), 1.81 (s, 12 H, CH$_3$), -0.07 (s, 2 H, NH) ppm. - **$^{13}$C NMR** (75 MHz, C$_6$D$_6$, 25 °C): $\delta$ = 160.6, 149.8, 149.2, 148.6, 137.0, 133.0, 132.2, 122.2, 121.6, 39.1, 36.1, 36.1, 35.9, 35.4, 35.3, 34.7, 32.9, 32.8, 32.7, 30.9 ppm. - not all quartary atoms could be detected. - **IR** (ATR, 400-4000 cm$^{-1}$): $\tilde{v}$ = 2956 (m), 2921 (s), 2855 (m), 1711 (w), 1688 (w), 1498 (m), 1301 (s), 1071 (s), 1022 (m), 755 (s), 722 (m) cm$^{-1}$. - **UV/Vis** (DCM): $\lambda$ = 687 (s), 619 (sh), 340 (s), 306 (s), 232 (s) nm. - **MS** (APCI-HRMS(+)): m/z = 957.5992 [M+H]$^+$, cal. for $C_{62}H_{72}N_{10}$+H: 957.6014.

**ABAB N$_4$-Pc*H$_2$: Yield:** <1%. - **R$_f$** (PE/EA 4:1) = 0.53. - **$^1$H NMR** (CDCl$_3$, 300 MHz): $\delta$ = 9.79 (s, 4 H, Ar-H), 2.22 (s, 8 H, CH$_2$), 2.11 (s, 8 H, CH$_2$), 1.90 (s, 24 H, CH$_3$), 1.85 (s, 24 H, CH$_3$), -0.68 (s, 2 H, NH) ppm. **$^{13}$C NMR** (C$_6$D$_6$, 75 MHz): $\delta$ = 29.8, 33.0, 36.5 ppm. - **UV/Vis** (DCM): $\lambda$ = 672 (s), 642 (sh), 607 (s), 347 (s), 309 (s) nm. - **MS** (APCI-HRMS(+)): m/z = 959.5922 [M+H]$^+$, cal. for $C_{60}H_{70}N_{12}$+H$_1$: 959.5919.

**A$_2$B$_2$ N$_4$-Pc*H$_2$: Yield:** 342 mg, 357 mmol, 17%. - **$^1$H NMR** (300 MHz, CDCl$_3$, 25 °C): $\delta$ = 9.56 (s, 2 H, Ar-H), 9.50 (s, 2 H, Ar-H), 2.20 (s, 8 H, CH$_2$), 2.05 (s, 8 H, CH$_2$), 1.91 (s, 24 H, CH$_3$), 1.81 (s, 12 H, CH$_3$), 1.80 (s, 12 H, CH$_3$), -0.04 (s, 2 H, NH) ppm. - **$^{13}$C NMR** (75 MHz, C$_6$D$_6$, 25 °C): $\delta$ = 162.4, 150.0, 149.6, 145.1, 144.4, 135.2, 134.8, 122.7, 122.0, 39.4, 39.3, 36.1, 36.0, 35.2, 34.6, 34.5, 32.7, 32.7, 21.1, 30.8 ppm. - **IR** (ATR, 400-4000 cm$^{-1}$): $\tilde{v}$ = 2915 (m), 2857 (m), 1358 (s), 1328 (s), 1148 (s), 1128 (s), 761 (m), 752 (m), 679 (s), 543 (w) cm$^{-1}$. - **UV/Vis** (DCM): $\lambda$ = 680 (s), 656 (s), 625 (sh), 348 (s), 232 (s) nm. - **MS** (APCI-HRMS(+)): m/z = 959.5921 [M+H]$^+$, cal. for $C_{60}H_{70}N_{12}$+H: 959.5919.

**N$_6$-Pc*H$_2$: Yield:** 178 mg, 185 mmol, 9%. - **$^1$H NMR** (300 MHz, CDCl$_3$, 25 °C): $\delta$ = 9.75 (s, 2 H, Ar-H), 2.22 (s, 4 H, CH$_2$), 2.19 (s, 12 H, CH$_2$), 1.93 (s, 12 H, CH$_3$), 1.91 (s, 12 H, CH$_3$), 1.90 (s, 12 H, CH$_3$), 1.83 (s, 12 H, CH$_3$), -0.53 (s, 2 H, NH) ppm. - **$^{13}$C NMR** (75 MHz, C$_6$D$_6$, 25 °C): $\delta$ = 162.0, 139.1, 131.3, 134.1 (w), 132.9 (w), 132.9 (w), 131.3 (w), 127.1 (w), 123.1, 46.0, 41.3 (w), 39.7, 39.3, 36.3, 34.5, 32.8, 31.8, 31.7, 31.1, 30.9, 30.8, 29.8, 29.5, 29.3, 29.2, 28.6, 28.1, 27.5, 27.5 ppm. - not all quartary atoms could be detected. - **IR** (ATR, 400-4000 cm$^{-1}$): $\tilde{v}$ = 2957 (s), 2924 (s), 1727 (s), 1693 (w), 1511 (s), 1457 (m), 1411 (m), 1072 (m), 1021 (s), 756 (w), 744 (m), 677 (s) cm$^{-1}$. - **UV/Vis** (DCM): $\lambda$ = 660 (s), 597 (sh), 343 (s), 234 (s) nm. - **MS** (APCI-HRMS(+)): m/z = 961.5825 [M+H]$^+$, cal. for $C_{58}H_{68}N_{14}$+H: 961.5824.

**$^{(1)}$Ppz*H$_2$: Yield:** 174 mg, 181 mmol, 9%. - **$^1$H NMR** (300 MHz, CDCl$_3$, 25 °C): $\delta$ = 2.22 (s, 16 H, CH$_2$), 1.93 (s, 48 H, CH$_2$), -1.02 (s, 2 H, NH) ppm. – **UV/Vis** (DCM): $\lambda$ = 655 (s), 621 (s), 576 (s), 343 (s), 231 (s) nm. - **MS** (APCI-HRMS(+)): m/z = 963.5732 [M+H]$^+$, cal. for $C_{56}H_{66}N_{16}$+H: 963.5729.

**Additional Information** (1): The characterization of the Pc*H$_2$ and Ppz*H$_2$ results of a separate synthesis; here the compounds only have been identified by APCI-HRMS and UV-Vis spectroscopy.

**Ring Expansion of a Spc** N$_6$-Pc*H$_2$ and N$_4$-Pc*H$_2$ could be obtained by inserting 1.1 eq isoindoline of PN* **1** in the corresponding Spc [Sppz*BCl] or N$_4$-[Spc*BCl]. Therefore, the solution was stirred in DMSO at 120 °C, 8 h. After removing the solvent under reduced pressure the product was purified by column chromatography PE/EE 1:1.

The N$_6$-/N$_4$-Pc*H$_2$ could be identified by using $^1$H NMR, UV-Vis spectroscopy and APCI-HRMS. All data correlated with the ones described above.

**Synthesis of [N$_x$-Pc*Zn]** 1 eq N$_x$-Pc*H$_2$ was dissolved in 2 mL toluene/5 *m*mol Pc. 1.1 eq [Zn(hmds)$_2$] was added at once. The solution was stirred for 30 min at 75 °C until the conversion to N$_x$-[Pc*Zn] was completed (NMR control). The solvent was evaporated as well as the rest of the HHMDS and [Zn(hmds)$_2$] and finally washed out with Et$_2$O. The compounds were dried in vacuum.

**N$_2$-[Pc*Zn]: Yield:** 80%. - **$^1$H NMR** (300 MHz, C$_6$D$_6$, 25 °C): $\delta$ = 9.97 (s, 4 H, Ar-H), 9.94 (s, 2 H, Ar-H), 1.92 (s, 4 H, CH$_2$), 1.86 (s, 12 H, CH$_3$), 1.80 (s, 12 H, CH$_2$), 1.56 (s, 24 H, CH$_3$), 1.46 (s, 12 H, CH$_3$) ppm. - **UV/Vis** (DCM): $\lambda$ = 675 (s), 613 (sh), 360 (s),

228 (s) nm. - **Fluorescence** (DCM, $\lambda_{ex}$ = 350 nm) $\lambda$ = 697 nm. - **MS** (APCI-HRMS(+)): $m/z$ = 1019.5139 [M+H]$^+$, cal. for $C_{62}H_{70}N_{10}$+H: 1019.5149.

**ABAB N$_4$-[Pc$^*$Zn]**: $R_f$ (Tol/THF 20:1) = 0.63. - **$^1$H NMR** ($C_6D_6$, 300 MHz): $\delta$ = 9.96 (s, 4 H, Ar-*H*), 1.91 (s, 8 H, C*H*$_2$), 1.85 (s, 24 H, C*H*$_3$), 1.78 (s, 8 H, C*H*$_2$), 1.45 (s, 25 H, C*H*$_3$) ppm. - **UV/Vis** (DCM): $\lambda$ = 687 (s), 657 (s), 631 (sh), 596 (sh), 357 (s) nm. - **MS** (APCI-HRMS(+)): $m/z$ = 1021.5073 [M+H]$^+$, cal. for $C_{60}H_{68}N_{12}Zn_1$+H$_1$: 1021.5054.

**A$_2$B$_2$ N$_4$-[Pc$^*$Zn]**: Yield: 71%. - **$^1$H NMR** (300 MHz, $C_6D_6$, 25 °C): $\delta$ = 9.93 (s, 2 H, Ar-H), 9.91 (s, 2 H, Ar-H), 1.91 (s, 8 H, C*H*$_2$), 1.83 (s, 8 H, C*H*$_2$), 1.78 (s, 24 H, C*H*$_3$), 1.53 (s, 12 H, C*H*$_3$), 1.43 (s, 12 H, C*H*$_3$) ppm. - **$^1$H NMR** (300 MHz, [D$_2$]DCM, 25 °C): $\delta$ = 9.59 (s, 2 H, Ar-H), 9.56 (s, 2 H, Ar-H), 2.24 (s, 8 H, C*H*$_2$), 2.08 (s, 8 H, C*H*$_2$), 1.93 (s, 12 H, C*H*$_3$), 1.92 (s, 12 H, C*H*$_3$), 1.83 (s, 12 H, C*H*$_3$), 1.82 (s, 12 H, C*H*$_3$) ppm. - **UV/Vis** (DCM): $\lambda$ = 666 (s), 601 (sh), 361 (s) nm. - **Fluorescence** (DCM, $\lambda_{ex}$ = 350 nm) $\lambda$ = 680 nm. - **MS** (APCI-HRMS(+)): $m/z$ = 1021.5043 [M+H]$^+$, cal. for $C_{60}H_{68}N_{12}$+H: 1021.5054.

**N$_6$-[Pc$^*$Zn]**: Yield: 78%. - **$^1$H NMR** (300 MHz, $C_6D_6$, 25 °C): $\delta$ = 9.96 (s, 2 H, Ar-H), 1.89 (s, 12 H, C*H*$_2$), 1.83 (s, 12 H, C*H*$_3$), 1.79 (s, 24 H, C*H*$_3$), 1.76 (s, 4 H, C*H*$_2$), 1.43 (s, 12 H, C*H*$_3$) ppm. - **UV/Vis** (DCM): $\lambda$ = 662 (s), 649 (s), 357 (s), 227 (s) nm. - **Fluorescence** (DCM, $\lambda_{ex}$ = 350 nm) $\lambda$ = 673 nm. - **MS** (APCI-HRMS(+)): $m/z$ = 1023.4959 [M+H]$^+$, cal. for $C_{58}H_{66}N_{14}$+H: 1023.4959.

**Photophysical measurements**. The steady-state fluorescence spectra were measured using an AMINCO-Bowman Series 2 luminescence spectrometer. For singlet oxygen and fluorescence determination as well as the titration experiments, the dye samples were purified by preparative TLC on Merck aluminum sheets coated with silica gel 60 F254 to ensure high purity. Both the $\Phi_\Delta$ and $\Phi_F$ values were determined by comparative methods using unsubstituted zinc phthalocyanine (ZnPc) as a reference ($\Phi_{\Delta(THF)}$ = 0.53, $\Phi_{F(THF)}$ = 0.32) for N$_x$-[Pc*M]. The photophysical data for N$_x$-[Spc$^*$BCl] series were determined with rhodamin G6 ($\Phi_{F(EtOH)}$ = 0.94) and bengal rose ($\Phi_{\Delta(MeOH)}$ = 0.76) as reference compounds and long-pass filter FGL455 instead of OG530 for $\Phi_\Delta$ determination. All of the experiments were performed in triplicate, and the data represent the means of the measurements (estimated error ±10%). For $\Phi_F$ determination, the sample and reference were excited at 598 nm and 490 nm for N$_x$-[Pc*M] and N$_x$-[Spc$^*$BCl], respectively, and the absorbance at the Q-band maximum was maintained at less than 0.05 to avoid the inner filter effect. The emission spectra were corrected for the instrument response. The $\Phi_F$ values were corrected for the refractive indices of the solvents. The time-resolved fluorescence measurements were performed on FS5 fluorometer (Edinburgh Instruments) with picosecond pulsed diode lasers excitation at 371 nm (EPL-375, pulse width ~ 68 ps) and 634 nm (EPL-635, pulse width ~ 81 ps). The decay curves were fitted to exponential functions with Fluoracle software (Edinburgh Instruments).

**Computational details.** Unconstrained structural optimizations of the compounds investigated were carried out with density functional theory (DFT), using the PBE[37] exchange-correlation functional together with the def2-TZVPP[38] basis set for all elements adding a semiempirical dispersion correction term (DFT-D3).[39] Orbital energies were derived on this level. The Gaussian09[40] optimizer together with Turbomole 6.5[41] energies and gradients was used. The resolution-of-identity[42] and the multipole accelerated RI-J[43] approximation was applied throughout. The excitation energies were derived from time-dependent DFT computations using B3LYP[44]/def2-TZVPP and the Tamm-Dancoff approximation as implemented in the ORCA[45] package. The RIJCOSX approximation was used. Polarizabilities and dipole moments were derived with PBE/def2-TZVPP with ORCA.

## Acknowledgements

This work was funded by DFG within SFB 1083. The singlet oxygen study was supported by LOEWE Priority Program SynChemBio. We also would like to thank J. Wallauer and rhd-instruments for their advice in our CV measurements.

Supporting Information

to

# Experimental and Computational Study of Isomerically Pure Soluble Azaphthalocyanines and Azasubphthalocyanines of Varying Number of Aza Units

Martin Liebold[a], Eugen Sharikow[a], Elisabeth Seikel[a], Lukas Trombach[b], Klaus Harms[b], Petr Zimčík[c], Veronika Nováková[c], Ralf Tonner[b,*], and Jörg Sundermeyer[a,*] [a] *Fachbereich Chemie and Wissenschaftliches Zentrum für Materialwissenschaften (WZMW), Philipps-Universität Marburg, Hans-Meerwein-Straße 4, 35032 Marburg, Germany.*
  *e-mail: jsu@chemie.uni-marburg.de*
[b] *Fachbereich Chemie, Philipps-Universität Marburg, Hans-Meerwein-Straße 4, 35032 Marburg, Germany.*
  *e-mail: tonner@chemie.uni-marburg.de*
[c] *Department of Pharmaceutical Chemistry and Drug Control, Faculty of Pharmacy in Hradec Kralove, Charles University in Prague, Heyrovskeho 1203, Hradec Kralove, 500 05 Czech Republic*
  *e-mail: petr.zimcik@faf.cuni.cz and veronika.novakova@faf.cuni.cz*

**Table of Contents**

## 1.1 Experimental Section

All reactions in which high pressure was needed, were carried out in an autoclave. Therefore, a glass (pressure <10 bar) or a steel autoclave (<100 bar) was used. The 100 mL stainless steel autoclaves *V4A-Edelstahlautoklaven* were constructed by the precision engineering department of the Philipps-Universität Marburg (Department of Chemistry). Solids and liquids were weighed out in a 100 mL teflon flask or in the corresponding glass flask, a magnetic stirrer was added and the autoclave was closed and sealed using EPDM-O-rings of CLEFF DICHTUNG company. The actual pressure was regulated using a manometer. The temperature was adjusted using a corresponding aluminium block in which the thermometer of the magnetic stirrer MR 3003 HEIDOLPH could be added (Pt-100-thermal element). If necessary, the heating block might be preheated (for Pc reactions). If necessary, the glass autoclaves were heated in an oil bath. A sketch of the steel autoclave is displayed in **Fehler! Verweisquelle konnte nicht gefunden werden.**.

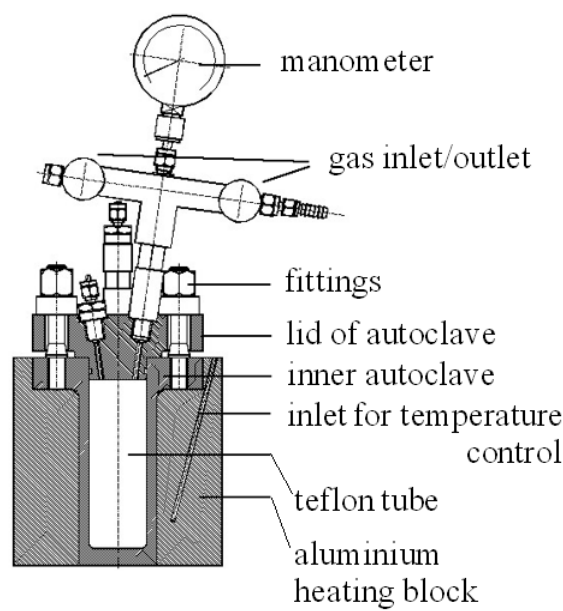

Sketch of the used steel autoclave: *V4A-stainless steel autoclave*.

## 1.1.1 Synthesis of PDN*

The synthesis of the PDN* **1** was carried out according to MIKHALENKO.[1] The synthesis of 1,1,4,4-tetramethyltetralin-6,7-dicarboxylic acid (VI) was modified by using an overpressure of $O_2$ in a $[Co(OAc)_2] \cdot 4H_2O$ catalytic autoclave reaction.

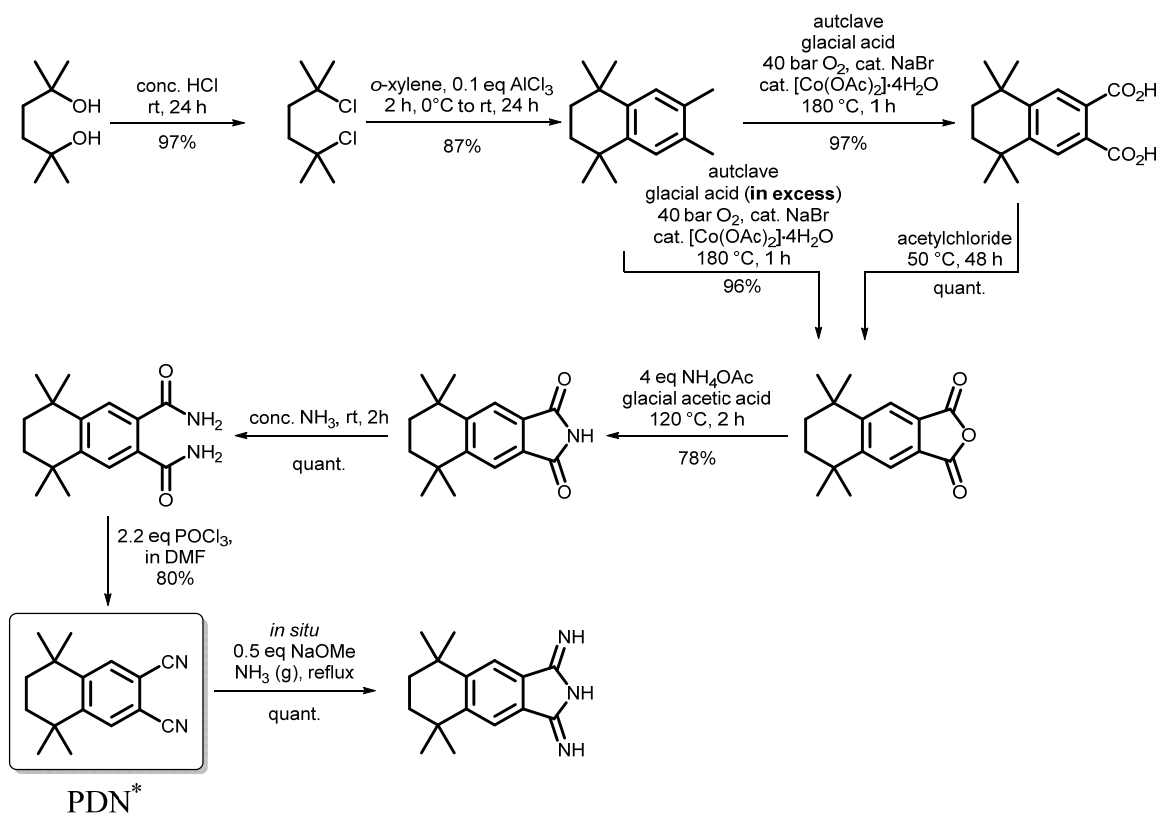

## 1.1.2 Alternative Synthesis of 1,1,4,4-Tetramethyltetralin-6,7-dicarboxylic acid

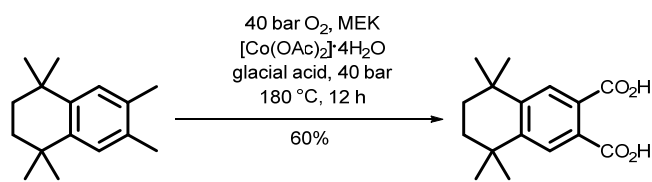

**Method a)** The reaction was carried out in an autoclave reaction as described in method b), using 50 bar $O_2$ and 0.1 eq $[Co(OAc)_2] \cdot 4H_2O$ as catalyst, 5 eq MEK as cocatalyst instead of NaBr, and 5 mL glacial acetic acid /g tetraline as solvent. The mixture was stirred for 12 h while the autoclave was recharged several times with $O_2$, until the pressure stayed constant. Workup was carried out according to method b).

**Yield**: 60% (only carboxylic acid was observed). - **$^1$H NMR**: according to analysis below.

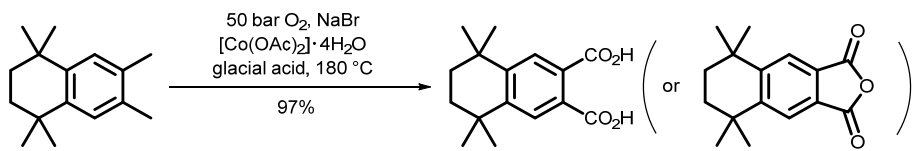

**Method b)** 15 g tetraline (69.3 mmol, 1 eq) were dissolved in 50 mL glacial acid in a 100 mL teflon tube and added into an autoclave (steel autoclave as visible in "Methods" section **Fehler! Verweisquelle konnte nicht gefunden werden.**). The teflon tube was additionally filled-up with 750 mg [Co(OAc)$_2$]·4H$_2$O (3 mmol, 4 mol%) and 300 mg NaBr (3 mmol, 4 mol%). The mixture was briefly stirred until everything was dissolved. The autoclave was charged with O$_2$ to a pressure of 20 bar. With stirring, the solution was heated to 180 °C. After a short period of initiation (<30 min) the reaction started and the pressure decreased. The autoclave was recharged with O$_2$ to 50 bar as often as necessary, until the pressure stayed constant at about 50 bar. After 2 h, the reaction was complete. The reaction may also be stirred overnight. The resulting solution was concentrated under reduced pressure and was washed with water to remove Co-salts and NaBr traces. The product can be additionally precipitated in 20% aq. HCl (500 mL/20 g acid), followed by filtration and washing with water.

**Yield**: 18.5 g, 67 mmol, 97%. - **$^1$H NMR** (DMSO-$d_6$, 300 MHz): $\delta$ = 13.07 (s, 2 H, COO$H$), 7.57 (s, 2 H, Ar-C$H$), 1.66 (s, 4 H, -C$H_2$), 1.25 (s, 12 H, -C$H_3$) ppm. - **MS** (ESI(+), MeOH): $m/z$ = 277.1436, cal. for C$_{16}$H$_{20}$O$_4$+H$_1$: 277.1434. - **Elemental analysis** (C$_{16}$H$_{20}$O$_4$, M = 276.33): fnd. (cal.): C 69.83% (69.55%), H 7.18% (7.30%), N 7.53% (0.00%).

**Additional information**: In some cases, the corresponding anhydride was detected by $^1$H NMR spectroscopy. The amount of anhydride is dependent on the amount of glacial acid used in the reaction in ratio to the tetraline amount. No other by-products could be observed. A complete turnover was determined. The high nitrogen value in elemental analysis is caused by the measurement in closed setup in the glovebox, neither in the synthesis nor in precursor steps were nitrogen containing chemicals used.

### 1.1.3 Alternative Synthesis of the Anhydride of 1,1,4,4-Tetramethyltetralin-6,7-dicarboxylic acid

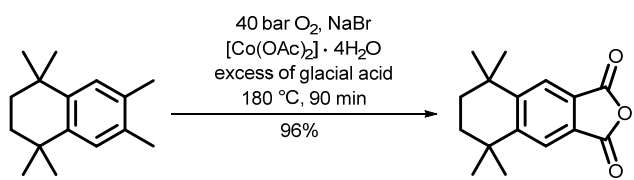

50 mg NaBr (486 µmol, ~3.5 mol%), 125 mg [Co(OAc)$_2$]·4H$_2$O (502 µmol, ~3.5 mol%), 3.00 g tetraline (13.9 mmol, 1 eq) were added into an autoclave and dissolved in 10 mL glacial acid. The autoclave was charged with 20 bar O$_2$ and heated to 180 °C for 90 min. The pressure increased to 30 bar, dropped down to 20 bar and was recharged to 30-40 bar two more times. The autoclave was recharged as often as needed, until the pressure stayed constant. Carboxylic acid anhydride was precipitated out of the solution and washed with water. A second/third portion was yielded overnight, filtered, washed with water and finally dried in vacuum.

**Yield**: 3.43 g, 13.3 mmol, 96%. - **$^1$H NMR** (CDCl$_3$, 300 MHz): $\delta$ = 7.95 (s, 2 H, Ar-C*H*), 1.76 (s, 4 H, -C*H$_2$*), 1.36 (s, 12 H, -C*H$_3$*) ppm.

### 1.1.4 Synthesis of PyzDN*

The synthesis of 3,3,6,6-tetramethylcyclohexane-1,2-dione (**XIV**) was carried out following the procedure of JONES.[2] The synthesis of PyzDN* **2** was carried out according to the procedure of SEIKEL.[3]

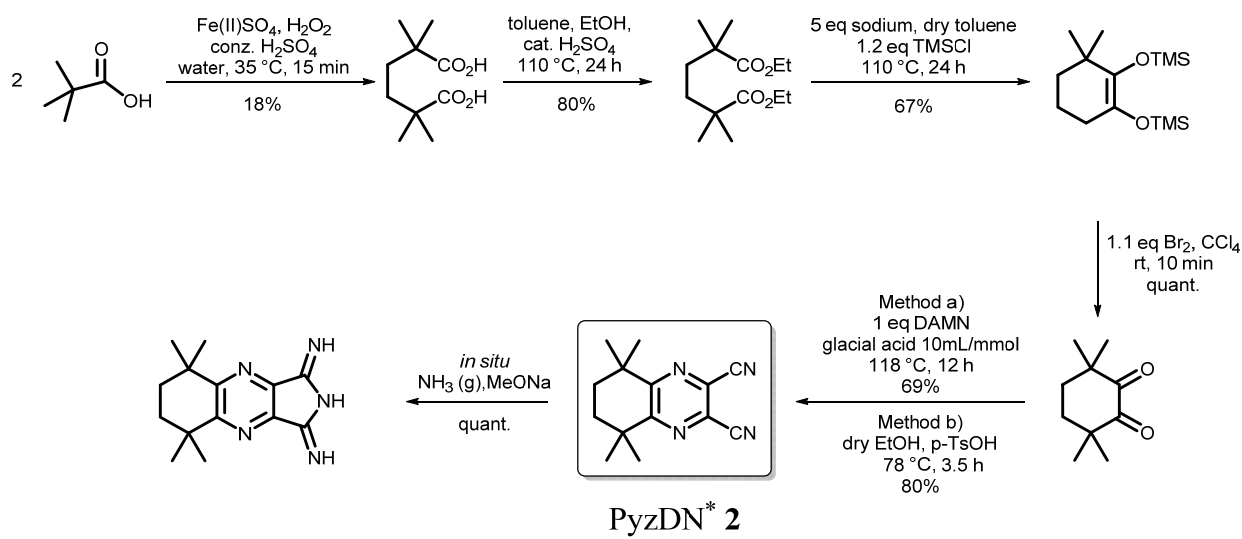

PyzDN* **2**

### 1.1.5 Synthesis of [Spc*BCl] [4,5]

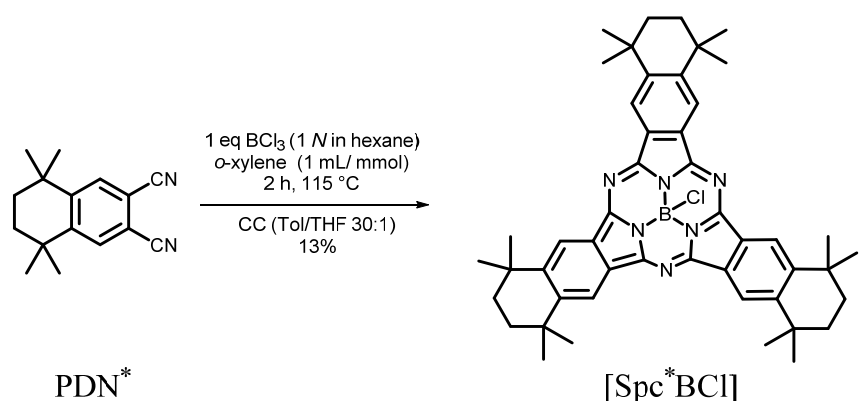

PDN*  [Spc*BCl]

General Procedure 1: Synthesis of [Spc(*)BX].

1 eq PDN was dissolved in 2 mL toluene /5 mmol PDN. Fresh BCl$_3$ solution (1 M in heptane or in *o*-xylene, 3 eq BCl$_3$ per intended [SpcBCl]) was added at once and then heated to 150 °C. The yellowish solution turned deep pink. After heating for about 5 h, the solution was loaded onto a short silica pluck, purified with DCM/EE to remove the boron compounds and then purified by CC (PE/EA 10:1).

**Yield:** ~5%. - $R_f$ (PE/EE 10:1) = 0.4. - **$^1$H NMR** (C$_6$D$_6$, 300 MHz): δ = 8.91 (s, 6 H, Ar-C*H*), 1.65-1.52 (m, 12 H, -C*H$_2$*), 1.42 (s, 18 H, -C*H$_3$*), 1.10 (s, 18 H, -C*H$_3$*) ppm. - **$^1$H NMR** (CD$_2$Cl$_2$, 300 MHz): δ = 8.84 (s, 6 H, Ar-C*H*), 1.94-1.80 (m, 12 H, -C*H$_2$*), 1.66 (s, 18 H, -C*H$_3$*), 1.44 (s, 18 H, -C*H$_3$*) ppm. - **$^{13}$C NMR** (CD$_2$Cl$_2$, 75 MHz): δ = 149.6, 149.2, 129.4, 120.4, 36.0, 35.3, 32.9, 32.3 ppm. - **UV-Vis** (DCM): λ = 582 (s), 528 (sh), 318 (s), 267 (s) nm. - **Fluorescence** (DCM, $λ_{ex}$ = 350 nm): λ = 592 nm. - $Φ_F$ ($λ_{ex}$ = 490 nm) = 0.18. - $Φ_Δ$ = 0.64. - **MS** (APCI-HRMS(+)): *m/z* = 761.4247 [M+H]$^+$, cal. for C$_{48}$H$_{54}$B$_1$Cl$_1$N$_6$+H$_1$: 761.4272. - **X-ray**: Crystals could be obtained out of CD$_2$Cl$_2$ in the NMR tube at rt.

**Additional information**: In some cases, degradation of [Spc*BCl] in the form of a fast decolouration was observed in acidic solvents.

### 1.1.6 Synthesis of [Sppz*BCl] [4,5]

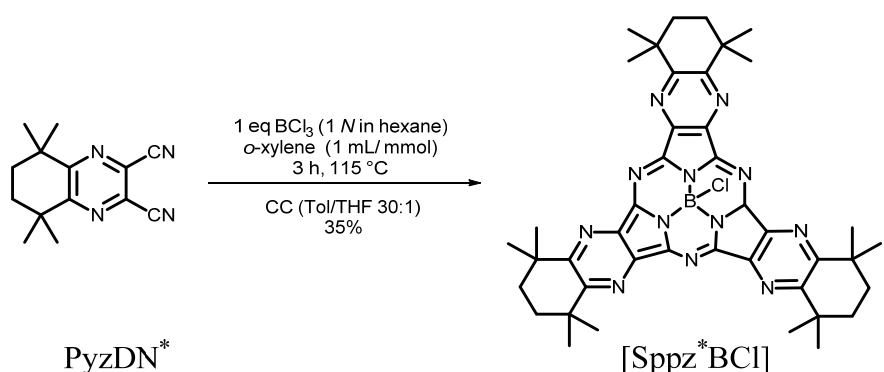

PyzDN*                                         [Sppz*BCl]

According to General Procedure 1 (p. 5). The pink solution was purified by CC (Tol/THF 30:1). The pink solid was dried in vacuum.

**Yield:** 35%. - $R_f$ (Tol/THF 30:1) = 0.35. - $R_f$ (DCM) = 0.6. - **$^1$H NMR** (CDCl$_3$, 300 MHz): δ = 1.94–2.07 (m, 12 H, -C$H_2$), 1.75 (s, 18 H, -C$H_3$), 1.52 (s, 18 H, -C$H_3$) ppm. - **$^1$H NMR** (CD$_2$Cl$_2$, 300 MHz): δ = 1.53 (s, 18 H, -C$H_3$), 1.78 (s, 18 H, -C$H_3$), 1.96–2.16 (m, 12 H, -C$H_2$) ppm. - **UV-Vis** (DCM): λ = 534 (s), 489 (sh), 334 (m), 303 (s) nm. - **Fluorescence** (DCM, $λ_{ex}$ = 350 nm): λ = 546 nm. - $Φ_F$ ($λ_{ex}$ = 490 nm) = 0.19. - $Φ_Δ$ = 0.88. - **MS** (APCI-HRMS(+)): m/z = 767.3964 [M+H]$^+$, cal. for C$_{42}$H$_{48}$B$_1$Cl$_1$N$_{12}$+H$_1$: 767.3980.

### 1.1.7 Synthesis of Azasubphthalocyanines N$_x$-[Spc*BCl]

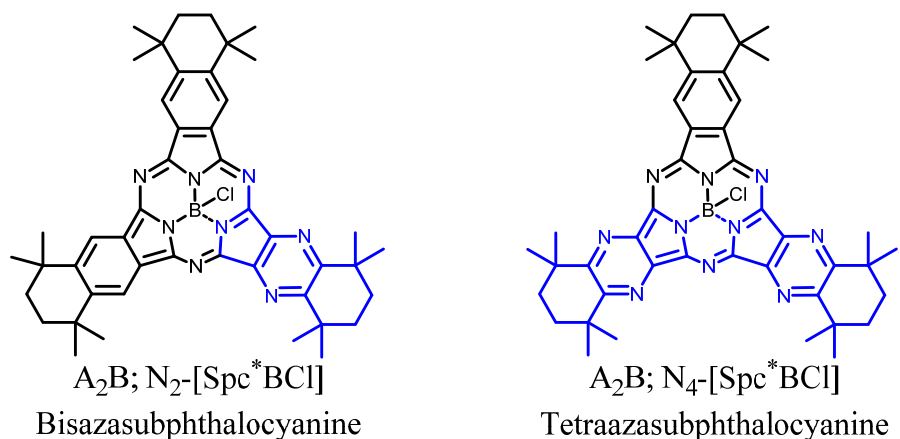

A$_2$B; N$_2$-[Spc*BCl]                    A$_2$B; N$_4$-[Spc*BCl]
Bisazasubphthalocyanine              Tetraazasubphthalocyanine

According to General Procedure 1 (p. 5), using PDN*/PyzDN* in a 1:1 ratio. After evaporation of the solvent, the solid residue was loaded onto silica and the crude product was purified via gradient CC (Tol → Tol/THF).

**[Spc*BCl]: Yield:** <1%. - The product was identified by using UV-Vis and MS. The analysis is in accordance to the one described in section 1.1.5.

**N$_2$-[Spc*BCl]: Yield:** 8 mg, 10.5 μmol, 1%. - **$^1$H NMR** (CDCl$_3$, 300 MHz): δ = 8.87 (s, 2 H, Ar-C$H$), 8.78 (s, 2 H, Ar-C$H$), 1.77-1.55 (m, 12 H, -C$H_2$), 1.38 (s, 9 H, -C$H_3$), 1.35 (s, 9 H, -

C$H_3$), 1.05 (s, 9 H, -C$H_3$), 1.03 (s, 9 H, -C$H_3$) ppm. - **UV-Vis** (DCM): $\lambda$ = 572 (s), 519 (sh), 327 (m), 303 (s) nm. - **Fluorescence** (DCM, $\lambda_{ex}$ = 350 nm): $\lambda$ = 577 nm. - $\Phi_F$ ($\lambda_{ex}$ = 490 nm) = 0.24. - $\Phi_\Delta$ = 0.67. - **MS** (APCI-HRMS(+)): *m/z* = 763.4162 [M+H]$^+$, cal. for C$_{46}$H$_{53}$B$_1$Cl$_1$N$_8$+H$_1$: 763.4177.

**N$_4$-[ Spc*BCl]: Yield:** 12 mg, 15.7 µmol, 2%. - **$^1$H NMR** (CDCl$_3$, 300 MHz): $\delta$ = 8.78 (s, 2 H, Ar-C$H$), 1.66-1.51 (m, 12 H, -C$H_2$), 1.38 (s, 9 H, -C$H_3$), 1.38 (s, 9 H, -C$H_3$), 1.35 (s, 9 H, -C$H_3$), 1.35 (s, 9 H, -C$H_3$) ppm. - **UV-Vis** (DCM): $\lambda$ = 557 (s), 506 (sh), 330 (s), 296 (s) nm. - **Fluorescence** (DCM, $\lambda_{ex}$ = 350 nm): $\lambda$ = 567 nm. - $\Phi_F$ ($\lambda_{ex}$ = 490 nm) = 0.22. - $\Phi_\Delta$ = 0.65. - **MS** (APCI-HRMS(+)): *m/z* = 765.4069 [M+H]$^+$, cal. for C$_{44}$H$_{51}$B$_1$Cl$_1$N$_{10}$+H$_1$: 765.4082.

**[Sppz*BCl]: Yield**: <1%. - The analysis is in accordance to the one described in section 1.1.6.

**Additional information**: For the synthesis of N$_x$-[Spc*BCl], depending on the desired product, the corresponding ratios of dinitriles was varied. In the $^1$H NMR spectra, dinitriles traces were observed.

## 1.1.8 Synthesis of Asymmetrical Phthalocyanines: $N_x$-$Pc^*H_2$

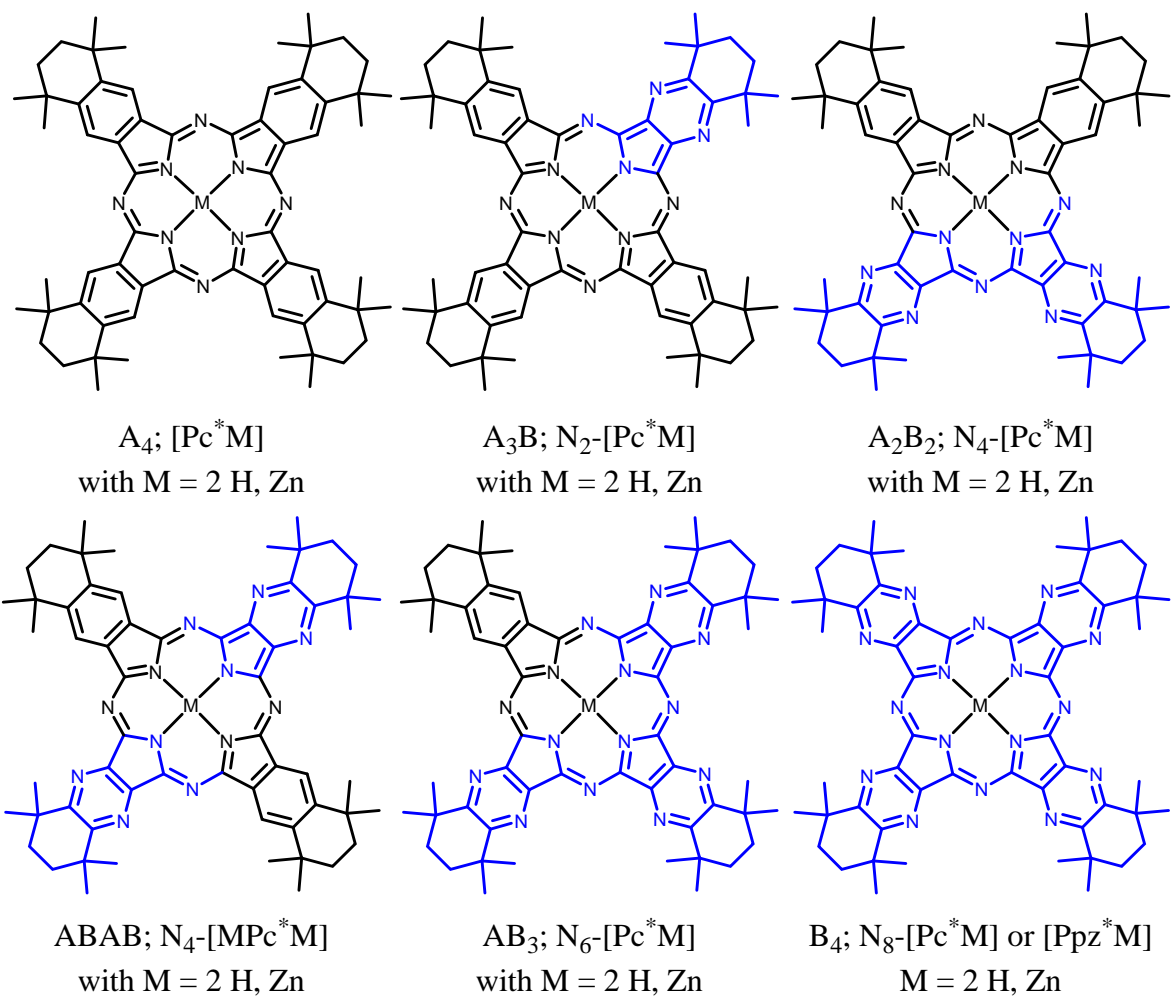

A$_4$; [Pc$^*$M]
with M = 2 H, Zn

A$_3$B; N$_2$-[Pc$^*$M]
with M = 2 H, Zn

A$_2$B$_2$; N$_4$-[Pc$^*$M]
with M = 2 H, Zn

ABAB; N$_4$-[MPc$^*$M]
with M = 2 H, Zn

AB$_3$; N$_6$-[Pc$^*$M]
with M = 2 H, Zn

B$_4$; N$_8$-[Pc$^*$M] or [Ppz$^*$M]
M = 2 H, Zn

## 1.1.9 Synthesis of $N_x$-$Pc^*H_2$ [4]

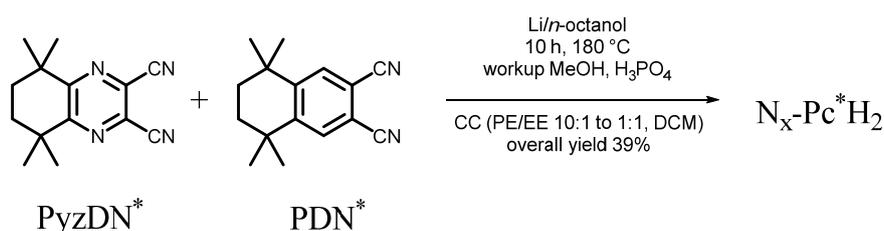

PyzDN$^*$    PDN$^*$

According to **Fehler! Verweisquelle konnte nicht gefunden werden.** (p. **Fehler! Textmarke nicht definiert.**), using PyzDN$^*$ **2**/PDN$^*$ **1** in a 1:1 ratio. The solution turned dark green. The isomers were separated by CC (PE(40/60)/EE gradient). At first, Pc$^*$H$_2$ and N$_2$-Pc$^*$H$_2$ were eluted, then ABAB and the symmetrical A$_2$B$_2$ N$_4$-Pc$^*$H$_2$ were eluted. After changing the solvent to PE/EE 1:1 a mixture of A$_2$B$_2$ N$_4$-Pc$^*$H$_2$ and AB$_3$ N$_6$-Pc$^*$H$_2$ was eluted. Finally, Ppz$^*$H$_2$ was eluted with pure DCM. In a second CC (DCM), A$_2$B$_2$ N$_4$-Pc$^*$H$_2$ and AB$_3$ N$_6$-Pc$^*$H$_2$ were separated.

**Pc*H₂**: **Yield:** 2%. - The analysis is in accordance with values in section **Fehler! Verweisquelle konnte nicht gefunden werden.**. - $\Phi_F$ ($\lambda_{ex}$ = 350 nm) = 0.44; - ($\lambda_{ex}$ = 598 nm) = 0.45. - $\Phi_\Delta$ = 0.16.

**N₂-Pc*H₂**: **Yield**: 4%. - **¹H NMR** (CDCl₃, 300 MHz): $\delta$ = 9.69 (s, 2 H, Ar-C*H*), 9.65 (s, 2 H, Ar-C*H*), 9.42 (s, 2 H, Ar-C*H*), 2.19 (s, 4 H, -C*H*₂), 2.07 (s, 8 H, -C*H*₂), 2.04 (s, 4 H, -C*H*₂), 1.90 (s, 12 H, -C*H*₃), 1.83 (s, 12 H, -C*H*₃), 1.81 (s, 12 H, -C*H*₃), 1.81 (s, 12 H, -C*H*₃), -0.07 (s, 2 H, -N*H*) ppm. - **¹³C NMR** (C₆D₆, 75 MHz): $\delta$ = 160.6, 149.8, 149.2, 148.6, 137.0, 133.0, 132.2, 122.2, 121.6, 39.1, 36.1, 36.1, 35.9, 35.4, 35.3, 34.7, 32.9, 32.8, 32.7, 30.9 ppm. - not all quartary atoms could be detected. - **IR** (ATR, 400-4000 cm⁻¹): $\tilde{v}$ = 3475 (vw), 3297 (vw), 2956 (m), 2921 (m), 2855 (m), 2555 (vw), 2348 (vw), 1711 (m), 1688 (m), 1621 (m), 1539 (m), 1498 (s), 1460 (m), 1382 (m), 1358 (m), 1328 (m), 1301 (s), 1258 (m), 1242 (m), 1188 (m), 1159 (m), 1140 (m), 1119 (m), 1089 (m), 1071 (m), 1022 (s), 999 (m), 980 (m), 891 (m), 848 (m), 804 (m), 755 (s), 722 (m), 681 (s), 622 (m), 541 (m) cm⁻¹. - **CV** (DCM, [TBA]PF₆, Fc): $E_{ox1}$ = 0.43, $E_{red1}$ = -1.23, $E_{red2}$ = -1.65 V. - **UV-Vis** (DCM): $\lambda$ = 687 (s), 619 (sh), 340 (s), 306 (s), 232 (s) nm. - $\Phi_F$ ($\lambda_{ex}$ = 350 nm) = 0.31; - ($\lambda_{ex}$ = 598 nm) = 0.33. - $\Phi_\Delta$ = 0.14. - **MS** (APCI-HRMS(+)): $m/z$ = 957.5992 [M+H]⁺, cal. for C₆₂H₇₂N₁₀+H₁: 957.6014. - **Elemental analysis** (C₆₂H₇₂N₁₀, M = 957.30 g/mol): fnd. (cal.): C: 75.29% (77.79%), H: 8.92% (7.58%), N: 10.14% (14.63%).

**A₂B₂ N₄-Pc*H₂**: **Yield**: 17%. - **¹H NMR** (CDCl₃, 300 MHz): $\delta$ = 9.56 (s, 2 H, Ar-C*H*), 9.50 (s, 2 H, Ar-C*H*), 2.20 (s, 8 H, -C*H*₂), 2.05 (s, 8 H, -C*H*₂), 1.91 (s, 24 H, -C*H*₃), 1.81 (s, 12 H, -C*H*₃), 1.80 (s, 12 H, -C*H*₃), -0.04 (s, 2 H, -N*H*) ppm. - **¹H NMR** (Pyridine-$d_5$, 300 MHz): $\delta$ = 9.87 (s, 2 H, Ar-C*H*), 9.68 (s, 2 H, Ar-C*H*), 1.94 (s, 8 H, -C*H*₂), 1.76 (s, 8 H, -C*H*₂), 1.72 (s, 12 H, -C*H*₃), 1.67 (s, 12 H, -C*H*₃), 1.46 (s, 12 H, -C*H*₃), 1.37 (s, 12 H, -C*H*₃), -0.26 (s, 2 H, -N*H*) ppm. - **¹H NMR** (C₆D₆, 300 MHz): $\delta$ = 9.83 (s, 2 H, Ar-C*H*), 9.82 (s, 2 H, Ar-C*H*), 1.86 (s, 8 H, -C*H*₂), 1.79 (s, 8 H, -C*H*₂), 1.75 (s, 24 H, -C*H*₃), 1.48 (s, 12 H, -C*H*₃), 1.38 (s, 12 H, -C*H*₃), -0.31 (s, 2 H, -N*H*) ppm. - **¹³C NMR** (C₆D₆, 75 MHz): $\delta$ = 162.4, 150.0, 149.6, 145.1, 144.4, 135.2, 134.8, 122.7, 122.0, 39.4, 39.3, 36.1, 36.0, 35.2, 34.6, 34.5, 32.7, 32.7, 21.1, 30.8 ppm. - **IR** (ATR, 400-4000 cm⁻¹): $\tilde{v}$ = 3526 (w), 3284 (w), 2915 (s), 2857 (s), 1736 (w), 1638 (w), 1553 (w), 1496 (w), 1455 (s), 1381 (w), 1358 (m), 1328 (m), 1300 (s), 1254 (m), 1242 (m), 1190 (m), 1162 (m), 1148 (m), 1128 (m), 1083 (w), 1048 (w), 1019 (s), 1001 (s), 983 (s), 949 (w), 936 (w), 893 (m), 851 (m), 828 (m), 761 (s), 752 (s), 717 (m), 700 (m), 679 (m), 623 (w), 560 (w), 543 (m), 507 (m), 443 (w), 429 (w) cm⁻¹. - **CV** (DCM, [TBA]PF₆, Fc): $E_{ox1}$ = 0.28, $E_{red1}$ = -1.47, $E_{red2}$ = -1.79 V. - **UV-Vis** (DCM): $\lambda$ = 680 (s), 656 (s), 625 (sh),

348 (s), 232 (s) nm. - $\Phi_F$ ($\lambda_{ex}$ = 350 nm) = 0.13; - ($\lambda_{ex}$ = 598 nm) = 0.14. - $\Phi_\Delta$ = 0.07. - **MS** (APCI-HRMS(+)): $m/z$ = 959.5921 [M+H]$^+$, cal. for $C_{60}H_{70}N_{12}+H_1$: 959.5919. - **Elemental analysis** ($C_{60}H_{70}N_{12}$, M = 959.28 g/mol): fnd. (cal.): C: 72.03% (75.12%), H: 7.36% (7.36%), N: 16.36% (17.52%).

**N$_6$-Pc*H$_2$**: Yield: 9%. - **$^1$H NMR** (CDCl$_3$, 300 MHz): $\delta$ = 9.75 (s, 2 H, Ar-C*H*), 2.22 (s, 4 H, -C*H$_2$*), 2.19 (s, 12 H, -C*H$_2$*), 1.93 (s, 12 H, -C*H$_3$*), 1.91 (s, 12 H, -C*H$_3$*), 1.90 (s, 12 H, -C*H$_3$*), 1.83 (s, 12 H, -C*H$_3$*), -0.53 (s, 2 H, -N*H*) ppm. - **$^{13}$C NMR** (C$_6$D$_6$, 75 MHz): $\delta$ = 162.0, 139.1, 131.3 134.1 (w), 132.9 (w), 132.9 (w), 131.3 (w), 127.1 (w), 123.1, 46.0, 41.3 (w), 39.7, 39.3, 36.3, 34.5, 32.8, 31.8, 31.7, 31.1, 30.9, 30.8, 29.8, 29.5, 29.3, 29.2, 28.6, 28.1, 27.5, 27.5 ppm. - 6 weak signal, marked with (w); not all quartary atoms could be detected. **IR** (ATR, 400-4000 cm$^{-1}$): $\tilde{\nu}$ = 3291 (w), 2957 (s), 2924 (s), 2856 (s), 1727 (s), 1693 (s), 1637 (m), 1540 (m), 1503 (m), 1457 (s), 1411 (m), 1381 (m), 1360 (m), 1330 (m), 1305 (m), 1281 (m), 1257 (s), 1192 (s), 1157 (w), 1139 (m), 1127 (s), 1086 (m), 1072 (m), 1021 (s), 995 (s), 952 (m), 935 (w), 893 (w), 852 (m), 828 (m), 804 (m), 756 (m), 744 (s), 719 (s), 700 (w), 677 (s), 631 (m), 542 (w), 506 (w), 466 (w), 429 (w) cm$^{-1}$. - **CV** (DCM, [TBA]PF$_6$, Fc): $E_{ox1}$ = 0.35, $E_{red1}$ = -1.47, $E_{red2}$ = -1.82 V. - **UV-Vis** (DCM): $\lambda$ = 660 (s), 597 (sh), 343 (s), 234 (s) nm. - $\Phi_F$ ($\lambda_{ex}$ = 350 nm) = 0.05; - ($\lambda_{ex}$ = 598 nm) = 0.03. - $\Phi_\Delta$ = 0.04. - **MS** (APCI-HRMS(+)): $m/z$ = 961.5825 [M+H]$^+$, cal. for $C_{58}H_{68}N_{14}+H_1$: 961.5824. - **Elemental analysis** ($C_{58}H_{68}N_{14}$, M = 961.25 g/mol): fnd. (cal.): C: 71.99% (72.47%), H: 9.49% (7.13%), N: 12.38% (20.40%).

**Ppz*H$_2$**: Yield: 9%. - The analysis is in accordance with values in section **Fehler! Verweisquelle konnte nicht gefunden werden.**. - $\Phi_F$ ($\lambda_{ex}$ = 350 nm) = 0.03; - ($\lambda_{ex}$ = 598 nm) = 0.03. - $\Phi_\Delta$ = 0.05.

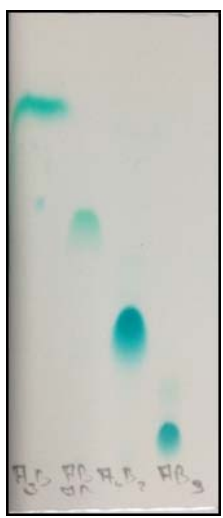

**1.1.10 Synthesis of N$_4$-[Pc$^*$Zn] using [Zn(OAc)$_2$]**

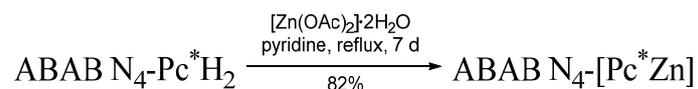

ABAB N$_4$-Pc$^*$H$_2$ $\xrightarrow[82\%]{\text{[Zn(OAc)}_2\text{]·2H}_2\text{O, pyridine, reflux, 7 d}}$ ABAB N$_4$-[Pc$^*$Zn]

General Procedure for the Synthesis of N$_x$-[Pc$^*$Zn] using [Zn(OAc)$_2$]:

1 eq ABAB or A$_2$B$_2$ N$_4$-Pc$^*$H$_2$ was dissolved in pyridine and 1.2 eq [Zn(OAc)$_2$]·2H$_2$O were added. After refluxing for 1 h, the reaction was monitored by TLC (DCM), and was stirred for another 1 h. The resulting blue solutions were concentrated in vacuum, filtered, and washed with an excess of water. After washing the product off the filter paper using CHCl$_3$/MeOH 3:1, the blue product was dried in vacuum and purified by preparative TLC or CC (Tol/THF 20:1).

**ABAB N$_4$-[Pc$^*$Zn]**: **Yield**: 82%. - $R_f$ (Tol/THF 20:1) = 0.63. - **$^1$H NMR** (C$_6$D$_6$, 300 MHz): δ = 9.96 (s, 4 H, Ar-C*H*), 1.91 (s, 8 H, -C*H*$_2$), 1.85 (s, 24 H, -C*H*$_3$), 1.78 (s, 8 H, -C*H*$_2$), 1.45 (s, 24 H, -C*H*$_3$) ppm. - **UV-Vis** (DCM): λ = 687 (s), 657 (s), 631 (sh), 596 (sh), 357 (s) nm. - Φ$_F$ ($λ_{ex}$ = 350 nm) = 0.16; - ($λ_{ex}$ = 598 nm) = 0.18. - Φ$_\Delta$ = 0.73. - **MS** (APCI-HRMS(+)): *m/z* = 1021.5073 [M+H]$^+$, cal. for C$_{60}$H$_{68}$N$_{12}$Zn$_1$+H$_1$: 1021.5054.

**A$_2$B$_2$ N$_4$-[Pc$^*$Zn]**: **Yield**: 76%. - The analysis is in accordance with the data described above.

All other N$_x$-[Pc$^*$Zn] are described in the Paper.

## 1.2 Suggested radical autoxidation mechanism of tetraline IV.

In literature, such Co(III) or Mn(III) assisted aerobic oxidations are discussed as radical chain reactions. Rate determining step is the selective benzylic C-H activation, H abstraction by Co(III)-coordinated acetate (quasi acetyl radical ligands) or by bromine radicals generated in the presence of bromide promotor. Instead of H atom abstraction, a sequence of electron transfer followed by deprotonation of the formed radical cation is discussed. The benzylic radicals Ar-$CH_2 \cdot$ are trapped by oxygen, the peroxy radical is reduced by Co(II). Hock cleavage leads to a cabaldehyde, which undergoes further metal catalysed autoxidation steps to the carbonic acid:

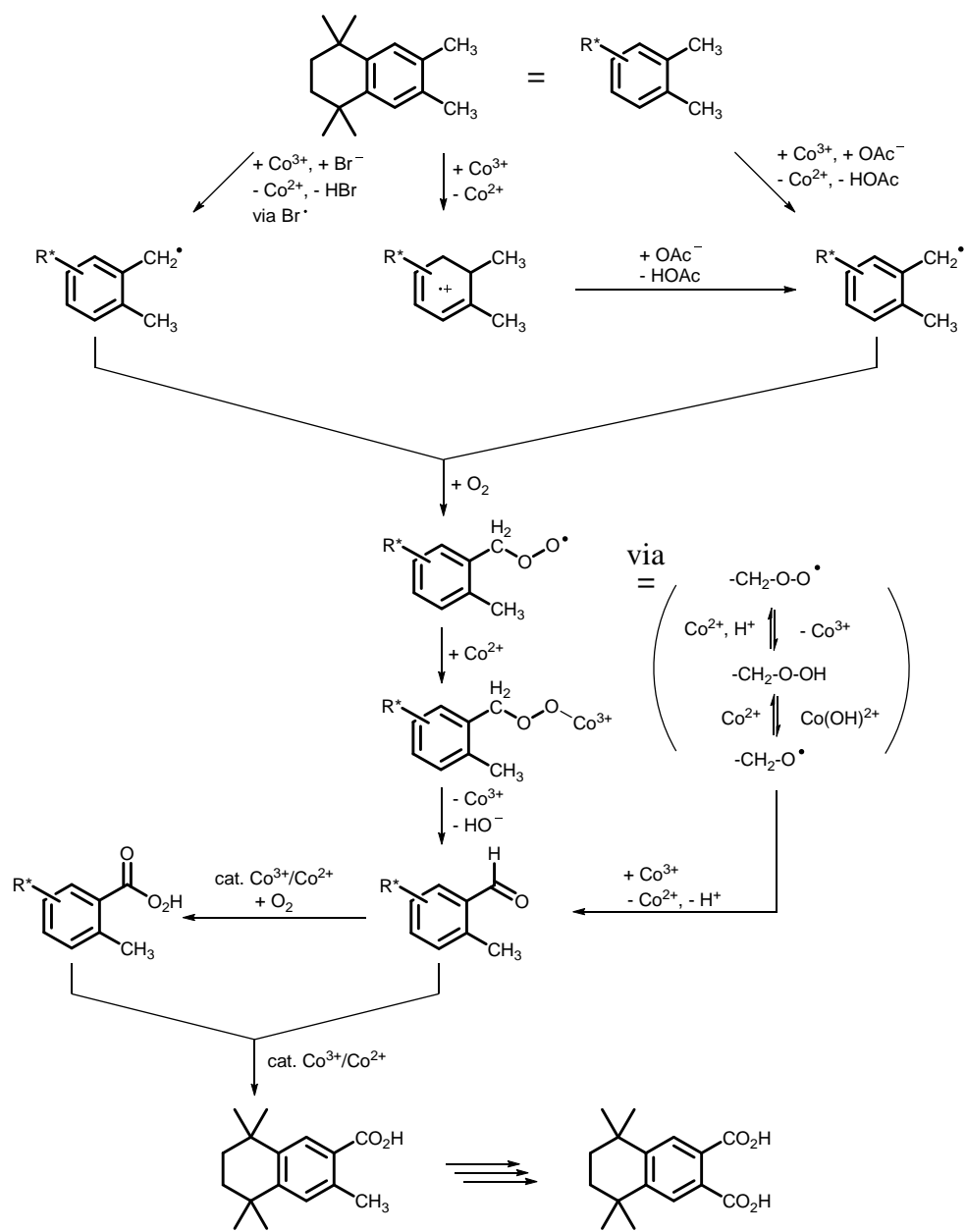

## 1.3 Absorbance and Emission Spectra

Absorption spectra of metal-free (full lines) and zinc complexes (dashed lines) in THF. The spectra were normalized to the same absorption in the B-band. a) **3a**, **3b**, b) **4a**, **4b**, c) **5a**, **5b**, d) **6a**, **6b**, e) **7a**, **7b**, f) **8a**, **8b**.

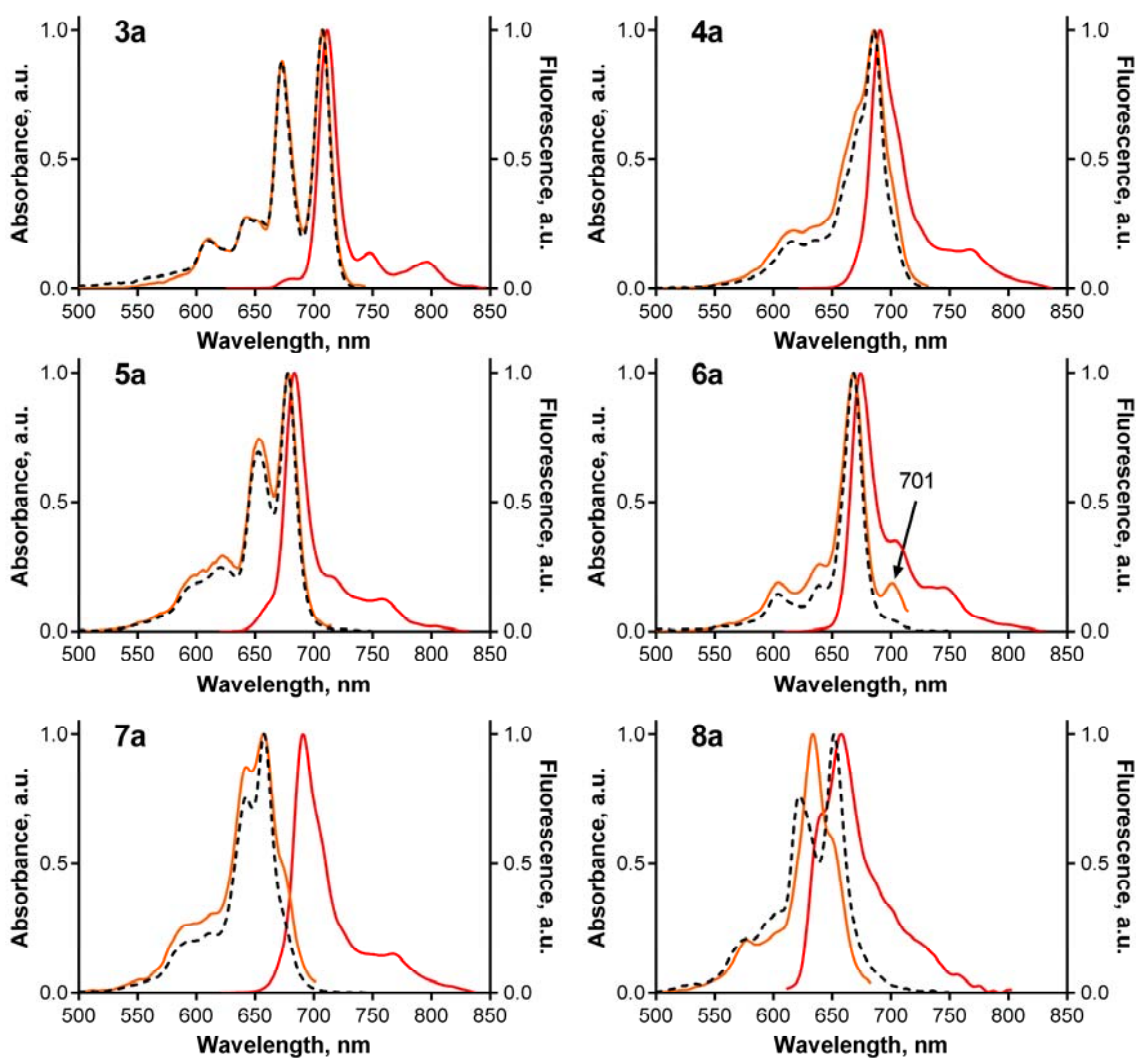

**Figure SI-1**. Normalized absorption (black, dashed), emission (red) and excitation (orange) spectra of $N_x$-Pc*$H_2$ in THF.

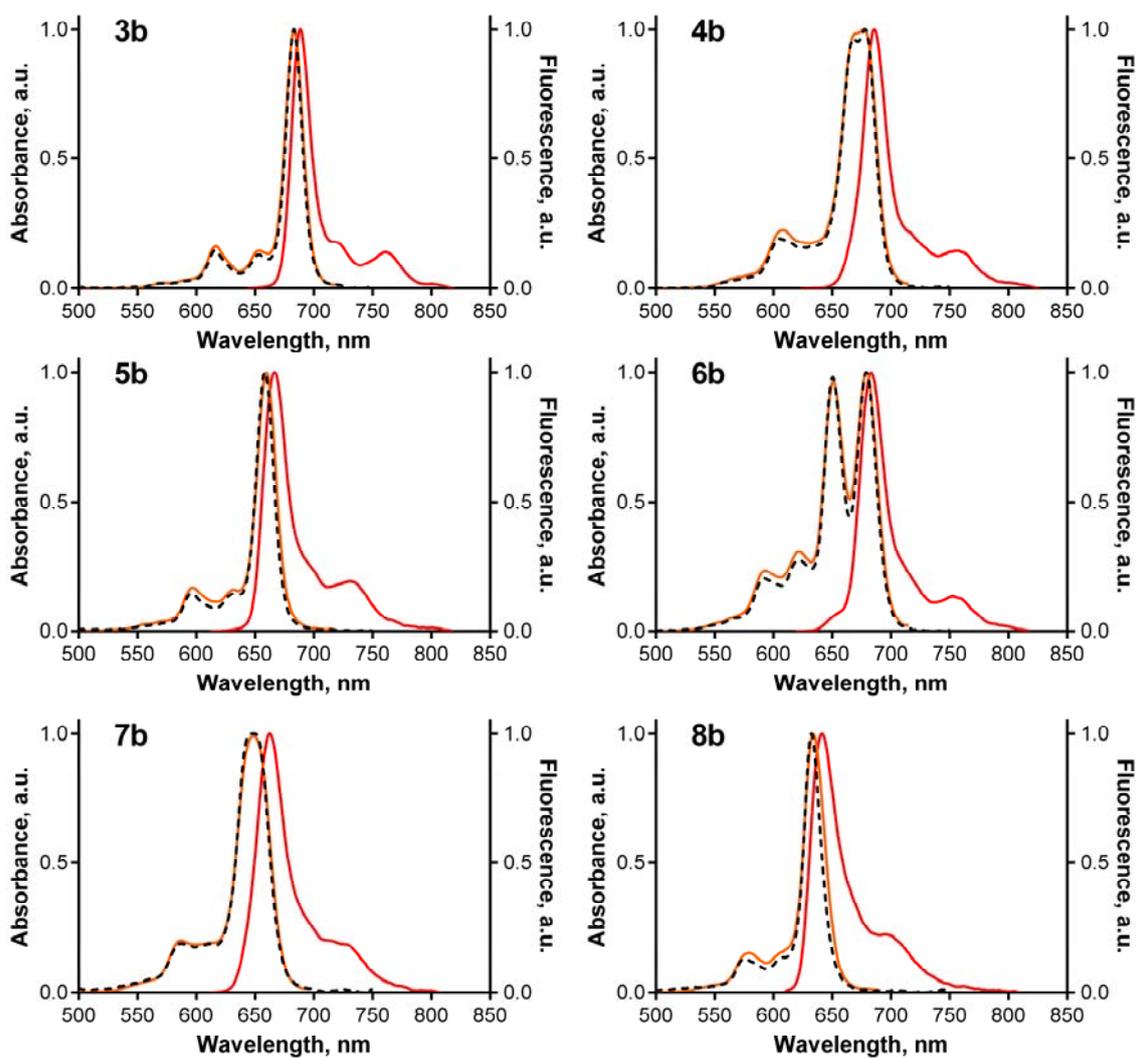

**Figure SI-2**. Normalized absorption (black, dashed), emission (red) and excitation (orange) spectra of $N_x$-Pc*Zn in THF.

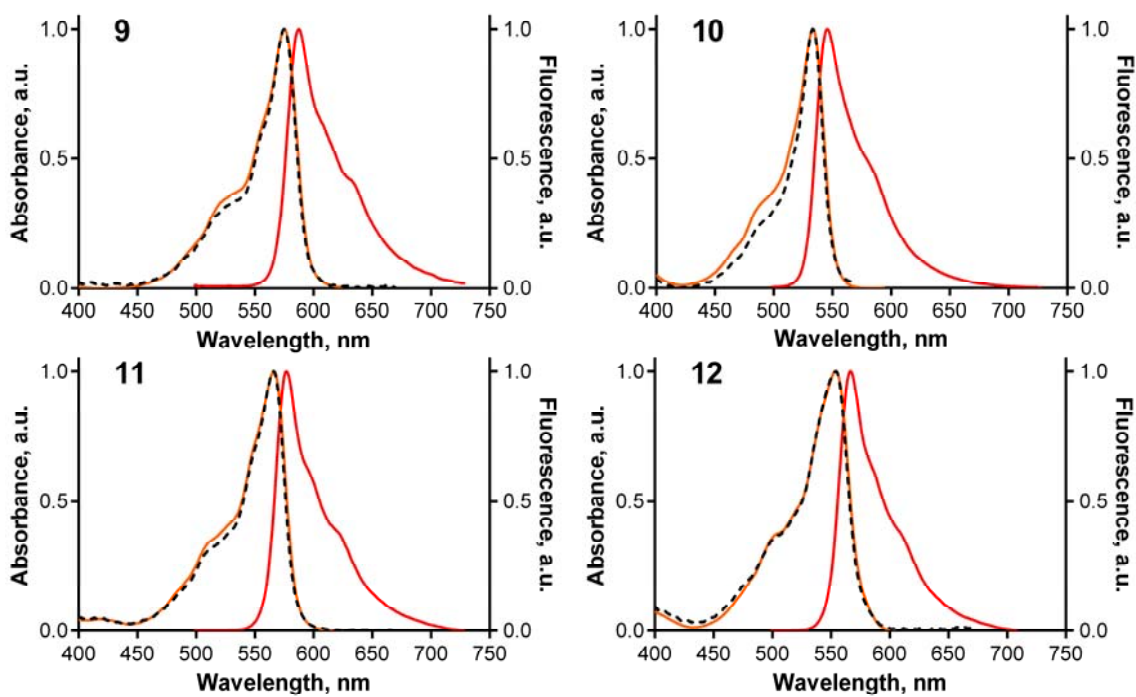

**Figure SI-3**. Normalized absorption (black, dashed), emission (red) and excitation (orange) spectra of $N_x$-Spc*BCl in MeOH.

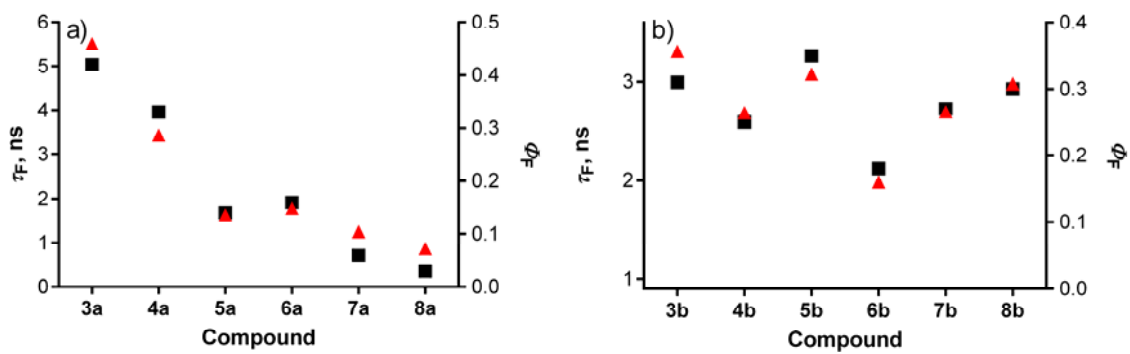

**Figure SI-4**. Correlation between $\Phi_F$ (black squares) and $\tau_F$ (red triangles) values for $N_x$-Pc*$H_2$ in THF (a) and $N_x$-Pc*Zn in THF (b). For purposes of this presentation, the relative contribution of each process to the average lifetime for compounds with two lifetime components (**7a** and **8a**) was calculated.

## 1.4 Computational results

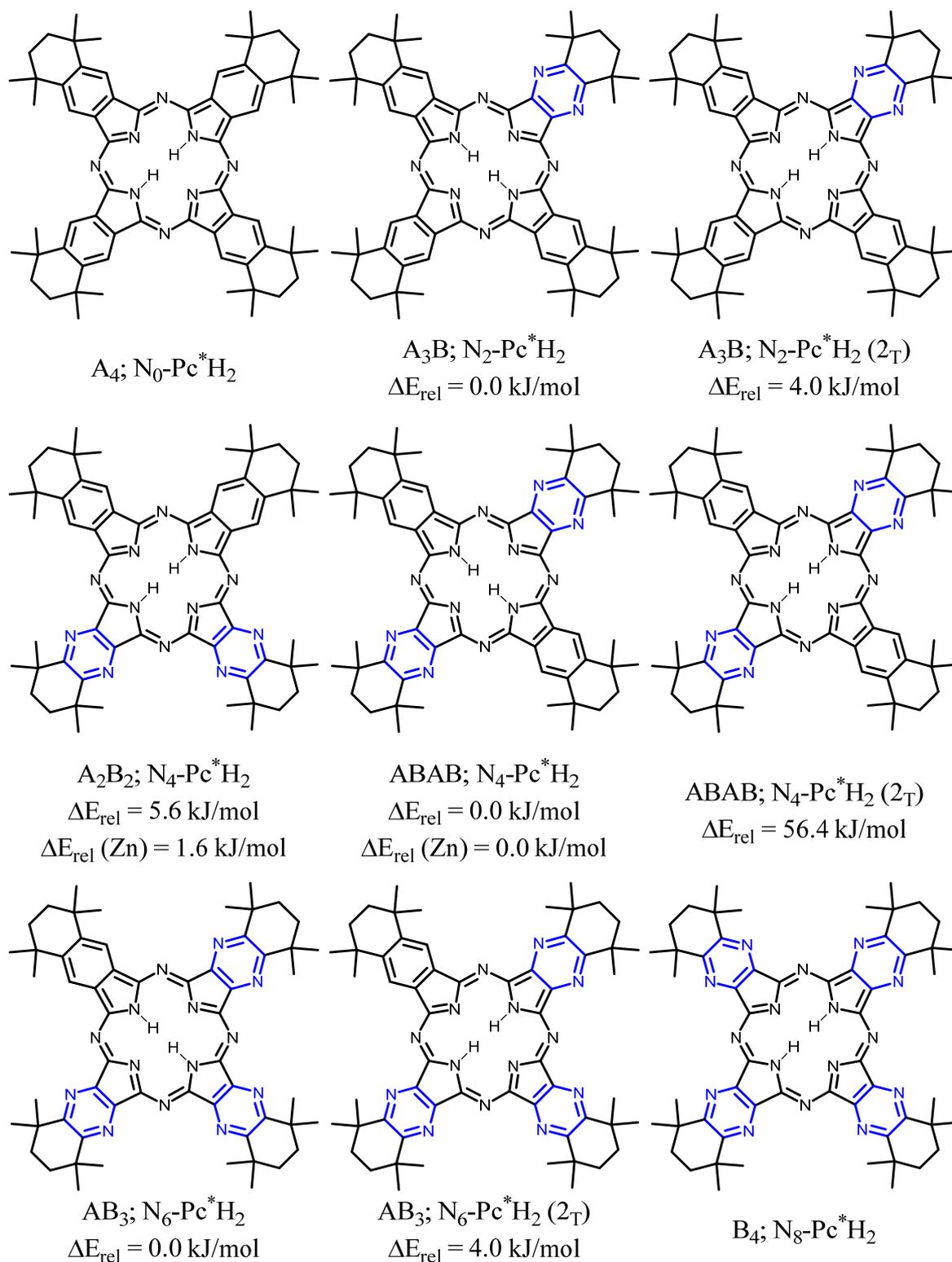

**Figure SI-5**. The phthalocyanines $N_x$-Pc*$H_2$ ($A_4H_2$, $A_3B_1$, $A_2B_2$, $A_1B_3$ and $B_4H_2$) computationally investigated. Regioisomers are numbered and tautomers marked (subscript T). Energies are given relative to the most stable isomer ($\Delta E_{rel}$). Zn-substituted phthalocyanines [Pc*Zn] are derived by substituting the two central H atoms by Zn. Relative isomer energies are given as $\Delta E_{rel}(Zn)$, no tautomers are found for [Pc*Zn].

**Table SI-1:** Computed excitation energies (λ in nm), oscillator strengths (*f*), frontier orbital gaps ($\Delta_{H\text{-}L/L+1}$ in nm) and molecular orbital character of the transitions for the Q-bands. Dipole moment (μ) and isotropic polarizability (α) in a.u.[a]

| Molecule | λ / nm | Δλ / nm | *f* | *Character* | $\Delta_{H\text{-}L/L+1}$ [c] | μ [d] | Δα [e] |
|---|---|---|---|---|---|---|---|
| N$_8$-Pc*H$_2$ | 541 | 0 | 0.42 | H → L+1 | 781 | 0.00 | 0.0 |
|  | 536 | 0 | 0.53 | H → L | 818 |  |  |
| N$_6$-Pc*H$_2$ | 550 | 9 | 0.56 | H → L | 861 | 0.87 | 23.6 |
|  | 549 | 13 | 0.49 | H → L+1 | 797 |  |  |
| N$_6$-Pc*H$_2$ (2$_T$) | 559 | 18 | 0.47 | H → L+1 | 839 | 0.92 | 26.6 |
|  | 552 | 16 | 0.60 | H → L | 848 |  |  |
| N$_4$-Pc*H$_2$ (A$_2$B$_2$) | 563 | 22 | 0.55 | H → L+1 (39%) | 845 | 1.25 | 28.3 |
|  |  |  |  | H → L (27%) |  |  |  |
|  |  |  |  | H → L (41%) |  |  |  |
|  | 559 | 23 | 0.63 | H → L+1 (18%) |  |  |  |
|  |  |  |  | H-2 → L+1 (15%) |  |  |  |
| N$_4$-Pc*H$_2$ (ABAB) | 561 | 20 | 0.62 | H → L | 910 | 0.00 | 47.8 |
|  | 560 | 24 | 0.55 | H → L+1 | 815 |  |  |
| N$_4$-Pc*H$_2$ (ABAB, 2$_T$) | 583 | 42 | 0.56 | H → L+1 | 892 | 0.06 | 61.8 |
|  | 580 | 44 | 0.54 | H → L | 911 |  |  |
| N$_2$-Pc*H$_2$ | 569 | 28 | 0.62 | H → L+1 | 855 | 0.90 | 71.1 |
|  | 564 | 28 | 0.72 | H → L | 915 |  |  |
| N$_2$-Pc*H$_2$ (2$_T$) | 577 | 36 | 0.61 | H → L+1 | 897 | 0.87 | 73.3 |
|  | 571 | 35 | 0.69 | H → L | 900 |  |  |
| N$_0$-Pc*H$_2$ | 580 | 39 | 0.69 | H → L+1 | 903 | 0.00 | 94.9 |
|  | 568 | 32 | 0.81 | H → L | 923 |  |  |
| N$_8$-[Pc*Zn] | 532 | 0 | 0.51 | H → L | 788 | 0.00 | 0.0 |
|  | 532 | 0 | 0.51 | H → L+1 | 788 |  |  |
| N$_6$-[Pc*Zn] | 549 | 17 | 0.55 | H → L | 840 | 0.89 | 24.6 |
|  | 544 | 12 | 0.59 | H → L+1 | 810 |  |  |
| N$_4$-[Pc*Zn] (A$_2$B$_2$) | 556 | 26 | 0.62 | H → L | 852 | 1.25 | 47.6 |
|  | 554 | 22 | 0.63 | H → L+1 | 848 |  |  |
| N$_4$-[Pc*Zn] (ABAB) | 565 | 33 | 0.60 | H → L | 898 | 0.00 | 49.5 |
|  | 557 | 26 | 0.63 | H → L+1 | 831 |  |  |
| N$_2$-[Pc*Zn] | 566 | 34 | 0.69 | H → L | 897 | 0.88 | 71.1 |
|  | 564 | 32 | 0.69 | H → L+1 | 863 |  |  |
| N$_0$-[Pc*Zn] | 570 | 38 | 0.77 | H → L | 902 | 0.00 | 93.7 |
|  | 570 | 38 | 0.77 | H → L+1 | 901 |  |  |

[a] Shift relative to Q1/Q2 band of B$_4$M. [b] dominant MO transitions for the Q-bands (H: HOMO, H-2: HOMO-2, L: LUMO, L+1: LUMO+1). For A$_2$B$_2$H$_{2\text{-}1}$, relative contributions are given in percent. [c] Frontier MO gap of the orbitals given in the "character" column HOMO/LUMO and HOMO/LUMO+1) in nm. [d] All compounds exhibit an in-plane dipole moment only pointing toward the nitrogen-rich parts of the molecule. [e] Polarizabilities are given relative to B$_4$H$_2$ (α = 1091.3 a.u.) and B$_4$Zn (α = 1097.1 a.u.), respectively.

## 1.5 ¹H NMR Spectra

### 1.5.1 Subphthalocyanines

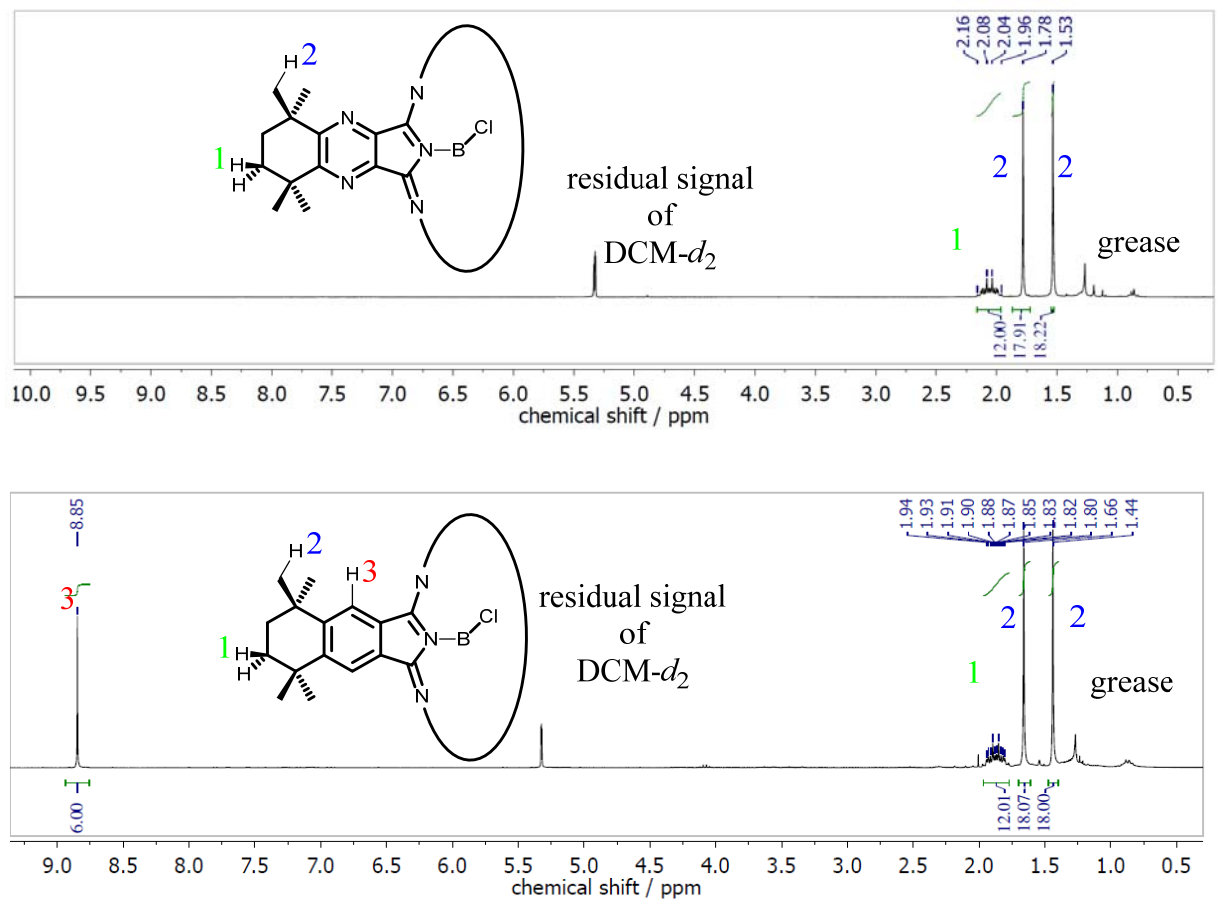

**Figure SI-6**. ¹H NMR spectrum of [Sppz*BCl] (above) / [Spc*BCl] (below) in DCM-$d_2$, 300 MHz.

## 1.5.2 Azaphthalocyanines

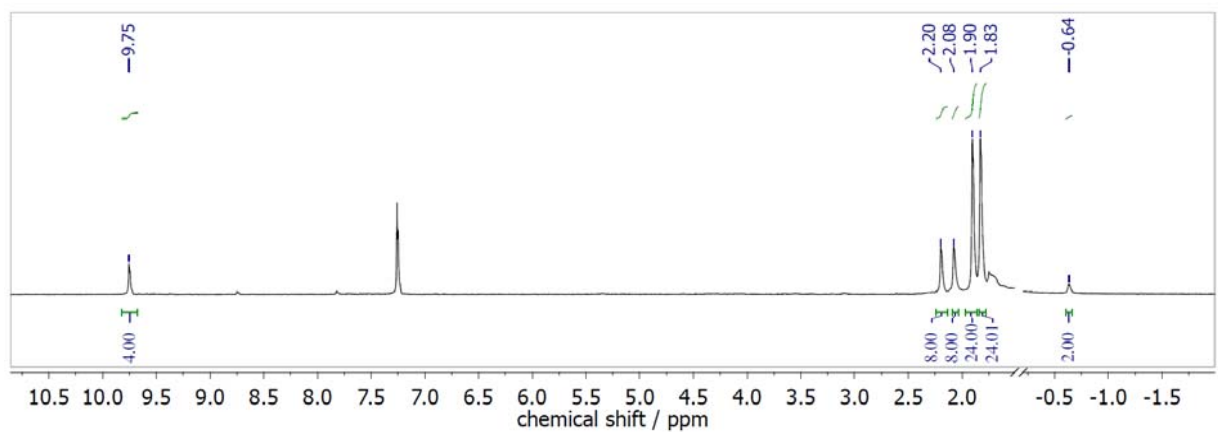

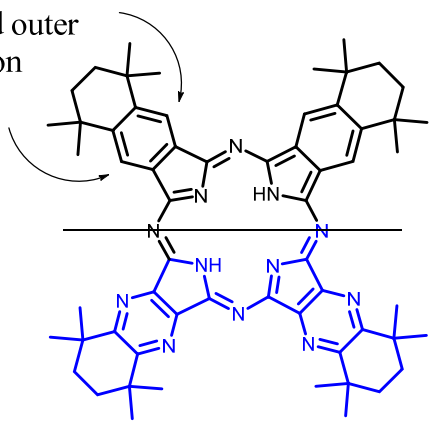

A$_2$B$_2$
Tetraazaphthalocyanine
two aromatic signals

inner and outer proton

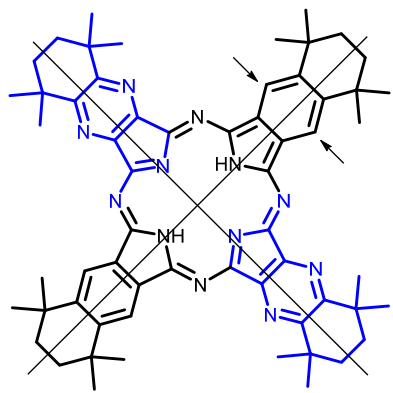

ABAB
Tetraazaphthalocyanine

only one aromatic signals

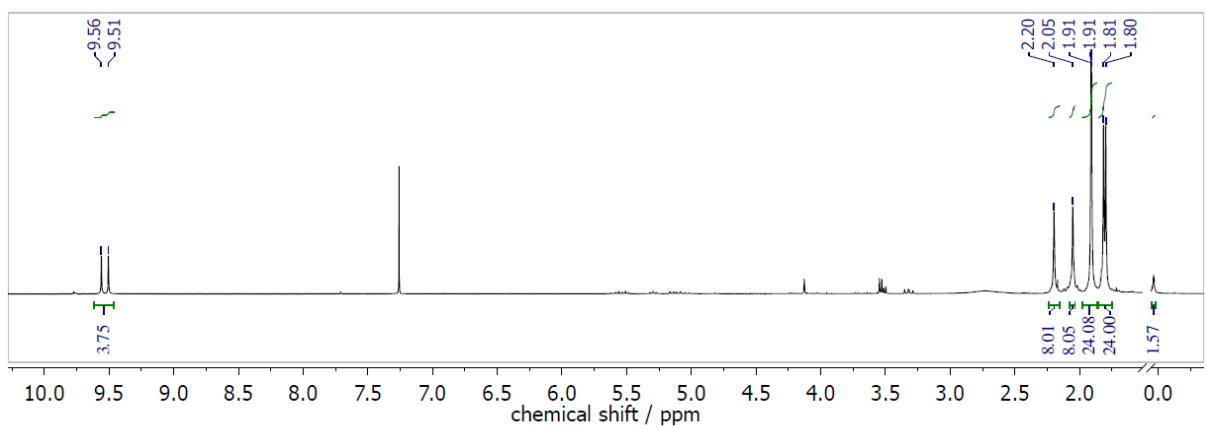

**Figure SI-7**. $^1$H NMR spectrum of ABAB (above) / A$_2$B$_2$ (below) N$_4$- H$_2$Pc* in CDCl$_3$, 300 MHz.

## 1.6 Cylovoltammetric measurements

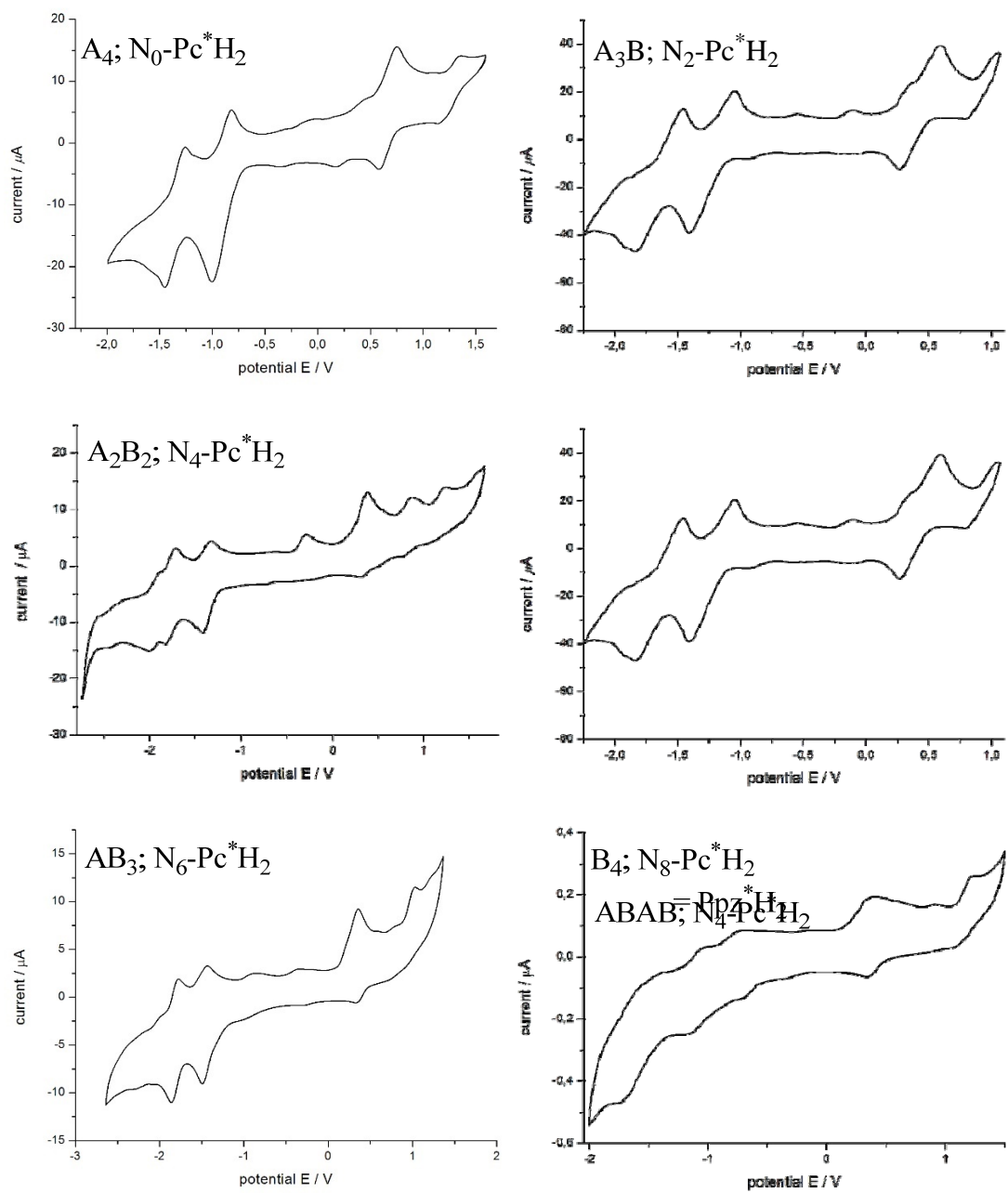

**Figure SI-8**. Cyclovoltammetric measurements of $N_x$-Pc$^*$H$_2$. All measurements were carried out in DCM, except of Ppz$^*$H$_2$, which was measured in THF because of solubility. The red circles mark impurities caused by used DCM. In all cases, different scan rates were tested, here, 20 mV/s measurements are shown. The spectra are calibrated using Fc in the final scan, measured in [TBA]PF$_6$.

## 1.7 Crystal Structures

Crystal were measured with a Bruker D8 QUEST area detector. Crystal structures were resolved and refined by Dr. K. Harms of the service Department of Chemistry of the Philipps-Universität Marburg.

| | |
|---|---|
| Diffractometer type | Bruker D8 QUEST area detector |
| Wavelength | 0.71073 Å |
| Temperature | 100(2) K |
| Index ranges | -14<=h<=14, -28<=k<=28, -15<=l<=15 |
| Data collection software | BRUKER APEX2 2014.9-0 [6] |
| Cell refinement software | BRUKER SAINT [7] |
| Data reduction software | SAINT V8.34A (Bruker AXS Inc., 2013) [7] |
| | |
| Programs used | XT V2014/1 (Bruker AXS Inc., 2014) [8,9] |
| | SHELXL-2014/7 (Sheldrick, 2014) [8,10] |
| | DIAMOND (Crystal Impact) [11] |

## 1.7.1 Crystal Structure of [Spc*BCl]

| | |
|---|---|
| Habitus, colour | prism, colourless |
| Crystal size | 0.22 x 0.22 x 0.09 mm$^3$ |
| Crystal system | Orthorhombic |
| Space group | Cmc2$_1$      Z = 4 |
| Unit cell dimensions | a = 24.8236(10) Å    α= 90°. |
| | b = 16.3090(6) Å    β= 90°. |
| | c = 10.9728(4) Å    γ = 90°. |
| Volume | 4442.3(3) Å$^3$ |
| Cell determination | 9874 peaks with Theta 2.4 to 25.4°. |
| Empirical formula | C$_{49}$ H$_{56}$ B Cl$_3$ N$_6$ |
| Moiety formula | C$_{48}$ H$_{54}$ B Cl N$_6$, CH$_2$Cl$_2$ |
| Formula weight | 846.15 |
| Density (calculated) | 1.265 Mg/m$^3$ |
| Absorption coefficient | 0.248 mm$^{-1}$ |
| F(000) | 1792 |
| | |
| Solution and refinement: | |
| Reflections collected | 28577 |
| Independent reflections | 4177 [R(int) = 0.0342] |
| Completeness to theta = 25.242° | 99.9 % |
| Observed reflections | 4031[I>2sigma(I) ] |
| Reflections used for refinement | 4177 |
| Absorption correction | Semi-empirical from equivalents [12] |
| Max. and min. transmission | 0.98 and 0.86 |
| Flack parameter (absolute struct.) | 0.001(14) |
| Largest diff. peak and hole | 0.283 and -0.192 e.Å$^{-3}$ |
| Solution | Direct methods [8] |
| Refinement | Full-matrix least-squares on F$^2$ [8] |
| Treatment of hydrogen atoms | Calculated positions, constr. ref. |
| Data / restraints / parameters | 4177 / 44 / 311 |
| Goodness-of-fit on F$^2$ | 1.049 |
| R index (all data) | wR2 = 0.0754 |
| R index conventional [I>2sigma(I)] | R1 = 0.0293 |

**Table SI-2**. Atomic coordinates and equivalent isotropic displacement parameters (Å$^2$) for ML153F_0m. U(eq) is defined as one third of the trace of the orthogonalized U$^{ij}$ tensor.

| | x | y | z | U(eq) | Occupancy |
|---|---|---|---|---|---|
| N1   | 0.5000      | 0.08718(16)  | 0.7070(3)    | 0.0163(6)   | 1 |
| N3   | 0.54732(7)  | 0.15242(11)  | 0.54643(17)  | 0.0130(4)   | 1 |
| N15  | 0.59330(7)  | 0.26327(11)  | 0.45407(16)  | 0.0136(4)   | 1 |
| N17  | 0.5000      | 0.24139(16)  | 0.4157(2)    | 0.0127(5)   | 1 |
| C2   | 0.54616(9)  | 0.11154(13)  | 0.6547(2)    | 0.0149(4)   | 1 |
| C4   | 0.59270(8)  | 0.19816(13)  | 0.5299(2)    | 0.0128(4)   | 1 |
| C5   | 0.62944(9)  | 0.17210(13)  | 0.6260(2)    | 0.0138(4)   | 1 |
| C6   | 0.68271(9)  | 0.19023(13)  | 0.6541(2)    | 0.0141(4)   | 1 |
| C7   | 0.70788(9)  | 0.15358(13)  | 0.7545(2)    | 0.0144(4)   | 1 |
| C8   | 0.76727(10) | 0.17460(14)  | 0.7780(2)    | 0.0186(5)   | 1 |
| C9   | 0.79202(10) | 0.11748(16)  | 0.8738(2)    | 0.0253(6)   | 1 |
| C10  | 0.75505(11) | 0.10051(17)  | 0.9799(2)    | 0.0257(6)   | 1 |
| C11  | 0.70338(10) | 0.05628(15)  | 0.9411(2)    | 0.0197(5)   | 1 |
| C12  | 0.67783(9)  | 0.10049(13)  | 0.8318(2)    | 0.0150(4)   | 1 |
| C13  | 0.62411(9)  | 0.08414(14)  | 0.8058(2)    | 0.0162(5)   | 1 |
| C14  | 0.60011(9)  | 0.11829(13)  | 0.7033(2)    | 0.0148(5)   | 1 |
| C16  | 0.54605(9)  | 0.28670(13)  | 0.40512(19)  | 0.0137(4)   | 1 |
| C18  | 0.52856(9)  | 0.36689(14)  | 0.3620(2)    | 0.0145(4)   | 1 |
| C19  | 0.55637(9)  | 0.43836(13)  | 0.3359(2)    | 0.0174(5)   | 1 |
| C20  | 0.52846(9)  | 0.51054(14)  | 0.3088(2)    | 0.0180(5)   | 1 |
| C21  | 0.56208(11) | 0.58749(16)  | 0.2815(3)    | 0.0287(6)   | 1 |
| C22  | 0.5284(2)   | 0.6650(3)    | 0.2921(7)    | 0.0259(11)  | 0.5 |
| C23  | 0.4778(2)   | 0.6530(3)    | 0.2168(6)    | 0.0242(11)  | 0.5 |
| C24  | 0.66584(12) | 0.05663(19)  | 1.0520(2)    | 0.0328(6)   | 1 |
| C25  | 0.71454(12) | -0.03249(16) | 0.9045(3)    | 0.0306(6)   | 1 |
| C26  | 0.80131(9)  | 0.16592(17)  | 0.6619(2)    | 0.0267(6)   | 1 |
| C27  | 0.77020(10) | 0.26457(15)  | 0.8206(2)    | 0.0242(5)   | 1 |
| C28  | 0.59861(12) | 0.57227(19)  | 0.1717(3)    | 0.0376(7)   | 1 |
| C29  | 0.5886(4)   | 0.6179(6)    | 0.3816(10)   | 0.033(2)    | 0.5 |
| C29A | 0.6058(3)   | 0.6001(6)    | 0.3970(9)    | 0.0251(17)  | 0.5 |
| B1   | 0.5000      | 0.1569(2)    | 0.4646(3)    | 0.0130(7)   | 1 |
| Cl1  | 0.5000      | 0.07386(5)   | 0.34930(7)   | 0.02119(19) | 1 |
| C1S  | 0.5136(3)   | 0.1520(5)    | 1.0477(6)    | 0.054(3)    | 0.5 |
| Cl1S | 0.56140(14) | 0.22218(19)  | 1.1014(2)    | 0.0507(6)   | 0.5 |
| Cl2S | 0.44951(15) | 0.1926(3)    | 1.0606(5)    | 0.1026(15)  | 0.5 |

**Table SI-3.** Bond lengths [Å] and angles [°] for ML153F_0m.

| | | | |
|---|---|---|---|
| N1-C2 | 1.342(3) | C20-C21 | 1.537(3) |
| N1-C2#1 | 1.342(3) | C21-C29 | 1.374(10) |
| N3-C2 | 1.363(3) | C21-C22 | 1.521(6) |
| N3-C4 | 1.363(3) | C21-C28 | 1.528(4) |
| N3-B1 | 1.480(3) | C21-C29A | 1.682(8) |
| N15-C16 | 1.345(3) | C22-C23 | 1.514(7) |
| N15-C4 | 1.349(3) | C22-C29 | 1.947(10) |
| N17-C16 | 1.366(3) | C22-H22A | 0.9700 |
| N17-C16#1 | 1.366(3) | C22-H22B | 0.9700 |
| N17-B1 | 1.479(4) | C23-H23A | 0.9700 |
| C2-C14 | 1.446(3) | C23-H23B | 0.9701 |
| C4-C5 | 1.457(3) | C24-H24A | 0.9800 |
| C5-C6 | 1.390(3) | C24-H24B | 0.9800 |
| C5-C14 | 1.421(3) | C24-H24C | 0.9800 |
| C6-C7 | 1.401(3) | C25-H25A | 0.9800 |
| C6-H6 | 0.9500 | C25-H25B | 0.9800 |
| C7-C12 | 1.423(3) | C25-H25C | 0.9800 |
| C7-C8 | 1.536(3) | C26-H26A | 0.9800 |
| C8-C9 | 1.533(3) | C26-H26B | 0.9800 |
| C8-C26 | 1.535(3) | C26-H26C | 0.9800 |
| C8-C27 | 1.542(3) | C27-H27A | 0.9800 |
| C9-C10 | 1.508(4) | C27-H27B | 0.9800 |
| C9-H9A | 0.9900 | C27-H27C | 0.9800 |
| C9-H9B | 0.9900 | C28-H28A | 0.9800 |
| C10-C11 | 1.532(3) | C28-H28B | 0.9800 |
| C10-H10A | 0.9900 | C28-H28C | 0.9800 |
| C10-H10B | 0.9900 | C29-H29A | 0.9800 |
| C11-C25 | 1.528(4) | C29-H29B | 0.9800 |
| C11-C24 | 1.533(4) | C29-H29C | 0.9800 |
| C11-C12 | 1.537(3) | C29A-H29D | 0.9800 |
| C12-C13 | 1.389(3) | C29A-H29E | 0.9800 |
| C13-C14 | 1.390(3) | C29A-H29F | 0.9800 |
| C13-H13 | 0.9500 | B1-N3#1 | 1.480(3) |
| C16-C18 | 1.457(3) | B1-Cl1 | 1.853(4) |
| C18-C19 | 1.385(3) | C1S-Cl2S | 1.730(9) |
| C18-C18#1 | 1.418(4) | C1S-Cl1S | 1.750(7) |
| C19-C20 | 1.398(3) | C1S-H1S1 | 0.9900 |
| C19-H19 | 0.9500 | C1S-H1S2 | 0.9900 |
| C20-C20#1 | 1.413(5) | | |
| | | | |
| C2-N1-C2#1 | 117.3(3) | C7-C6-H6 | 119.7 |
| C2-N3-C4 | 113.63(19) | C6-C7-C12 | 119.6(2) |
| C2-N3-B1 | 122.4(2) | C6-C7-C8 | 117.7(2) |
| C4-N3-B1 | 123.3(2) | C12-C7-C8 | 122.6(2) |
| C16-N15-C4 | 117.41(18) | C9-C8-C26 | 107.0(2) |
| C16-N17-C16#1 | 113.6(3) | C9-C8-C7 | 111.37(19) |
| C16-N17-B1 | 122.34(13) | C26-C8-C7 | 111.61(19) |
| C16#1-N17-B1 | 122.34(13) | C9-C8-C27 | 110.59(19) |
| N1-C2-N3 | 122.4(2) | C26-C8-C27 | 108.3(2) |
| N1-C2-C14 | 131.0(2) | C7-C8-C27 | 107.97(19) |
| N3-C2-C14 | 105.34(19) | C10-C9-C8 | 113.4(2) |
| N15-C4-N3 | 121.43(19) | C10-C9-H9A | 108.9 |
| N15-C4-C5 | 131.98(19) | C8-C9-H9A | 108.9 |
| N3-C4-C5 | 105.18(18) | C10-C9-H9B | 108.9 |
| C6-C5-C14 | 119.1(2) | C8-C9-H9B | 108.9 |
| C6-C5-C4 | 133.9(2) | H9A-C9-H9B | 107.7 |
| C14-C5-C4 | 106.92(19) | C9-C10-C11 | 112.4(2) |
| C5-C6-C7 | 120.5(2) | C9-C10-H10A | 109.1 |
| C5-C6-H6 | 119.7 | C11-C10-H10A | 109.1 |
| C9-C10-H10B | 109.1 | H24A-C24-H24C | 109.5 |
| C11-C10-H10B | 109.1 | H24B-C24-H24C | 109.5 |
| H10A-C10-H10B | 107.9 | C11-C25-H25A | 109.5 |
| C25-C11-C10 | 111.6(2) | C11-C25-H25B | 109.5 |
| C25-C11-C24 | 108.8(2) | H25A-C25-H25B | 109.5 |
| C10-C11-C24 | 106.7(2) | C11-C25-H25C | 109.5 |
| C25-C11-C12 | 108.3(2) | H25A-C25-H25C | 109.5 |
| C10-C11-C12 | 110.0(2) | H25B-C25-H25C | 109.5 |
| C24-C11-C12 | 111.6(2) | C8-C26-H26A | 109.5 |
| C13-C12-C7 | 119.9(2) | C8-C26-H26B | 109.5 |
| C13-C12-C11 | 117.8(2) | H26A-C26-H26B | 109.5 |
| C7-C12-C11 | 122.30(19) | C8-C26-H26C | 109.5 |
| C12-C13-C14 | 120.0(2) | H26A-C26-H26C | 109.5 |
| C12-C13-H13 | 120.0 | H26B-C26-H26C | 109.5 |
| C14-C13-H13 | 120.0 | C8-C27-H27A | 109.5 |
| C13-C14-C5 | 120.7(2) | C8-C27-H27B | 109.5 |
| C13-C14-C2 | 131.7(2) | H27A-C27-H27B | 109.5 |
| C5-C14-C2 | 107.56(19) | C8-C27-H27C | 109.5 |
| N15-C16-N17 | 122.8(2) | H27A-C27-H27C | 109.5 |
| N15-C16-C18 | 130.1(2) | H27B-C27-H27C | 109.5 |
| N17-C16-C18 | 105.30(19) | C21-C28-H28A | 109.5 |
| C19-C18-C18#1 | 119.90(13) | C21-C28-H28B | 109.5 |
| C19-C18-C16 | 132.4(2) | H28A-C28-H28B | 109.5 |
| C18#1-C18-C16 | 107.34(12) | C21-C28-H28C | 109.5 |
| C18-C19-C20 | 120.4(2) | H28A-C28-H28C | 109.5 |
| C18-C19-H19 | 119.8 | H28B-C28-H28C | 109.5 |
| C20-C19-H19 | 119.8 | C21-C29-C22 | 51.0(3) |
| C19-C20-C20#1 | 119.71(13) | C21-C29-H29A | 109.5 |
| C19-C20-C21 | 117.4(2) | C22-C29-H29A | 159.2 |
| C20#1-C20-C21 | 122.90(13) | C21-C29-H29B | 109.5 |
| C29-C21-C22 | 84.4(5) | C22-C29-H29B | 75.8 |
| C29-C21-C28 | 113.9(5) | H29A-C29-H29B | 109.5 |
| C22-C21-C28 | 121.5(3) | C21-C29-H29C | 109.5 |
| C29-C21-C20 | 113.5(5) | C22-C29-H29C | 86.5 |
| C22-C21-C20 | 111.4(3) | H29A-C29-H29C | 109.5 |
| C28-C21-C20 | 110.1(2) | H29B-C29-H29C | 109.5 |
| C28-C21-C29A | 103.4(4) | C21-C29A-H29D | 109.5 |
| C20-C21-C29A | 107.6(4) | C21-C29A-H29E | 109.5 |
| C23-C22-C21 | 107.8(4) | H29D-C29A-H29E | 109.5 |
| C23-C22-C29 | 149.3(5) | C21-C29A-H29F | 109.5 |
| C21-C22-C29 | 44.6(4) | H29D-C29A-H29F | 109.5 |
| C23-C22-H22A | 110.5 | H29E-C29A-H29F | 109.5 |
| C21-C22-H22A | 110.3 | N17-B1-N3#1 | 105.42(18) |
| C29-C22-H22A | 94.7 | N17-B1-N3 | 105.42(18) |
| C23-C22-H22B | 109.7 | N3#1-B1-N3 | 105.0(3) |
| C21-C22-H22B | 109.9 | N17-B1-Cl1 | 115.7(2) |
| C29-C22-H22B | 76.7 | N3#1-B1-Cl1 | 112.22(16) |
| H22A-C22-H22B | 108.6 | N3-B1-Cl1 | 112.22(16) |
| C22-C23-H23A | 109.0 | Cl2S-C1S-Cl1S | 110.2(4) |
| C22-C23-H23B | 109.8 | Cl2S-C1S-H1S1 | 109.6 |
| H23A-C23-H23B | 108.2 | Cl1S-C1S-H1S1 | 109.6 |
| C11-C24-H24A | 109.5 | Cl2S-C1S-H1S2 | 109.6 |
| C11-C24-H24B | 109.5 | Cl1S-C1S-H1S2 | 109.6 |
| H24A-C24-H24B | 109.5 | H1S1-C1S-H1S2 | 108.1 |
| C11-C24-H24C | 109.5 | | |

___________________________________________________________

Symmetry transformations used to generate equivalent atoms:
#1 -x+1,y,z

**Table SI-4.** Anisotropic displacement parameters (Å$^2$) for ML153F_0m.
The anisotropic displacement factor exponent takes the form: $-2\pi^2[\, h^2 a^{*2} U^{11} + ... + 2\, h\, k\, a^*\, b^*\, U^{12}\,]$

|       | U$^{11}$      | U$^{22}$      | U$^{33}$      | U$^{23}$       | U$^{13}$       | U$^{12}$       |
|-------|---------------|---------------|---------------|----------------|----------------|----------------|
| N1    | 0.0136(13)    | 0.0147(13)    | 0.0207(14)    | 0.0030(11)     | 0.000          | 0.000          |
| N3    | 0.0135(9)     | 0.0100(9)     | 0.0155(9)     | 0.0009(7)      | 0.0006(7)      | 0.0013(7)      |
| N15   | 0.0156(9)     | 0.0129(9)     | 0.0123(9)     | -0.0002(7)     | 0.0014(7)      | 0.0011(7)      |
| N17   | 0.0153(13)    | 0.0113(12)    | 0.0114(12)    | 0.0010(10)     | 0.000          | 0.000          |
| C2    | 0.0173(11)    | 0.0102(10)    | 0.0173(11)    | 0.0029(9)      | 0.0010(9)      | 0.0025(9)      |
| C4    | 0.0122(10)    | 0.0117(10)    | 0.0147(10)    | -0.0022(9)     | 0.0022(9)      | 0.0010(8)      |
| C5    | 0.0167(11)    | 0.0100(10)    | 0.0148(10)    | 0.0005(8)      | 0.0019(8)      | 0.0025(9)      |
| C6    | 0.0153(10)    | 0.0122(10)    | 0.0148(10)    | -0.0007(9)     | 0.0028(9)      | 0.0001(9)      |
| C7    | 0.0164(11)    | 0.0109(10)    | 0.0161(10)    | -0.0032(9)     | 0.0000(9)      | 0.0017(8)      |
| C8    | 0.0185(11)    | 0.0173(11)    | 0.0199(12)    | 0.0001(9)      | -0.0031(9)     | -0.0025(9)     |
| C9    | 0.0220(12)    | 0.0222(12)    | 0.0318(15)    | 0.0045(11)     | -0.0094(11)    | -0.0037(10)    |
| C10   | 0.0293(13)    | 0.0249(13)    | 0.0228(13)    | 0.0037(10)     | -0.0088(11)    | -0.0007(11)    |
| C11   | 0.0214(12)    | 0.0212(12)    | 0.0165(12)    | 0.0062(9)      | -0.0038(9)     | -0.0004(10)    |
| C12   | 0.0191(10)    | 0.0128(10)    | 0.0132(11)    | -0.0013(8)     | -0.0010(9)     | 0.0024(8)      |
| C13   | 0.0188(11)    | 0.0141(10)    | 0.0159(10)    | 0.0025(9)      | 0.0013(9)      | 0.0019(9)      |
| C14   | 0.0157(11)    | 0.0118(10)    | 0.0168(11)    | 0.0001(8)      | 0.0017(9)      | 0.0014(8)      |
| C16   | 0.0168(11)    | 0.0143(11)    | 0.0100(10)    | -0.0016(9)     | 0.0027(9)      | -0.0014(9)     |
| C18   | 0.0201(11)    | 0.0139(10)    | 0.0094(9)     | 0.0008(8)      | 0.0016(9)      | 0.0012(9)      |
| C19   | 0.0190(10)    | 0.0170(10)    | 0.0163(11)    | 0.0021(9)      | 0.0024(10)     | -0.0019(9)     |
| C20   | 0.0263(12)    | 0.0137(11)    | 0.0139(10)    | 0.0023(9)      | 0.0025(9)      | -0.0004(10)    |
| C21   | 0.0269(14)    | 0.0163(12)    | 0.0429(16)    | 0.0090(11)     | 0.0101(12)     | -0.0015(11)    |
| C22   | 0.032(3)      | 0.013(2)      | 0.033(3)      | 0.004(2)       | -0.007(2)      | -0.004(2)      |
| C23   | 0.028(2)      | 0.015(2)      | 0.030(3)      | 0.010(2)       | -0.003(2)      | -0.0007(18)    |
| C24   | 0.0364(15)    | 0.0440(17)    | 0.0179(13)    | 0.0104(12)     | -0.0024(11)    | -0.0013(13)    |
| C25   | 0.0389(15)    | 0.0209(13)    | 0.0318(13)    | 0.0056(11)     | -0.0133(13)    | 0.0037(11)     |
| C26   | 0.0145(11)    | 0.0363(14)    | 0.0293(14)    | -0.0023(12)    | -0.0007(11)    | -0.0002(10)    |
| C27   | 0.0251(12)    | 0.0206(12)    | 0.0268(13)    | -0.0015(10)    | -0.0076(10)    | -0.0041(10)    |
| C28   | 0.0354(16)    | 0.0410(16)    | 0.0362(17)    | 0.0160(14)     | 0.0075(13)     | -0.0115(13)    |
| C29   | 0.030(5)      | 0.024(4)      | 0.046(5)      | 0.008(3)       | -0.010(4)      | -0.006(3)      |
| C29A  | 0.022(4)      | 0.018(4)      | 0.035(4)      | 0.002(3)       | -0.016(3)      | -0.005(3)      |
| B1    | 0.0137(17)    | 0.0111(16)    | 0.0141(17)    | -0.0016(14)    | 0.000          | 0.000          |
| Cl1   | 0.0270(4)     | 0.0151(4)     | 0.0215(4)     | -0.0066(3)     | 0.000          | 0.000          |
| C1S   | 0.081(8)      | 0.047(3)      | 0.035(3)      | 0.002(3)       | -0.009(3)      | -0.023(4)      |
| Cl1S  | 0.0548(13)    | 0.0534(15)    | 0.0440(11)    | 0.0083(9)      | -0.0010(9)     | -0.0125(11)    |
| Cl2S  | 0.062(2)      | 0.095(3)      | 0.151(4)      | 0.058(3)       | -0.045(2)      | -0.032(2)      |

Table SI-5. Hydrogen coordinates and isotropic displacement parameters (Å$^2$) for ML153F_0m.

|      | x       | y       | z       | U(eq)  | Occupancy |
|------|---------|---------|---------|--------|-----------|
| H6   | 0.7022  | 0.2278  | 0.6048  | 0.017  | 1         |
| H9A  | 0.8016  | 0.0648  | 0.8344  | 0.030  | 1         |
| H9B  | 0.8257  | 0.1425  | 0.9048  | 0.030  | 1         |
| H10A | 0.7454  | 0.1530  | 1.0196  | 0.031  | 1         |
| H10B | 0.7743  | 0.0665  | 1.0405  | 0.031  | 1         |
| H13  | 0.6038  | 0.0496  | 0.8582  | 0.019  | 1         |
| H19  | 0.5946  | 0.4383  | 0.3364  | 0.021  | 1         |
| H22A | 0.5486  | 0.7120  | 0.2629  | 0.031  | 0.5       |
| H22B | 0.5187  | 0.6744  | 0.3766  | 0.031  | 0.5       |
| H23A | 0.4878  | 0.6332  | 0.1366  | 0.029  | 0.5       |
| H23B | 0.4593  | 0.7050  | 0.2071  | 0.029  | 0.5       |
| H24A | 0.6331  | 0.0256  | 1.0330  | 0.049  | 1         |
| H24B | 0.6562  | 0.1133  | 1.0725  | 0.049  | 1         |
| H24C | 0.6842  | 0.0313  | 1.1215  | 0.049  | 1         |
| H25A | 0.6806  | -0.0596 | 0.8834  | 0.046  | 1         |
| H25B | 0.7315  | -0.0614 | 0.9727  | 0.046  | 1         |
| H25C | 0.7387  | -0.0334 | 0.8339  | 0.046  | 1         |
| H26A | 0.8392  | 0.1772  | 0.6810  | 0.040  | 1         |
| H26B | 0.7886  | 0.2050  | 0.6005  | 0.040  | 1         |
| H26C | 0.7979  | 0.1100  | 0.6302  | 0.040  | 1         |
| H27A | 0.8079  | 0.2797  | 0.8350  | 0.036  | 1         |
| H27B | 0.7497  | 0.2709  | 0.8963  | 0.036  | 1         |
| H27C | 0.7549  | 0.3003  | 0.7577  | 0.036  | 1         |
| H28A | 0.6183  | 0.6227  | 0.1519  | 0.056  | 1         |
| H28B | 0.5767  | 0.5558  | 0.1016  | 0.056  | 1         |
| H28C | 0.6243  | 0.5286  | 0.1911  | 0.056  | 1         |
| H29A | 0.6135  | 0.5764  | 0.4130  | 0.050  | 0.5       |
| H29B | 0.5623  | 0.6320  | 0.4449  | 0.050  | 0.5       |
| H29C | 0.6088  | 0.6671  | 0.3584  | 0.050  | 0.5       |
| H29D | 0.6335  | 0.5572  | 0.3933  | 0.038  | 0.5       |
| H29E | 0.5866  | 0.5963  | 0.4748  | 0.038  | 0.5       |
| H29F | 0.6230  | 0.6540  | 0.3904  | 0.038  | 0.5       |
| H1S1 | 0.5212  | 0.1388  | 0.9613  | 0.065  | 0.5       |
| H1S2 | 0.5160  | 0.1006  | 1.0955  | 0.065  | 0.5       |

**Table SI-6.** Torsion angles [°] for ML153F_0m.
______________________________________________________________________________

| | | | |
|---|---|---|---|
| C2#1-N1-C2-N3 | -6.4(4) | C16#1-N17-C16-C18 | -10.8(3) |
| C2#1-N1-C2-C14 | 158.84(17) | B1-N17-C16-C18 | -176.2(2) |
| C4-N3-C2-N1 | 156.6(2) | N15-C16-C18-C19 | 14.2(4) |
| B1-N3-C2-N1 | -14.2(3) | N17-C16-C18-C19 | 179.0(3) |
| C4-N3-C2-C14 | -11.9(2) | N15-C16-C18-C18#1 | -158.6(2) |
| B1-N3-C2-C14 | 177.3(2) | N17-C16-C18-C18#1 | 6.19(19) |
| C16-N15-C4-N3 | 6.9(3) | C18#1-C18-C19-C20 | -0.4(3) |
| C16-N15-C4-C5 | -157.4(2) | C16-C18-C19-C20 | -172.5(2) |
| C2-N3-C4-N15 | -156.0(2) | C18-C19-C20-C20#1 | 0.4(3) |
| B1-N3-C4-N15 | 14.7(3) | C18-C19-C20-C21 | 179.7(2) |
| C2-N3-C4-C5 | 12.0(2) | C19-C20-C21-C29 | -69.4(5) |
| B1-N3-C4-C5 | -177.4(2) | C20#1-C20-C21-C29 | 109.9(4) |
| N15-C4-C5-C6 | -19.3(4) | C19-C20-C21-C22 | -162.6(4) |
| N3-C4-C5-C6 | 174.6(2) | C20#1-C20-C21-C22 | 16.7(4) |
| N15-C4-C5-C14 | 159.3(2) | C19-C20-C21-C28 | 59.7(3) |
| N3-C4-C5-C14 | -6.9(2) | C20#1-C20-C21-C28 | -121.05(18) |
| C14-C5-C6-C7 | 2.4(3) | C19-C20-C21-C29A | -52.3(4) |
| C4-C5-C6-C7 | -179.2(2) | C20#1-C20-C21-C29A | 127.0(4) |
| C5-C6-C7-C12 | -3.1(3) | C29-C21-C22-C23 | -163.8(6) |
| C5-C6-C7-C8 | 178.3(2) | C28-C21-C22-C23 | 81.5(5) |
| C6-C7-C8-C9 | -168.6(2) | C20-C21-C22-C23 | -50.7(5) |
| C12-C7-C8-C9 | 12.9(3) | C28-C21-C22-C29 | -114.6(6) |
| C6-C7-C8-C26 | -49.0(3) | C20-C21-C22-C29 | 113.1(5) |
| C12-C7-C8-C26 | 132.4(2) | C28-C21-C29-C22 | 122.0(4) |
| C6-C7-C8-C27 | 69.8(3) | C20-C21-C29-C22 | -110.9(4) |
| C12-C7-C8-C27 | -108.7(2) | C16-N17-B1-N3#1 | 137.5(2) |
| C26-C8-C9-C10 | -163.7(2) | C16#1-N17-B1-N3#1 | -26.6(3) |
| C7-C8-C9-C10 | -41.5(3) | C16-N17-B1-N3 | 26.6(3) |
| C27-C8-C9-C10 | 78.6(3) | C16#1-N17-B1-N3 | -137.5(2) |
| C8-C9-C10-C11 | 62.5(3) | C16-N17-B1-Cl1 | -98.0(2) |
| C9-C10-C11-C25 | 71.0(3) | C16#1-N17-B1-Cl1 | 98.0(2) |
| C9-C10-C11-C24 | -170.3(2) | C2-N3-B1-N17 | 140.7(2) |
| C9-C10-C11-C12 | -49.2(3) | C4-N3-B1-N17 | -29.2(3) |
| C6-C7-C12-C13 | 1.2(3) | C2-N3-B1-N3#1 | 29.6(3) |
| C8-C7-C12-C13 | 179.7(2) | C4-N3-B1-N3#1 | -140.25(18) |
| C6-C7-C12-C11 | 178.1(2) | C2-N3-B1-Cl1 | -92.6(2) |
| C8-C7-C12-C11 | -3.4(3) | C4-N3-B1-Cl1 | 97.6(2) |
| C25-C11-C12-C13 | 75.7(3) | | |
| C10-C11-C12-C13 | -162.2(2) | | |
| C24-C11-C12-C13 | -44.1(3) | | |
| C25-C11-C12-C7 | -101.3(3) | | |
| C10-C11-C12-C7 | 20.8(3) | | |
| C24-C11-C12-C7 | 139.0(2) | | |
| C7-C12-C13-C14 | 1.3(3) | | |
| C11-C12-C13-C14 | -175.7(2) | | |
| C12-C13-C14-C5 | -2.0(3) | | |
| C12-C13-C14-C2 | 179.6(2) | | |
| C6-C5-C14-C13 | 0.1(3) | | |
| C4-C5-C14-C13 | -178.7(2) | | |
| C6-C5-C14-C2 | 178.88(19) | | |
| C4-C5-C14-C2 | 0.1(2) | | |
| N1-C2-C14-C13 | 18.2(4) | | |
| N3-C2-C14-C13 | -174.7(2) | | |
| N1-C2-C14-C5 | -160.4(3) | | |
| N3-C2-C14-C5 | 6.7(2) | | |
| C4-N15-C16-N17 | -9.1(3) | | |
| C4-N15-C16-C18 | 153.4(2) | | |
| C16#1-N17-C16-N15 | 155.38(15) | | |
| B1-N17-C16-N15 | -10.0(4) | | |

## 1.7.2 Crystal Structure of N$_2$-Pc$^*$H$_2$

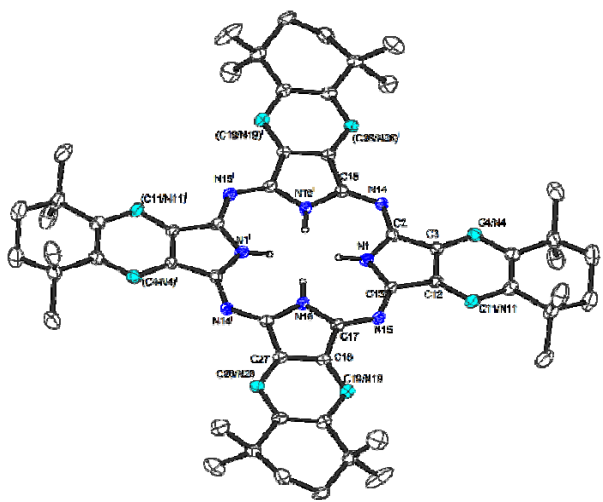

| | |
|---|---|
| Habitus, colour | plate, dark |
| Crystal size | 0.29 x 0.22 x 0.05 mm$^3$ |
| Crystal system | Monoclinic |
| Space group | P2$_1$/c          Z = 2 |
| Unit cell dimensions | a = 11.7078(4) Å          α = 90°. |
| | b = 23.2759(7) Å          β = 90.620(3)°. |
| | c = 12.9884(5) Å          γ = 90°. |
| Volume | 3539.3(2) Å$^3$ |
| Cell determination | 9908 peaks with Theta 2.3 to 25.2°. |
| Empirical formula | C$_{66}$ H$_{76}$ Cl$_{12}$ N$_{10}$ |
| Moiety formula | C$_{62}$ H$_{72}$ N$_{10}$, 4(CHCl$_3$) |
| Formula weight | 1434.76 |
| Density (calculated) | 1.346 Mg/m$^3$ |
| Absorption coefficient | 0.516 mm$^{-1}$ |
| F(000) | 1492 |
| | |
| Solution and refinement: | |
| Reflections collected | 35003 |
| Independent reflections | 6438 [R(int) = 0.0640] |
| Completeness to theta = 25.242° | 99.9 % |
| Observed reflections | 4815 [I>2sigma(I)] |
| Reflections used for refinement | 6438 |
| Absorption correction | Semi-empirical from equivalents [12] |
| Max. and min. transmission | 0.97 and 0.84 |
| Largest diff. peak and hole | 0.491 and -0.364 e.Å$^{-3}$ |
| Solution | Direct methods [8] |
| Refinement | Full-matrix least-squares on F$^2$ [8] |
| Treatment of hydrogen atoms | Mixture of constr. and indep. refinement |
| Data / restraints / parameters | 6438 / 203 / 602 |
| Goodness-of-fit on F$^2$ | 1.023 |
| R index (all data) | wR2 = 0.1178 |
| R index conventional [I>2sigma(I)] | R1 = 0.0489 |

**Table SI-7**. Atomic coordinates and equivalent isotropic displacement parameters (Å$^2$) for ML026F1_0m. U(eq) is defined as one third of the trace of the orthogonalized U$^{ij}$ tensor.

|      | x            | y            | z            | U(eq)       | Occupancy |
|------|--------------|--------------|--------------|-------------|-----------|
| N1   | 0.46545(16)  | 0.57666(8)   | 0.44525(14)  | 0.0153(4)   | 1         |
| C2   | 0.3647(2)    | 0.60621(10)  | 0.45428(17)  | 0.0157(5)   | 1         |
| C3   | 0.3755(2)    | 0.66129(10)  | 0.40227(17)  | 0.0156(5)   | 1         |
| C4   | 0.30198(19)  | 0.70643(10)  | 0.38898(17)  | 0.0184(5)   | 0.75      |
| N4   | 0.30198(19)  | 0.70643(10)  | 0.38898(17)  | 0.0184(5)   | 0.25      |
| C5   | 0.3388(2)    | 0.75406(11)  | 0.33488(19)  | 0.0204(5)   | 1         |
| C6   | 0.2532(2)    | 0.80284(11)  | 0.3170(2)    | 0.0268(6)   | 1         |
| C7   | 0.3150(2)    | 0.85787(12)  | 0.2849(2)    | 0.0333(7)   | 1         |
| C8   | 0.4002(2)    | 0.84750(12)  | 0.2011(2)    | 0.0338(7)   | 1         |
| C9   | 0.4967(2)    | 0.80731(11)  | 0.2343(2)    | 0.0268(6)   | 1         |
| C10  | 0.4506(2)    | 0.75531(11)  | 0.29354(18)  | 0.0202(5)   | 1         |
| C11  | 0.5243(2)    | 0.70961(10)  | 0.30888(17)  | 0.0193(5)   | 0.75      |
| N11  | 0.5243(2)    | 0.70961(10)  | 0.30888(17)  | 0.0193(5)   | 0.25      |
| C12  | 0.4863(2)    | 0.66329(10)  | 0.36324(17)  | 0.0160(5)   | 1         |
| C13  | 0.54049(19)  | 0.60941(10)  | 0.39127(17)  | 0.0154(5)   | 1         |
| N14  | 0.27091(16)  | 0.59019(8)   | 0.50347(14)  | 0.0148(4)   | 1         |
| N15  | 0.64678(16)  | 0.59652(8)   | 0.36288(15)  | 0.0171(4)   | 1         |
| C15  | 0.26385(19)  | 0.54088(10)  | 0.55360(17)  | 0.0151(5)   | 1         |
| N16  | 0.65576(16)  | 0.50203(8)   | 0.44055(14)  | 0.0152(4)   | 1         |
| C17  | 0.6968(2)    | 0.54665(10)  | 0.38416(17)  | 0.0161(5)   | 1         |
| C18  | 0.81001(19)  | 0.53126(10)  | 0.34750(17)  | 0.0154(5)   | 1         |
| C19  | 0.88411(19)  | 0.55965(10)  | 0.28424(17)  | 0.0177(5)   | 0.75      |
| N19  | 0.88411(19)  | 0.55965(10)  | 0.28424(17)  | 0.0177(5)   | 0.25      |
| C20  | 0.9858(2)    | 0.53270(10)  | 0.25919(18)  | 0.0178(5)   | 1         |
| C21  | 1.0653(2)    | 0.56474(11)  | 0.18526(19)  | 0.0230(6)   | 1         |
| C22  | 1.1584(4)    | 0.5251(2)    | 0.1448(4)    | 0.0275(11)  | 0.8       |
| C23  | 1.2118(3)    | 0.48884(18)  | 0.2312(3)    | 0.0250(9)   | 0.8       |
| C22A | 1.1861(17)   | 0.5330(11)   | 0.1886(15)   | 0.039(6)    | 0.2       |
| C23A | 1.1811(17)   | 0.4745(8)    | 0.1838(17)   | 0.040(4)    | 0.2       |
| C24  | 1.1240(2)    | 0.44669(11)  | 0.27591(18)  | 0.0190(5)   | 1         |
| C25  | 1.0129(2)    | 0.47795(10)  | 0.30069(17)  | 0.0170(5)   | 1         |
| C26  | 0.93612(18)  | 0.45022(10)  | 0.36436(17)  | 0.0162(5)   | 0.75      |
| N26  | 0.93612(18)  | 0.45022(10)  | 0.36436(17)  | 0.0162(5)   | 0.25      |
| C27  | 0.83524(19)  | 0.47700(10)  | 0.38659(17)  | 0.0158(5)   | 1         |
| C28  | 0.9988(3)    | 0.58457(17)  | 0.0908(2)    | 0.0527(10)  | 1         |
| C29  | 1.1156(3)    | 0.61711(13)  | 0.2400(3)    | 0.0416(8)   | 1         |
| C30  | 1.1785(5)    | 0.4210(2)    | 0.3759(4)    | 0.0294(12)  | 0.8       |
| C31  | 1.1000(5)    | 0.3983(2)    | 0.2006(4)    | 0.0241(11)  | 0.8       |
| C30A | 1.193(2)     | 0.4464(8)    | 0.3646(18)   | 0.035(6)    | 0.2       |
| C31A | 1.0955(19)   | 0.3834(11)   | 0.2407(18)   | 0.031(5)    | 0.2       |
| C33  | 0.1689(2)    | 0.78408(13)  | 0.2317(2)    | 0.0380(7)   | 1         |
| C34  | 0.1873(3)    | 0.81629(13)  | 0.4147(3)    | 0.0427(8)   | 1         |
| C35  | 0.5593(3)    | 0.78816(14)  | 0.1376(2)    | 0.0475(9)   | 1         |
| C36  | 0.5807(3)    | 0.83935(13)  | 0.3052(3)    | 0.0411(8)   | 1         |
| C1S  | 0.4189(4)    | 1.07467(18)  | 0.3759(3)    | 0.0589(10)  | 1         |
| Cl1  | 0.3082(13)   | 0.9987(7)    | 0.4049(13)   | 0.087(4)    | 0.089(3)  |
| Cl2  | 0.384(2)     | 0.9986(8)    | 0.3634(16)   | 0.080(4)    | 0.079(3)  |
| Cl3  | 0.4567(5)    | 1.00427(14)  | 0.3302(4)    | 0.0668(13)  | 0.635(3)  |
| Cl4  | 0.4849(10)   | 1.0172(5)    | 0.3321(10)   | 0.069(3)    | 0.234(3)  |
| Cl5  | 0.5264(15)   | 1.0658(9)    | 0.3633(14)   | 0.080(4)    | 0.070(2)  |
| Cl6  | 0.54339(12)  | 1.11200(7)   | 0.41534(12)  | 0.0615(4)   | 0.843(3)  |
| Cl7  | 0.488(3)     | 1.1245(12)   | 0.411(2)     | 0.077(6)    | 0.054(3)  |
| Cl8  | 0.345(2)     | 1.1101(9)    | 0.4684(17)   | 0.137(8)    | 0.103(3)  |
| Cl9  | 0.31974(12)  | 1.07437(15)  | 0.47132(11)  | 0.0788(8)   | 0.794(3)  |
| Cl10 | 0.3195(16)   | 1.0394(10)   | 0.4564(17)   | 0.098(5)    | 0.099(3)  |
| C2S  | 0.8490(3)    | 0.70760(18)  | 0.4147(3)    | 0.0404(10)  | 0.9       |
| C2SA | 0.962(4) | 0.733(2) | 0.560(4) | 0.078(6) | 0.1 |
| Cl11 | 0.844(2) | 0.7483(13) | 0.650(2) | 0.092(6) | 0.049(2) |
| Cl12 | 0.788(2) | 0.8059(13) | 0.5125(17) | 0.090(5) | 0.058(3) |
| Cl13 | 0.80543(14) | 0.76779(9) | 0.49037(10) | 0.0609(5) | 0.785(3) |
| Cl14 | 0.7662(10) | 0.7308(7) | 0.4884(9) | 0.088(3) | 0.123(3) |
| Cl15 | 0.887(2) | 0.7915(12) | 0.4584(19) | 0.097(5) | 0.058(2) |
| Cl16 | 0.7871(13) | 0.6961(7) | 0.4299(14) | 0.090(5) | 0.095(3) |
| Cl17 | 0.85946(7) | 0.72932(4) | 0.28604(6) | 0.0343(3) | 0.884(2) |
| Cl18 | 0.9371(18) | 0.6712(10) | 0.4653(18) | 0.068(5) | 0.079(3) |
| Cl19 | 0.98215(13) | 0.68209(7) | 0.46156(13) | 0.0437(4) | 0.809(3) |
| Cl20 | 0.9905(18) | 0.7094(11) | 0.524(2) | 0.081(5) | 0.061(3) |

**Table SI-8.** Bond lengths [Å] and angles [°] for ML026F1_0m.
___________________________________________

| | | | |
|---|---|---|---|
| N1-C13 | 1.363(3) | C22A-H22C | 0.9900 |
| N1-C2 | 1.371(3) | C22A-H22D | 0.9900 |
| N1-H1 | 0.8800 | C23A-C24 | 1.521(19) |
| C2-N14 | 1.330(3) | C23A-H23C | 0.9900 |
| C2-C3 | 1.455(3) | C23A-H23D | 0.9900 |
| C3-N4 | 1.368(3) | C24-C30A | 1.40(3) |
| C3-C4 | 1.368(3) | C24-C31 | 1.516(7) |
| C3-C12 | 1.398(3) | C24-C25 | 1.528(3) |
| C4-C5 | 1.384(3) | C24-C30 | 1.561(6) |
| C4-H4 | 0.9500 | C24-C31A | 1.58(3) |
| N4-C5 | 1.384(3) | C25-N26 | 1.387(3) |
| C5-C10 | 1.421(3) | C25-C26 | 1.387(3) |
| C5-C6 | 1.531(3) | C26-C27 | 1.369(3) |
| C6-C34 | 1.524(4) | C26-H26 | 0.9500 |
| C6-C7 | 1.532(4) | N26-C27 | 1.369(3) |
| C6-C33 | 1.539(4) | C27-C15#1 | 1.464(3) |
| C7-C8 | 1.503(4) | C28-H28A | 0.9800 |
| C7-H7A | 0.9900 | C28-H28B | 0.9800 |
| C7-H7B | 0.9900 | C28-H28C | 0.9800 |
| C8-C9 | 1.525(4) | C29-H29A | 0.9800 |
| C8-H8A | 0.9900 | C29-H29B | 0.9800 |
| C8-H8B | 0.9900 | C29-H29C | 0.9800 |
| C9-C35 | 1.527(4) | C30-H30A | 0.9800 |
| C9-C36 | 1.534(4) | C30-H30B | 0.9800 |
| C9-C10 | 1.535(3) | C30-H30C | 0.9800 |
| C10-N11 | 1.382(3) | C31-H31A | 0.9800 |
| C10-C11 | 1.382(3) | C31-H31B | 0.9800 |
| C11-C12 | 1.366(3) | C31-H31C | 0.9800 |
| C11-H11 | 0.9500 | C30A-H30D | 0.9800 |
| N11-C12 | 1.366(3) | C30A-H30E | 0.9800 |
| C12-C13 | 1.450(3) | C30A-H30F | 0.9800 |
| C13-N15 | 1.336(3) | C31A-H31D | 0.9800 |
| N14-C15 | 1.323(3) | C31A-H31E | 0.9800 |
| N15-C17 | 1.328(3) | C31A-H31F | 0.9800 |
| C15-N16#1 | 1.374(3) | C33-H33A | 0.9800 |
| C15-C27#1 | 1.464(3) | C33-H33B | 0.9800 |
| N16-C17 | 1.362(3) | C33-H33C | 0.9800 |
| N16-C15#1 | 1.374(3) | C34-H34A | 0.9800 |
| N16-H16 | 0.8800 | C34-H34B | 0.9800 |
| C17-C18 | 1.458(3) | C34-H34C | 0.9800 |
| C18-N19 | 1.371(3) | C35-H35A | 0.9800 |
| C18-C19 | 1.371(3) | C35-H35B | 0.9800 |
| C18-C27 | 1.392(3) | C35-H35C | 0.9800 |
| C19-C20 | 1.388(3) | C36-H36A | 0.9800 |
| C19-H19 | 0.9500 | C36-H36B | 0.9800 |
| N19-C20 | 1.388(3) | C36-H36C | 0.9800 |
| C20-C25 | 1.418(3) | C1S-Cl5 | 1.287(17) |
| C20-C21 | 1.537(3) | C1S-Cl7 | 1.48(3) |
| C21-C28 | 1.518(4) | C1S-Cl4 | 1.648(12) |
| C21-C29 | 1.527(4) | C1S-Cl8 | 1.701(18) |
| C21-C22 | 1.527(6) | C1S-Cl9 | 1.708(4) |
| C21-C22A | 1.60(2) | C1S-Cl6 | 1.768(5) |
| C22-C23 | 1.531(6) | C1S-Cl10 | 1.774(18) |
| C22-H22A | 0.9900 | C1S-Cl3 | 1.799(6) |
| C22-H22B | 0.9900 | C1S-Cl2 | 1.82(2) |
| C23-C24 | 1.539(5) | C1S-Cl1 | 2.227(18) |
| C23-H23A | 0.9900 | C1S-H1S | 1.09(5) |
| C23-H23B | 0.9900 | Cl1-Cl2 | 1.04(3) |
| C22A-C23A | 1.36(3) | Cl1-Cl10 | 1.17(3) |
| Cl1-Cl4 | 2.32(2) | C2SA-Cl18 | 1.92(6) |
| Cl2-Cl4 | 1.33(3) | C2SA-Cl15 | 2.08(6) |
| Cl2-Cl10 | 1.72(3) | C2SA-H2SA | 1.0000 |
| Cl2-Cl5 | 2.29(3) | Cl11-Cl14 | 2.32(3) |
| Cl4-Cl5 | 1.29(2) | Cl11-Cl12 | 2.32(4) |
| Cl5-Cl7 | 1.57(4) | Cl12-Cl15 | 1.40(3) |
| Cl7-Cl8 | 1.87(4) | Cl12-Cl14 | 1.79(3) |
| Cl8-Cl10 | 1.68(3) | Cl14-Cl16 | 1.138(18) |
| C2S-Cl17 | 1.751(4) | Cl14-Cl15 | 2.04(3) |
| C2S-Cl19 | 1.770(4) | Cl14-Cl18 | 2.46(2) |
| C2S-Cl13 | 1.789(5) | Cl15-Cl20 | 2.41(3) |
| C2S-H2S | 1.0000 | Cl16-Cl18 | 1.90(2) |
| C2SA-Cl11 | 1.85(6) | Cl18-Cl20 | 1.32(3) |
| | | | |
| C13-N1-C2 | 108.86(19) | C11-C10-C9 | 117.2(2) |
| C13-N1-H1 | 125.6 | C5-C10-C9 | 122.4(2) |
| C2-N1-H1 | 125.6 | C12-C11-C10 | 118.4(2) |
| N14-C2-N1 | 128.0(2) | C12-C11-H11 | 120.8 |
| N14-C2-C3 | 123.1(2) | C10-C11-H11 | 120.8 |
| N1-C2-C3 | 108.85(19) | C12-N11-C10 | 118.4(2) |
| N4-C3-C12 | 120.9(2) | N11-C12-C3 | 121.5(2) |
| C4-C3-C12 | 120.9(2) | C11-C12-C3 | 121.5(2) |
| N4-C3-C2 | 132.7(2) | N11-C12-C13 | 131.9(2) |
| C4-C3-C2 | 132.7(2) | C11-C12-C13 | 131.9(2) |
| C12-C3-C2 | 106.4(2) | C3-C12-C13 | 106.58(19) |
| C3-C4-C5 | 118.7(2) | N15-C13-N1 | 128.6(2) |
| C3-C4-H4 | 120.6 | N15-C13-C12 | 122.1(2) |
| C5-C4-H4 | 120.6 | N1-C13-C12 | 109.27(19) |
| C3-N4-C5 | 118.7(2) | C15-N14-C2 | 122.4(2) |
| N4-C5-C10 | 120.1(2) | C17-N15-C13 | 123.3(2) |
| C4-C5-C10 | 120.1(2) | N14-C15-N16#1 | 127.7(2) |
| N4-C5-C6 | 117.7(2) | N14-C15-C27#1 | 124.2(2) |
| C4-C5-C6 | 117.7(2) | N16#1-C15-C27#1 | 108.09(19) |
| C10-C5-C6 | 122.2(2) | C17-N16-C15#1 | 109.80(19) |
| C34-C6-C5 | 111.3(2) | C17-N16-H16 | 125.1 |
| C34-C6-C7 | 107.4(2) | C15#1-N16-H16 | 125.1 |
| C5-C6-C7 | 110.5(2) | N15-C17-N16 | 128.4(2) |
| C34-C6-C33 | 109.4(2) | N15-C17-C18 | 123.2(2) |
| C5-C6-C33 | 108.2(2) | N16-C17-C18 | 108.4(2) |
| C7-C6-C33 | 110.1(2) | N19-C18-C27 | 121.5(2) |
| C8-C7-C6 | 112.5(2) | C19-C18-C27 | 121.5(2) |
| C8-C7-H7A | 109.1 | N19-C18-C17 | 131.3(2) |
| C6-C7-H7A | 109.1 | C19-C18-C17 | 131.3(2) |
| C8-C7-H7B | 109.1 | C27-C18-C17 | 107.1(2) |
| C6-C7-H7B | 109.1 | C18-C19-C20 | 118.2(2) |
| H7A-C7-H7B | 107.8 | C18-C19-H19 | 120.9 |
| C7-C8-C9 | 112.9(2) | C20-C19-H19 | 120.9 |
| C7-C8-H8A | 109.0 | C18-N19-C20 | 118.2(2) |
| C9-C8-H8A | 109.0 | N19-C20-C25 | 120.4(2) |
| C7-C8-H8B | 109.0 | C19-C20-C25 | 120.4(2) |
| C9-C8-H8B | 109.0 | N19-C20-C21 | 117.0(2) |
| H8A-C8-H8B | 107.8 | C19-C20-C21 | 117.0(2) |
| C8-C9-C35 | 107.9(2) | C25-C20-C21 | 122.6(2) |
| C8-C9-C36 | 109.8(2) | C28-C21-C29 | 109.0(3) |
| C35-C9-C36 | 109.0(3) | C28-C21-C22 | 105.5(3) |
| C8-C9-C10 | 111.2(2) | C29-C21-C22 | 111.7(3) |
| C35-C9-C10 | 110.9(2) | C28-C21-C20 | 110.1(2) |
| C36-C9-C10 | 108.0(2) | C29-C21-C20 | 109.2(2) |
| N11-C10-C5 | 120.4(2) | C22-C21-C20 | 111.1(3) |
| C11-C10-C5 | 120.4(2) | C28-C21-C22A | 127.5(7) |
| N11-C10-C9 | 117.2(2) | C29-C21-C22A | 91.1(10) |
| C20-C21-C22A | 107.5(8) | H29A-C29-H29C | 109.5 |
| C21-C22-C23 | 111.6(3) | H29B-C29-H29C | 109.5 |
| C21-C22-H22A | 109.3 | C24-C30-H30A | 109.5 |
| C23-C22-H22A | 109.3 | C24-C30-H30B | 109.5 |
| C21-C22-H22B | 109.3 | H30A-C30-H30B | 109.5 |
| C23-C22-H22B | 109.3 | C24-C30-H30C | 109.5 |
| H22A-C22-H22B | 108.0 | H30A-C30-H30C | 109.5 |
| C22-C23-C24 | 111.0(3) | H30B-C30-H30C | 109.5 |
| C22-C23-H23A | 109.4 | C24-C31-H31A | 109.5 |
| C24-C23-H23A | 109.4 | C24-C31-H31B | 109.5 |
| C22-C23-H23B | 109.4 | H31A-C31-H31B | 109.5 |
| C24-C23-H23B | 109.4 | C24-C31-H31C | 109.5 |
| H23A-C23-H23B | 108.0 | H31A-C31-H31C | 109.5 |
| C23A-C22A-C21 | 115.0(18) | H31B-C31-H31C | 109.5 |
| C23A-C22A-H22C | 108.5 | C24-C30A-H30D | 109.5 |
| C21-C22A-H22C | 108.5 | C24-C30A-H30E | 109.5 |
| C23A-C22A-H22D | 108.5 | H30D-C30A-H30E | 109.5 |
| C21-C22A-H22D | 108.5 | C24-C30A-H30F | 109.5 |
| H22C-C22A-H22D | 107.5 | H30D-C30A-H30F | 109.5 |
| C22A-C23A-C24 | 114.1(16) | H30E-C30A-H30F | 109.5 |
| C22A-C23A-H23C | 108.7 | C24-C31A-H31D | 109.5 |
| C24-C23A-H23C | 108.7 | C24-C31A-H31E | 109.5 |
| C22A-C23A-H23D | 108.7 | H31D-C31A-H31E | 109.5 |
| C24-C23A-H23D | 108.7 | C24-C31A-H31F | 109.5 |
| H23C-C23A-H23D | 107.6 | H31D-C31A-H31F | 109.5 |
| C30A-C24-C23A | 113.4(13) | H31E-C31A-H31F | 109.5 |
| C30A-C24-C25 | 108.2(9) | C6-C33-H33A | 109.5 |
| C31-C24-C25 | 109.7(3) | C6-C33-H33B | 109.5 |
| C23A-C24-C25 | 110.2(8) | H33A-C33-H33B | 109.5 |
| C31-C24-C23 | 110.5(3) | C6-C33-H33C | 109.5 |
| C25-C24-C23 | 110.5(2) | H33A-C33-H33C | 109.5 |
| C31-C24-C30 | 108.9(3) | H33B-C33-H33C | 109.5 |
| C25-C24-C30 | 110.4(3) | C6-C34-H34A | 109.5 |
| C23-C24-C30 | 106.9(3) | C6-C34-H34B | 109.5 |
| C30A-C24-C31A | 110.6(12) | H34A-C34-H34B | 109.5 |
| C23A-C24-C31A | 105.2(11) | C6-C34-H34C | 109.5 |
| C25-C24-C31A | 109.1(8) | H34A-C34-H34C | 109.5 |
| N26-C25-C20 | 120.1(2) | H34B-C34-H34C | 109.5 |
| C26-C25-C20 | 120.1(2) | C9-C35-H35A | 109.5 |
| N26-C25-C24 | 117.6(2) | C9-C35-H35B | 109.5 |
| C26-C25-C24 | 117.6(2) | H35A-C35-H35B | 109.5 |
| C20-C25-C24 | 122.4(2) | C9-C35-H35C | 109.5 |
| C27-C26-C25 | 118.7(2) | H35A-C35-H35C | 109.5 |
| C27-C26-H26 | 120.7 | H35B-C35-H35C | 109.5 |
| C25-C26-H26 | 120.7 | C9-C36-H36A | 109.5 |
| C27-N26-C25 | 118.7(2) | C9-C36-H36B | 109.5 |
| N26-C27-C18 | 121.1(2) | H36A-C36-H36B | 109.5 |
| C26-C27-C18 | 121.1(2) | C9-C36-H36C | 109.5 |
| N26-C27-C15#1 | 132.3(2) | H36A-C36-H36C | 109.5 |
| C26-C27-C15#1 | 132.3(2) | H36B-C36-H36C | 109.5 |
| C18-C27-C15#1 | 106.6(2) | Cl5-C1S-Cl7 | 68.5(16) |
| C21-C28-H28A | 109.5 | Cl5-C1S-Cl4 | 50.5(11) |
| C21-C28-H28B | 109.5 | Cl7-C1S-Cl4 | 118.9(14) |
| H28A-C28-H28B | 109.5 | Cl5-C1S-Cl8 | 132.2(13) |
| C21-C28-H28C | 109.5 | Cl7-C1S-Cl8 | 71.5(15) |
| H28A-C28-H28C | 109.5 | Cl4-C1S-Cl8 | 151.7(9) |
| H28B-C28-H28C | 109.5 | Cl9-C1S-Cl6 | 110.9(2) |
| C21-C29-H29A | 109.5 | Cl5-C1S-Cl10 | 130.5(11) |
| C21-C29-H29B | 109.5 | Cl7-C1S-Cl10 | 122.5(14) |
| H29A-C29-H29B | 109.5 | Cl4-C1S-Cl10 | 98.1(9) |
| C21-C29-H29C | 109.5 | Cl8-C1S-Cl10 | 57.7(9) |
| Cl9-C1S-Cl3 | 114.0(3) | Cl5-Cl7-Cl8 | 105.2(19) |
| Cl6-C1S-Cl3 | 109.8(3) | Cl10-Cl8-C1S | 63.3(10) |
| Cl5-C1S-Cl2 | 92.9(13) | Cl10-Cl8-Cl7 | 107.3(15) |
| Cl7-C1S-Cl2 | 155.0(14) | C1S-Cl8-Cl7 | 48.8(10) |
| Cl4-C1S-Cl2 | 44.5(9) | Cl1-Cl10-Cl8 | 150(2) |
| Cl8-C1S-Cl2 | 114.7(11) | Cl1-Cl10-Cl2 | 36.5(12) |
| Cl10-C1S-Cl2 | 57.1(11) | Cl8-Cl10-Cl2 | 121.9(15) |
| Cl5-C1S-Cl1 | 117.8(11) | Cl1-Cl10-C1S | 96.2(15) |
| Cl7-C1S-Cl1 | 152.2(11) | Cl8-Cl10-C1S | 59.0(9) |
| Cl4-C1S-Cl1 | 72.0(6) | Cl2-Cl10-C1S | 62.9(10) |
| Cl8-C1S-Cl1 | 87.9(9) | Cl17-C2S-Cl19 | 110.8(2) |
| Cl10-C1S-Cl1 | 31.4(8) | Cl17-C2S-Cl13 | 108.8(2) |
| Cl2-C1S-Cl1 | 27.7(8) | Cl19-C2S-Cl13 | 109.2(2) |
| Cl5-C1S-H1S | 116(3) | Cl17-C2S-H2S | 109.4 |
| Cl7-C1S-H1S | 99(3) | Cl19-C2S-H2S | 109.4 |
| Cl4-C1S-H1S | 109(2) | Cl13-C2S-H2S | 109.4 |
| Cl8-C1S-H1S | 94(3) | Cl11-C2SA-Cl18 | 116(3) |
| Cl9-C1S-H1S | 103(2) | Cl11-C2SA-Cl15 | 88(3) |
| Cl6-C1S-H1S | 114(2) | Cl18-C2SA-Cl15 | 91(2) |
| Cl10-C1S-H1S | 109(3) | Cl11-C2SA-H2SA | 118.1 |
| Cl3-C1S-H1S | 106(2) | Cl18-C2SA-H2SA | 118.1 |
| Cl2-C1S-H1S | 105(2) | Cl15-C2SA-H2SA | 118.1 |
| Cl1-C1S-H1S | 101(2) | C2SA-Cl11-Cl14 | 71.6(19) |
| Cl2-Cl1-Cl10 | 102(2) | C2SA-Cl11-Cl12 | 80(2) |
| Cl2-Cl1-C1S | 54.2(13) | Cl14-Cl11-Cl12 | 45.5(9) |
| Cl10-Cl1-C1S | 52.4(11) | Cl15-Cl12-Cl14 | 78.4(18) |
| Cl2-Cl1-Cl4 | 12.7(13) | Cl15-Cl12-Cl11 | 91.2(18) |
| Cl10-Cl1-Cl4 | 89.2(13) | Cl14-Cl12-Cl11 | 67.0(12) |
| C1S-Cl1-Cl4 | 42.4(4) | Cl16-Cl14-Cl12 | 140.9(17) |
| Cl1-Cl2-Cl4 | 157(2) | Cl16-Cl14-Cl15 | 102.1(14) |
| Cl1-Cl2-Cl10 | 41.6(14) | Cl12-Cl14-Cl15 | 42.2(11) |
| Cl4-Cl2-Cl10 | 115.7(15) | Cl16-Cl14-Cl11 | 130.2(13) |
| Cl1-Cl2-C1S | 98.2(16) | Cl12-Cl14-Cl11 | 67.4(12) |
| Cl4-Cl2-C1S | 60.7(9) | Cl15-Cl14-Cl11 | 77.6(11) |
| Cl10-Cl2-C1S | 60.0(9) | Cl16-Cl14-Cl18 | 48.4(10) |
| Cl1-Cl2-Cl5 | 128.5(18) | Cl12-Cl14-Cl18 | 117.0(11) |
| Cl4-Cl2-Cl5 | 28.8(8) | Cl15-Cl14-Cl18 | 78.4(9) |
| Cl10-Cl2-Cl5 | 87.0(12) | Cl11-Cl14-Cl18 | 84.1(10) |
| C1S-Cl2-Cl5 | 34.2(6) | Cl12-Cl15-Cl14 | 59.4(17) |
| Cl5-Cl4-Cl2 | 121.5(13) | Cl12-Cl15-C2SA | 101(2) |
| Cl5-Cl4-C1S | 50.1(8) | Cl14-Cl15-C2SA | 73.6(18) |
| Cl2-Cl4-C1S | 74.8(9) | Cl12-Cl15-Cl20 | 115(2) |
| Cl5-Cl4-Cl1 | 111.6(11) | Cl14-Cl15-Cl20 | 74.4(11) |
| Cl2-Cl4-Cl1 | 10.0(10) | C2SA-Cl15-Cl20 | 18.9(14) |
| C1S-Cl4-Cl1 | 65.6(6) | Cl14-Cl16-Cl18 | 105.0(13) |
| C1S-Cl5-Cl4 | 79.4(13) | Cl20-Cl18-Cl16 | 111.4(17) |
| C1S-Cl5-Cl7 | 61.7(14) | Cl20-Cl18-C2SA | 19.4(18) |
| Cl4-Cl5-Cl7 | 141.0(19) | Cl16-Cl18-C2SA | 93.5(19) |
| C1S-Cl5-Cl2 | 52.8(9) | Cl20-Cl18-Cl14 | 86.0(15) |
| Cl4-Cl5-Cl2 | 29.6(9) | Cl16-Cl18-Cl14 | 26.6(6) |
| Cl7-Cl5-Cl2 | 112.6(16) | C2SA-Cl18-Cl14 | 67.4(17) |
| C1S-Cl7-Cl5 | 49.8(11) | Cl18-Cl20-Cl15 | 95.7(17) |
| C1S-Cl7-Cl8 | 59.6(13) | | |

___________________________________________________________________

Symmetry transformations used to generate equivalent atoms:
#1 -x+1,-y+1,-z+1

**Table SI-9.** Anisotropic displacement parameters ($Å^2$) for ML026F1_0m.
The anisotropic displacement factor exponent takes the form: $-2\pi^2[ h^2 a^{*2}U^{11} + ... + 2 h k a^* b^* U^{12}]$

|      | $U^{11}$    | $U^{22}$    | $U^{33}$    | $U^{23}$     | $U^{13}$     | $U^{12}$     |
|------|-------------|-------------|-------------|--------------|--------------|--------------|
| N1   | 0.0131(10)  | 0.0183(10)  | 0.0147(10)  | 0.0016(8)    | 0.0015(8)    | 0.0013(8)    |
| C2   | 0.0179(12)  | 0.0154(12)  | 0.0139(12)  | -0.0022(9)   | -0.0003(9)   | 0.0006(10)   |
| C3   | 0.0167(12)  | 0.0150(12)  | 0.0149(12)  | -0.0002(9)   | 0.0002(9)    | 0.0008(10)   |
| C4   | 0.0167(12)  | 0.0184(12)  | 0.0202(12)  | -0.0007(10)  | 0.0023(9)    | 0.0010(10)   |
| N4   | 0.0167(12)  | 0.0184(12)  | 0.0202(12)  | -0.0007(10)  | 0.0023(9)    | 0.0010(10)   |
| C5   | 0.0208(13)  | 0.0205(13)  | 0.0199(13)  | -0.0022(10)  | 0.0001(10)   | -0.0006(11)  |
| C6   | 0.0256(14)  | 0.0201(14)  | 0.0348(15)  | 0.0019(12)   | 0.0036(12)   | 0.0048(11)   |
| C7   | 0.0317(16)  | 0.0253(15)  | 0.0430(17)  | 0.0067(13)   | -0.0007(13)  | 0.0037(12)   |
| C8   | 0.0312(16)  | 0.0294(16)  | 0.0407(17)  | 0.0159(13)   | 0.0001(13)   | 0.0002(13)   |
| C9   | 0.0256(14)  | 0.0273(15)  | 0.0277(14)  | 0.0121(12)   | 0.0007(11)   | 0.0007(12)   |
| C10  | 0.0206(13)  | 0.0229(13)  | 0.0170(12)  | 0.0025(10)   | -0.0023(10)  | -0.0023(11)  |
| C11  | 0.0176(12)  | 0.0224(13)  | 0.0181(12)  | 0.0059(10)   | 0.0013(9)    | 0.0007(10)   |
| N11  | 0.0176(12)  | 0.0224(13)  | 0.0181(12)  | 0.0059(10)   | 0.0013(9)    | 0.0007(10)   |
| C12  | 0.0161(12)  | 0.0179(12)  | 0.0140(12)  | 0.0021(9)    | -0.0001(9)   | 0.0012(10)   |
| C13  | 0.0151(12)  | 0.0184(12)  | 0.0128(11)  | 0.0030(9)    | 0.0009(9)    | -0.0005(10)  |
| N14  | 0.0148(10)  | 0.0149(10)  | 0.0146(10)  | 0.0012(8)    | 0.0010(8)    | 0.0003(8)    |
| N15  | 0.0151(10)  | 0.0193(11)  | 0.0171(10)  | 0.0047(8)    | 0.0033(8)    | 0.0016(8)    |
| C15  | 0.0126(12)  | 0.0183(12)  | 0.0145(12)  | -0.0012(10)  | 0.0004(9)    | -0.0008(10)  |
| N16  | 0.0138(10)  | 0.0161(10)  | 0.0156(10)  | 0.0023(8)    | 0.0006(8)    | 0.0005(8)    |
| C17  | 0.0156(12)  | 0.0182(13)  | 0.0145(12)  | 0.0014(9)    | 0.0011(9)    | -0.0004(10)  |
| C18  | 0.0141(12)  | 0.0197(13)  | 0.0124(11)  | -0.0010(9)   | 0.0011(9)    | -0.0004(10)  |
| C19  | 0.0185(12)  | 0.0186(12)  | 0.0160(12)  | 0.0022(9)    | -0.0004(9)   | -0.0016(10)  |
| N19  | 0.0185(12)  | 0.0186(12)  | 0.0160(12)  | 0.0022(9)    | -0.0004(9)   | -0.0016(10)  |
| C20  | 0.0158(12)  | 0.0227(13)  | 0.0147(12)  | -0.0005(10)  | -0.0006(10)  | -0.0052(10)  |
| C21  | 0.0219(13)  | 0.0247(14)  | 0.0225(13)  | 0.0026(11)   | 0.0043(11)   | -0.0045(11)  |
| C22  | 0.030(3)    | 0.031(2)    | 0.022(3)    | 0.005(2)     | 0.012(2)     | -0.0058(18)  |
| C23  | 0.013(2)    | 0.024(2)    | 0.038(3)    | -0.0029(19)  | 0.0085(17)   | -0.0014(16)  |
| C22A | 0.017(10)   | 0.084(18)   | 0.018(11)   | 0.024(11)    | 0.009(8)     | 0.016(10)    |
| C23A | 0.029(11)   | 0.038(12)   | 0.053(13)   | 0.002(10)    | 0.007(9)     | -0.002(8)    |
| C24  | 0.0164(12)  | 0.0230(13)  | 0.0178(13)  | -0.0017(10)  | 0.0024(10)   | 0.0000(10)   |
| C25  | 0.0148(12)  | 0.0231(13)  | 0.0130(11)  | -0.0027(10)  | -0.0014(9)   | -0.0030(10)  |
| C26  | 0.0145(11)  | 0.0177(12)  | 0.0164(11)  | -0.0011(9)   | 0.0010(9)    | -0.0007(9)   |
| N26  | 0.0145(11)  | 0.0177(12)  | 0.0164(11)  | -0.0011(9)   | 0.0010(9)    | -0.0007(9)   |
| C27  | 0.0151(12)  | 0.0192(12)  | 0.0130(11)  | -0.0003(9)   | 0.0005(9)    | -0.0003(10)  |
| C28  | 0.0424(19)  | 0.086(3)    | 0.0302(17)  | 0.0246(17)   | 0.0032(14)   | -0.0227(19)  |
| C29  | 0.0400(18)  | 0.0344(17)  | 0.051(2)    | -0.0028(15)  | 0.0123(15)   | -0.0174(14)  |
| C30  | 0.020(2)    | 0.043(3)    | 0.026(2)    | 0.007(3)     | -0.0024(17)  | 0.009(3)     |
| C31  | 0.022(2)    | 0.023(3)    | 0.027(3)    | -0.004(2)    | 0.004(2)     | -0.0021(19)  |
| C30A | 0.051(13)   | 0.031(12)   | 0.023(9)    | 0.015(10)    | 0.017(8)     | 0.032(12)    |
| C31A | 0.009(8)    | 0.038(14)   | 0.047(16)   | -0.009(10)   | 0.007(10)    | 0.002(8)     |
| C33  | 0.0272(16)  | 0.0336(17)  | 0.053(2)    | 0.0007(14)   | -0.0085(14)  | 0.0079(13)   |
| C34  | 0.0494(19)  | 0.0290(16)  | 0.050(2)    | 0.0006(14)   | 0.0182(16)   | 0.0109(15)   |
| C35  | 0.056(2)    | 0.047(2)    | 0.0390(19)  | 0.0223(15)   | 0.0205(16)   | 0.0119(17)   |
| C36  | 0.0308(16)  | 0.0363(17)  | 0.056(2)    | 0.0178(15)   | -0.0056(15)  | -0.0115(14)  |
| C1S  | 0.064(2)    | 0.076(3)    | 0.0364(19)  | -0.0010(17)  | -0.0009(17)  | 0.0133(19)   |
| Cl1  | 0.076(6)    | 0.081(7)    | 0.103(9)    | 0.023(5)     | -0.001(5)    | 0.029(5)     |
| Cl2  | 0.067(6)    | 0.080(6)    | 0.094(9)    | 0.007(5)     | -0.005(6)    | 0.020(5)     |
| Cl3  | 0.084(3)    | 0.0396(11)  | 0.0765(17)  | 0.0016(10)   | -0.0378(18)  | 0.0096(13)   |
| Cl4  | 0.049(4)    | 0.098(6)    | 0.058(4)    | -0.012(4)    | -0.008(3)    | 0.023(4)     |
| Cl5  | 0.064(5)    | 0.101(7)    | 0.075(10)   | -0.004(6)    | -0.002(5)    | 0.015(5)     |
| Cl6  | 0.0392(7)   | 0.0820(10)  | 0.0634(8)   | 0.0098(7)    | 0.0074(6)    | -0.0053(7)   |
| Cl7  | 0.086(10)   | 0.088(7)    | 0.057(11)   | 0.009(7)     | 0.005(9)     | 0.005(6)     |
| Cl8  | 0.136(12)   | 0.112(8)    | 0.164(15)   | -0.018(8)    | 0.074(11)    | -0.011(7)    |
| Cl9  | 0.0354(7)   | 0.154(2)    | 0.0474(8)   | -0.0111(11)  | 0.0083(5)    | -0.0282(11)  |
| Cl10 | 0.084(8)    | 0.099(8)    | 0.112(9)    | 0.004(6)     | 0.031(7)     | 0.008(6)     |
| C2S  | 0.027(2)    | 0.052(2)    | 0.043(2)    | 0.0160(17)   | 0.0086(17)   | -0.0024(18)  |
| C2SA | 0.067(9) | 0.091(12) | 0.075(11) | -0.003(8) | 0.003(7) | 0.003(8) |
| Cl11 | 0.072(11) | 0.119(14) | 0.084(8) | -0.015(8) | 0.004(7) | 0.000(10) |
| Cl12 | 0.094(10) | 0.103(9) | 0.072(10) | -0.015(7) | 0.009(8) | 0.022(7) |
| Cl13 | 0.0700(10) | 0.0714(11) | 0.0415(7) | -0.0011(7) | 0.0193(6) | 0.0237(9) |
| Cl14 | 0.069(6) | 0.107(8) | 0.088(7) | -0.021(6) | 0.004(5) | 0.028(5) |
| Cl15 | 0.093(9) | 0.112(9) | 0.087(10) | -0.009(7) | 0.010(7) | 0.022(7) |
| Cl16 | 0.068(7) | 0.091(8) | 0.110(9) | -0.026(7) | -0.021(6) | 0.030(6) |
| Cl17 | 0.0291(5) | 0.0373(5) | 0.0367(5) | 0.0076(4) | 0.0024(3) | 0.0030(3) |
| Cl18 | 0.055(8) | 0.083(9) | 0.065(9) | 0.003(7) | 0.005(7) | 0.006(7) |
| Cl19 | 0.0368(8) | 0.0535(8) | 0.0407(7) | 0.0124(6) | -0.0020(7) | 0.0067(7) |
| Cl20 | 0.064(8) | 0.096(10) | 0.084(11) | -0.005(8) | 0.004(7) | 0.000(7) |

**Table SI-10**. Hydrogen coordinates and isotropic displacement parameters (Å$^2$) for ML026F1_0m.

|      | x         | y          | z        | U(eq)     | Occupancy |
|------|-----------|------------|----------|-----------|-----------|
| H1   | 0.4794    | 0.5422     | 0.4702   | 0.018     | 0.5       |
| H4   | 0.2271    | 0.7051     | 0.4164   | 0.022     | 0.75      |
| H7A  | 0.2580    | 0.8864     | 0.2607   | 0.040     | 1         |
| H7B  | 0.3550    | 0.8743     | 0.3457   | 0.040     | 1         |
| H1S  | 0.373(4)  | 1.0953(19) | 0.312(4) | 0.102(15) | 1         |
| H8A  | 0.4332    | 0.8848     | 0.1799   | 0.041     | 1         |
| H8B  | 0.3602    | 0.8308     | 0.1407   | 0.041     | 1         |
| H11  | 0.5995    | 0.7104     | 0.2823   | 0.023     | 0.75      |
| H16  | 0.5879    | 0.5008     | 0.4690   | 0.018     | 0.5       |
| H19  | 0.8663    | 0.5968     | 0.2582   | 0.021     | 0.75      |
| H22A | 1.2186    | 0.5484     | 0.1120   | 0.033     | 0.8       |
| H22B | 1.1253    | 0.4993     | 0.0917   | 0.033     | 0.8       |
| H23A | 1.2405    | 0.5145     | 0.2865   | 0.030     | 0.8       |
| H23B | 1.2775    | 0.4670     | 0.2040   | 0.030     | 0.8       |
| H22C | 1.2323    | 0.5470     | 0.1303   | 0.047     | 0.2       |
| H22D | 1.2265    | 0.5440     | 0.2529   | 0.047     | 0.2       |
| H23C | 1.2597    | 0.4592     | 0.1783   | 0.048     | 0.2       |
| H23D | 1.1388    | 0.4633     | 0.1205   | 0.048     | 0.2       |
| H26  | 0.9532    | 0.4134     | 0.3920   | 0.019     | 0.75      |
| H28A | 1.0514    | 0.6018     | 0.0414   | 0.079     | 1         |
| H28B | 0.9417    | 0.6131     | 0.1111   | 0.079     | 1         |
| H28C | 0.9603    | 0.5516     | 0.0589   | 0.079     | 1         |
| H29A | 1.1697    | 0.6364     | 0.1945   | 0.062     | 1         |
| H29B | 1.1553    | 0.6048     | 0.3031   | 0.062     | 1         |
| H29C | 1.0540    | 0.6438     | 0.2577   | 0.062     | 1         |
| H30A | 1.1319    | 0.3888     | 0.3999   | 0.044     | 0.8       |
| H30B | 1.1820    | 0.4506     | 0.4294   | 0.044     | 0.8       |
| H30C | 1.2559    | 0.4074     | 0.3613   | 0.044     | 0.8       |
| H31A | 1.0472    | 0.3708     | 0.2317   | 0.036     | 0.8       |
| H31B | 1.1716    | 0.3788     | 0.1839   | 0.036     | 0.8       |
| H31C | 1.0655    | 0.4141     | 0.1375   | 0.036     | 0.8       |
| H30D | 1.1548    | 0.4244     | 0.4188   | 0.052     | 0.2       |
| H30E | 1.2052    | 0.4859     | 0.3880   | 0.052     | 0.2       |
| H30F | 1.2664    | 0.4285     | 0.3491   | 0.052     | 0.2       |
| H31D | 1.0643    | 0.3619     | 0.2988   | 0.047     | 0.2       |
| H31E | 1.1654    | 0.3646     | 0.2169   | 0.047     | 0.2       |
| H31F | 1.0392    | 0.3844     | 0.1844   | 0.047     | 0.2       |
| H33A | 0.1170    | 0.8159     | 0.2154   | 0.057     | 1         |
| H33B | 0.1247    | 0.7510     | 0.2555   | 0.057     | 1         |
| H33C | 0.2114    | 0.7734     | 0.1700   | 0.057     | 1         |
| H34A | 0.1391    | 0.8502     | 0.4031   | 0.064     | 1         |
| H34B | 0.2413    | 0.8240     | 0.4712   | 0.064     | 1         |
| H34C | 0.1392    | 0.7834     | 0.4325   | 0.064     | 1         |
| H35A | 0.6271    | 0.7659     | 0.1574   | 0.071     | 1         |
| H35B | 0.5826    | 0.8220     | 0.0982   | 0.071     | 1         |
| H35C | 0.5084    | 0.7643     | 0.0952   | 0.071     | 1         |
| H36A | 0.6457    | 0.8143     | 0.3217   | 0.062     | 1         |
| H36B | 0.5421    | 0.8502     | 0.3689   | 0.062     | 1         |
| H36C | 0.6081    | 0.8740     | 0.2704   | 0.062     | 1         |
| H2S  | 0.7907    | 0.6764     | 0.4201   | 0.048     | 0.9       |
| H2SA | 1.0422    | 0.7411     | 0.5841   | 0.093     | 0.1       |

**Table SI-11.** Torsion angles [°] for ML026F1_0m.
_________________________________________________________________

| | | | |
|---|---|---|---|
| C13-N1-C2-N14 | -178.1(2) | C2-C3-C12-C11 | 179.5(2) |
| C13-N1-C2-C3 | 0.4(3) | N4-C3-C12-C13 | 179.6(2) |
| N14-C2-C3-N4 | -1.0(4) | C4-C3-C12-C13 | 179.6(2) |
| N1-C2-C3-N4 | -179.5(2) | C2-C3-C12-C13 | 0.5(2) |
| N14-C2-C3-C4 | -1.0(4) | C2-N1-C13-N15 | -178.1(2) |
| N1-C2-C3-C4 | -179.5(2) | C2-N1-C13-C12 | -0.1(3) |
| N14-C2-C3-C12 | 178.0(2) | N11-C12-C13-N15 | -1.0(4) |
| N1-C2-C3-C12 | -0.5(3) | C11-C12-C13-N15 | -1.0(4) |
| C12-C3-C4-C5 | 0.8(3) | C3-C12-C13-N15 | 177.9(2) |
| C2-C3-C4-C5 | 179.7(2) | N11-C12-C13-N1 | -179.2(2) |
| C12-C3-N4-C5 | 0.8(3) | C11-C12-C13-N1 | -179.2(2) |
| C2-C3-N4-C5 | 179.7(2) | C3-C12-C13-N1 | -0.3(3) |
| C3-N4-C5-C10 | 0.5(3) | N1-C2-N14-C15 | 0.8(4) |
| C3-N4-C5-C6 | 177.4(2) | C3-C2-N14-C15 | -177.5(2) |
| C3-C4-C5-C10 | 0.5(3) | N1-C13-N15-C17 | 0.5(4) |
| C3-C4-C5-C6 | 177.4(2) | C12-C13-N15-C17 | -177.4(2) |
| N4-C5-C6-C34 | 44.4(3) | C2-N14-C15-N16#1 | -3.8(4) |
| C4-C5-C6-C34 | 44.4(3) | C2-N14-C15-C27#1 | 176.5(2) |
| C10-C5-C6-C34 | -138.8(3) | C13-N15-C17-N16 | -3.7(4) |
| N4-C5-C6-C7 | 163.7(2) | C13-N15-C17-C18 | 176.0(2) |
| C4-C5-C6-C7 | 163.7(2) | C15#1-N16-C17-N15 | -178.2(2) |
| C10-C5-C6-C7 | -19.5(3) | C15#1-N16-C17-C18 | 2.1(3) |
| N4-C5-C6-C33 | -75.7(3) | N15-C17-C18-N19 | -3.4(4) |
| C4-C5-C6-C33 | -75.7(3) | N16-C17-C18-N19 | 176.3(2) |
| C10-C5-C6-C33 | 101.0(3) | N15-C17-C18-C19 | -3.4(4) |
| C34-C6-C7-C8 | 169.3(3) | N16-C17-C18-C19 | 176.3(2) |
| C5-C6-C7-C8 | 47.8(3) | N15-C17-C18-C27 | 178.6(2) |
| C33-C6-C7-C8 | -71.7(3) | N16-C17-C18-C27 | -1.7(3) |
| C6-C7-C8-C9 | -62.9(3) | C27-C18-C19-C20 | 0.3(3) |
| C7-C8-C9-C35 | 166.0(3) | C17-C18-C19-C20 | -177.4(2) |
| C7-C8-C9-C36 | -75.4(3) | C27-C18-N19-C20 | 0.3(3) |
| C7-C8-C9-C10 | 44.1(3) | C17-C18-N19-C20 | -177.4(2) |
| N4-C5-C10-N11 | -1.4(4) | C18-N19-C20-C25 | -1.7(3) |
| C6-C5-C10-N11 | -178.1(2) | C18-N19-C20-C21 | 178.0(2) |
| C4-C5-C10-C11 | -1.4(4) | C18-C19-C20-C25 | -1.7(3) |
| C6-C5-C10-C11 | -178.1(2) | C18-C19-C20-C21 | 178.0(2) |
| N4-C5-C10-C9 | -179.1(2) | N19-C20-C21-C28 | -50.0(3) |
| C4-C5-C10-C9 | -179.1(2) | C19-C20-C21-C28 | -50.0(3) |
| C6-C5-C10-C9 | 4.2(4) | C25-C20-C21-C28 | 129.7(3) |
| C8-C9-C10-N11 | 166.5(2) | N19-C20-C21-C29 | 69.7(3) |
| C35-C9-C10-N11 | 46.4(3) | C19-C20-C21-C29 | 69.7(3) |
| C36-C9-C10-N11 | -72.9(3) | C25-C20-C21-C29 | -110.6(3) |
| C8-C9-C10-C11 | 166.5(2) | C19-C20-C21-C22 | -166.6(3) |
| C35-C9-C10-C11 | 46.4(3) | C25-C20-C21-C22 | 13.1(4) |
| C36-C9-C10-C11 | -72.9(3) | N19-C20-C21-C22A | 167.3(9) |
| C8-C9-C10-C5 | -15.7(3) | C25-C20-C21-C22A | -13.0(10) |
| C35-C9-C10-C5 | -135.9(3) | C28-C21-C22-C23 | -164.4(3) |
| C36-C9-C10-C5 | 104.8(3) | C29-C21-C22-C23 | 77.3(4) |
| C5-C10-C11-C12 | 0.9(4) | C20-C21-C22-C23 | -45.0(4) |
| C9-C10-C11-C12 | 178.7(2) | C21-C22-C23-C24 | 65.6(4) |
| C5-C10-N11-C12 | 0.9(4) | C28-C21-C22A-C23A | -89.0(17) |
| C9-C10-N11-C12 | 178.7(2) | C29-C21-C22A-C23A | 155.8(16) |
| C10-N11-C12-C3 | 0.5(4) | C20-C21-C22A-C23A | 45.2(18) |
| C10-N11-C12-C13 | 179.3(2) | C21-C22A-C23A-C24 | -65(2) |
| C10-C11-C12-C3 | 0.5(4) | C22A-C23A-C24-C30A | -75(2) |
| C10-C11-C12-C13 | 179.3(2) | C22A-C23A-C24-C25 | 46.6(19) |
| N4-C3-C12-N11 | -1.4(4) | C22A-C23A-C24-C31A | 164.1(18) |
| C2-C3-C12-N11 | 179.5(2) | C22-C23-C24-C31 | 72.6(4) |
| C4-C3-C12-C11 | -1.4(4) | C22-C23-C24-C25 | -48.9(4) |

S39

| | | | |
|---|---|---|---|
| C22-C23-C24-C30 | -169.0(4) | Cl8-C1S-Cl2-Cl5 | -139.9(13) |
| N19-C20-C25-N26 | 1.9(3) | Cl10-C1S-Cl2-Cl5 | -137.8(12) |
| C21-C20-C25-N26 | -177.8(2) | Cl1-C1S-Cl2-Cl5 | -155.2(16) |
| C19-C20-C25-C26 | 1.9(3) | Cl1-Cl2-Cl4-Cl5 | -4(6) |
| C21-C20-C25-C26 | -177.8(2) | Cl10-Cl2-Cl4-Cl5 | -6(2) |
| N19-C20-C25-C24 | -179.7(2) | C1S-Cl2-Cl4-Cl5 | 18.9(14) |
| C19-C20-C25-C24 | -179.7(2) | Cl1-Cl2-Cl4-C1S | -23(5) |
| C21-C20-C25-C24 | 0.6(3) | Cl10-Cl2-Cl4-C1S | -24.9(13) |
| C30A-C24-C25-N26 | -71.5(9) | Cl5-Cl2-Cl4-C1S | -18.9(14) |
| C23A-C24-C25-N26 | 164.0(9) | Cl10-Cl2-Cl4-Cl1 | -2(4) |
| C31A-C24-C25-N26 | 49.0(9) | C1S-Cl2-Cl4-Cl1 | 23(5) |
| C31-C24-C25-C26 | 73.7(3) | Cl5-Cl2-Cl4-Cl1 | 4(6) |
| C23-C24-C25-C26 | -164.2(3) | Cl7-C1S-Cl4-Cl5 | -2.7(17) |
| C30-C24-C25-C26 | -46.3(3) | Cl8-C1S-Cl4-Cl5 | -108(2) |
| C30A-C24-C25-C20 | 110.0(9) | Cl10-C1S-Cl4-Cl5 | -137.0(14) |
| C31-C24-C25-C20 | -104.8(3) | Cl2-C1S-Cl4-Cl5 | -158.9(16) |
| C23A-C24-C25-C20 | -14.4(9) | Cl1-C1S-Cl4-Cl5 | -154.6(12) |
| C23-C24-C25-C20 | 17.3(3) | Cl5-C1S-Cl4-Cl2 | 158.9(16) |
| C30-C24-C25-C20 | 135.2(3) | Cl7-C1S-Cl4-Cl2 | 156.2(15) |
| C31A-C24-C25-C20 | -129.5(9) | Cl8-C1S-Cl4-Cl2 | 51(2) |
| C20-C25-C26-C27 | -0.6(3) | Cl10-C1S-Cl4-Cl2 | 21.8(13) |
| C24-C25-C26-C27 | -179.1(2) | Cl1-C1S-Cl4-Cl2 | 4.2(10) |
| C20-C25-N26-C27 | -0.6(3) | Cl5-C1S-Cl4-Cl1 | 154.6(12) |
| C24-C25-N26-C27 | -179.1(2) | Cl7-C1S-Cl4-Cl1 | 152.0(13) |
| C25-N26-C27-C18 | -0.8(3) | Cl8-C1S-Cl4-Cl1 | 47(2) |
| C25-N26-C27-C15#1 | 177.3(2) | Cl10-C1S-Cl4-Cl1 | 17.6(9) |
| C25-C26-C27-C18 | -0.8(3) | Cl2-C1S-Cl4-Cl1 | -4.2(10) |
| C25-C26-C27-C15#1 | 177.3(2) | Cl7-C1S-Cl5-Cl4 | 177.5(16) |
| N19-C18-C27-N26 | 0.9(4) | Cl8-C1S-Cl5-Cl4 | 142.4(14) |
| C17-C18-C27-N26 | 179.1(2) | Cl10-C1S-Cl5-Cl4 | 62.4(18) |
| C19-C18-C27-C26 | 0.9(4) | Cl2-C1S-Cl5-Cl4 | 14.7(11) |
| C17-C18-C27-C26 | 179.1(2) | Cl1-C1S-Cl5-Cl4 | 27.4(13) |
| N19-C18-C27-C15#1 | -177.6(2) | Cl4-C1S-Cl5-Cl7 | -177.5(16) |
| C19-C18-C27-C15#1 | -177.6(2) | Cl8-C1S-Cl5-Cl7 | -35.0(18) |
| C17-C18-C27-C15#1 | 0.6(2) | Cl10-C1S-Cl5-Cl7 | -115.0(18) |
| Cl10-Cl1-Cl2-Cl4 | -3(6) | Cl2-C1S-Cl5-Cl7 | -162.8(13) |
| C1S-Cl1-Cl2-Cl4 | 20(4) | Cl1-C1S-Cl5-Cl7 | -150.1(12) |
| C1S-Cl1-Cl2-Cl10 | 22.8(17) | Cl7-C1S-Cl5-Cl2 | 162.8(13) |
| Cl4-Cl1-Cl2-Cl10 | 3(6) | Cl4-C1S-Cl5-Cl2 | -14.7(11) |
| Cl10-Cl1-Cl2-C1S | -22.8(17) | Cl8-C1S-Cl5-Cl2 | 127.8(15) |
| Cl4-Cl1-Cl2-C1S | -20(4) | Cl10-C1S-Cl5-Cl2 | 47.8(15) |
| Cl10-Cl1-Cl2-Cl5 | -5(2) | Cl1-C1S-Cl5-Cl2 | 12.7(9) |
| C1S-Cl1-Cl2-Cl5 | 17.6(10) | Cl2-Cl4-Cl5-C1S | -24.1(18) |
| Cl4-Cl1-Cl2-Cl5 | -2(4) | Cl1-Cl4-Cl5-C1S | -24.8(11) |
| Cl5-C1S-Cl2-Cl1 | 155.2(16) | Cl2-Cl4-Cl5-Cl7 | -21(4) |
| Cl7-C1S-Cl2-Cl1 | 115(3) | C1S-Cl4-Cl5-Cl7 | 4(2) |
| Cl4-C1S-Cl2-Cl1 | 171(2) | Cl1-Cl4-Cl5-Cl7 | -21(3) |
| Cl8-C1S-Cl2-Cl1 | 15.3(19) | C1S-Cl4-Cl5-Cl2 | 24.1(18) |
| Cl10-C1S-Cl2-Cl1 | 17.3(14) | Cl1-Cl4-Cl5-Cl2 | -0.7(11) |
| Cl5-C1S-Cl2-Cl4 | -16.2(12) | Cl4-C1S-Cl7-Cl5 | 2.2(14) |
| Cl7-C1S-Cl2-Cl4 | -57(3) | Cl8-C1S-Cl7-Cl5 | 153.4(14) |
| Cl8-C1S-Cl2-Cl4 | -156.1(13) | Cl10-C1S-Cl7-Cl5 | 125.1(14) |
| Cl10-C1S-Cl2-Cl4 | -154.0(15) | Cl2-C1S-Cl7-Cl5 | 44(3) |
| Cl1-C1S-Cl2-Cl4 | -171(2) | Cl1-C1S-Cl7-Cl5 | 109(3) |
| Cl5-C1S-Cl2-Cl10 | 137.8(12) | Cl5-C1S-Cl7-Cl8 | -153.4(14) |
| Cl7-C1S-Cl2-Cl10 | 97(3) | Cl4-C1S-Cl7-Cl8 | -151.2(11) |
| Cl4-C1S-Cl2-Cl10 | 154.0(15) | Cl10-C1S-Cl7-Cl8 | -28.2(17) |
| Cl8-C1S-Cl2-Cl10 | -2.1(15) | Cl2-C1S-Cl7-Cl8 | -109(3) |
| Cl1-C1S-Cl2-Cl10 | -17.3(14) | Cl1-C1S-Cl7-Cl8 | -44(3) |
| Cl7-C1S-Cl2-Cl5 | -41(3) | Cl4-Cl5-Cl7-C1S | -4(2) |
| Cl4-C1S-Cl2-Cl5 | 16.2(12) | Cl2-Cl5-Cl7-C1S | -14.8(11) |
| C1S-Cl5-Cl7-Cl8 | 23.6(13) | Cl15-Cl12-Cl14-Cl18 | 26.1(18) |
| Cl4-Cl5-Cl7-Cl8 | 20(3) | Cl11-Cl12-Cl14-Cl18 | -70.3(12) |
| Cl2-Cl5-Cl7-Cl8 | 9(2) | Cl11-Cl12-Cl15-Cl14 | 66.3(11) |
| Cl5-C1S-Cl8-Cl10 | -117.6(17) | Cl14-Cl12-Cl15-C2SA | -64(2) |
| Cl7-C1S-Cl8-Cl10 | -151.8(17) | Cl11-Cl12-Cl15-C2SA | 3(2) |
| Cl4-C1S-Cl8-Cl10 | -35(3) | Cl14-Cl12-Cl15-Cl20 | -51.3(17) |
| Cl2-C1S-Cl8-Cl10 | 2.1(15) | Cl11-Cl12-Cl15-Cl20 | 15(2) |
| Cl1-C1S-Cl8-Cl10 | 9.1(11) | Cl12-Cl14-Cl16-Cl18 | 83(2) |
| Cl5-C1S-Cl8-Cl7 | 34.3(19) | Cl15-Cl14-Cl16-Cl18 | 62.3(15) |
| Cl4-C1S-Cl8-Cl7 | 117(2) | Cl11-Cl14-Cl16-Cl18 | -22(3) |
| Cl10-C1S-Cl8-Cl7 | 151.8(17) | Cl16-Cl18-Cl20-Cl15 | -27.6(18) |
| Cl2-C1S-Cl8-Cl7 | 153.9(14) | C2SA-Cl18-Cl20-Cl15 | -52(6) |
| Cl1-C1S-Cl8-Cl7 | 160.9(13) | Cl14-Cl18-Cl20-Cl15 | -35.9(11) |
| C1S-Cl7-Cl8-Cl10 | 26.2(16) | | |
| Cl5-Cl7-Cl8-Cl10 | 5(2) | | |
| Cl5-Cl7-Cl8-C1S | -20.8(11) | | |
| Cl2-Cl1-Cl10-Cl8 | 55(4) | | |
| C1S-Cl1-Cl10-Cl8 | 31(3) | | |
| Cl4-Cl1-Cl10-Cl8 | 54(4) | | |
| C1S-Cl1-Cl10-Cl2 | -23.4(18) | | |
| Cl4-Cl1-Cl10-Cl2 | -0.6(13) | | |
| Cl2-Cl1-Cl10-C1S | 23.4(17) | | |
| Cl4-Cl1-Cl10-C1S | 22.8(9) | | |
| C1S-Cl8-Cl10-Cl1 | -37(4) | | |
| Cl7-Cl8-Cl10-Cl1 | -59(5) | | |
| C1S-Cl8-Cl10-Cl2 | -2.3(17) | | |
| Cl7-Cl8-Cl10-Cl2 | -24(3) | | |
| Cl7-Cl8-Cl10-C1S | -21.8(13) | | |
| Cl4-Cl2-Cl10-Cl1 | 179(2) | | |
| C1S-Cl2-Cl10-Cl1 | 154(2) | | |
| Cl5-Cl2-Cl10-Cl1 | 175.9(19) | | |
| Cl1-Cl2-Cl10-Cl8 | -151(3) | | |
| Cl4-Cl2-Cl10-Cl8 | 27(3) | | |
| C1S-Cl2-Cl10-Cl8 | 2.2(17) | | |
| Cl5-Cl2-Cl10-Cl8 | 24(2) | | |
| Cl1-Cl2-Cl10-C1S | -154(2) | | |
| Cl4-Cl2-Cl10-C1S | 25.1(14) | | |
| Cl5-Cl2-Cl10-C1S | 22.2(6) | | |
| Cl5-C1S-Cl10-Cl1 | -77(2) | | |
| Cl7-C1S-Cl10-Cl1 | -165.6(18) | | |
| Cl4-C1S-Cl10-Cl1 | -33.5(15) | | |
| Cl8-C1S-Cl10-Cl1 | 162(2) | | |
| Cl2-C1S-Cl10-Cl1 | -15.4(12) | | |
| Cl5-C1S-Cl10-Cl8 | 120.4(18) | | |
| Cl7-C1S-Cl10-Cl8 | 32(2) | | |
| Cl4-C1S-Cl10-Cl8 | 164.1(13) | | |
| Cl2-C1S-Cl10-Cl8 | -177.8(16) | | |
| Cl1-C1S-Cl10-Cl8 | -162(2) | | |
| Cl5-C1S-Cl10-Cl2 | -61.8(18) | | |
| Cl7-C1S-Cl10-Cl2 | -150.2(17) | | |
| Cl4-C1S-Cl10-Cl2 | -18.1(10) | | |
| Cl8-C1S-Cl10-Cl2 | 177.8(16) | | |
| Cl1-C1S-Cl10-Cl2 | 15.4(12) | | |
| Cl18-C2SA-Cl11-Cl14 | -42(3) | | |
| Cl15-C2SA-Cl11-Cl14 | 48.2(15) | | |
| Cl18-C2SA-Cl11-Cl12 | -89(3) | | |
| Cl15-C2SA-Cl11-Cl12 | 1.9(17) | | |
| Cl15-Cl12-Cl14-Cl16 | -30(3) | | |
| Cl11-Cl12-Cl14-Cl16 | -127(2) | | |
| Cl11-Cl12-Cl14-Cl15 | -96.4(16) | | |
| Cl15-Cl12-Cl14-Cl11 | 96.4(16) | | |

## 1.7.3 Crystal Structure of A$_2$B$_2$ N$_4$-[Pc*Zn·H$_2$O]

| | |
|---|---|
| Habitus, colour | prism, dark |
| Crystal size | 0.27 x 0.06 x 0.03 mm$^3$ |
| Crystal system | Monoclinic |
| Space group | P2$_1$/c            Z = 2 |
| Unit cell dimensions | a = 6.0419(5) Å        α= 90°. |
| | b = 16.2175(13) Å     β= 95.305(3)°. |
| | c = 31.008(2) Å         γ = 90°. |
| Volume | 3025.3(4) Å$^3$ |
| Cell determination | 9556 peaks with Theta 2.3 to 25.2°. |
| Empirical formula | C$_{60}$ H$_{70}$ N$_{12}$ O Zn |
| Moiety formula | C$_{60}$ H$_{70}$ N$_{12}$ O Zn |
| Formula weight | 1040.65 |
| Density (calculated) | 1.142 Mg/m$^3$ |
| Absorption coefficient | 0.454 mm$^{-1}$ |
| F(000) | 1104 |
| | |
| Solution and refinement: | |
| Reflections collected | 32818 |
| Independent reflections | 5598 [R(int) = 0.0711] |
| Completeness to theta = 25.242° | 99.9 % |
| Observed reflections | 4276[I > 2(I)] |
| Reflections used for refinement | 5598 |
| Absorption correction | Numerical Mu Calculated[12] |
| Max. and min. transmission | 0.99 and 0.89 |
| Largest diff. peak and hole | 0.288 and -0.347 e.Å$^{-3}$ |
| Solution | Direct methods |
| Refinement | Full-matrix least-squares on F$^2$ |
| Treatment of hydrogen atoms | Calculated positions, constr. ref. |
| Programs used | XT V2014/1 (Bruker AXS Inc., 2014)[13] |
| | SHELXL-2014/7 (Sheldrick, 2014)[14] |
| | DIAMOND (Crystal Impact)[11] |
| | ShelXle (Hübschle, Sheldrick, Dittrich, 2011)[15] |
| Data / restraints / parameters | 5598 / 0 / 353 |
| Goodness-of-fit on F$^2$ | 1.021 |
| R index (all data) | wR2 = 0.1467 |
| R index conventional [I>2sigma(I)] | R1 = 0.0580 |

**Table SI-12.** Atomic coordinates and equivalent isotropic displacement parameters (Å$^2$) for mL105c2_0m_sq. U(eq) is defined as one third of the trace of the orthogonalized U$^{ij}$ tensor.

|      | x           | y           | z           | U(eq)      | Occupancy |
|------|-------------|-------------|-------------|------------|-----------|
| Zn1  | 1.49975(18) | 0.51807(4)  | 0.50622(3)  | 0.0242(2)  | 0.5       |
| N1   | 1.5282(5)   | 0.39227(14) | 0.59501(8)  | 0.0354(6)  | 1         |
| C2   | 1.3557(6)   | 0.37918(17) | 0.56599(9)  | 0.0336(7)  | 1         |
| N3   | 1.3123(5)   | 0.41696(14) | 0.52645(8)  | 0.0330(6)  | 1         |
| C4   | 1.1271(6)   | 0.38191(18) | 0.50470(10) | 0.0357(8)  | 1         |
| C5   | 1.0459(5)   | 0.31799(17) | 0.53226(9)  | 0.0324(7)  | 1         |
| C6   | 0.8816(5)   | 0.26266(17) | 0.52335(9)  | 0.0369(7)  | 0.5       |
| N6   | 0.8816(5)   | 0.26266(17) | 0.52335(9)  | 0.0369(7)  | 0.5       |
| C7   | 0.8475(6)   | 0.20533(19) | 0.55472(10) | 0.0378(8)  | 1         |
| C8   | 0.6829(6)   | 0.1355(2)   | 0.54251(11) | 0.0459(9)  | 1         |
| C9   | 0.6011(7)   | 0.0982(3)   | 0.58286(13) | 0.0594(11) | 1         |
| C10  | 0.7865(8)   | 0.0790(2)   | 0.61729(13) | 0.0609(11) | 1         |
| C11  | 0.9179(6)   | 0.1542(2)   | 0.63326(11) | 0.0446(9)  | 1         |
| C12  | 0.9761(6)   | 0.20836(19) | 0.59537(10) | 0.0362(8)  | 1         |
| C13  | 1.1473(5)   | 0.26387(16) | 0.60328(9)  | 0.0362(7)  | 0.5       |
| N13  | 1.1473(5)   | 0.26387(16) | 0.60328(9)  | 0.0362(7)  | 0.5       |
| C14  | 1.1822(5)   | 0.31757(16) | 0.57052(9)  | 0.0306(7)  | 1         |
| N15  | 1.0358(5)   | 0.39883(15) | 0.46520(8)  | 0.0350(6)  | 1         |
| C16  | 1.1103(6)   | 0.45652(18) | 0.43907(9)  | 0.0345(7)  | 1         |
| N17  | 1.2882(5)   | 0.50688(15) | 0.44710(8)  | 0.0348(6)  | 1         |
| C18  | 1.3091(6)   | 0.55471(17) | 0.41116(9)  | 0.0335(7)  | 1         |
| C19  | 1.1260(6)   | 0.53481(17) | 0.37847(9)  | 0.0339(7)  | 1         |
| C20  | 1.0650(5)   | 0.56503(17) | 0.33820(9)  | 0.0357(7)  | 0.5       |
| N20  | 1.0650(5)   | 0.56503(17) | 0.33820(9)  | 0.0357(7)  | 0.5       |
| C21  | 0.8899(6)   | 0.52815(19) | 0.31471(10) | 0.0392(8)  | 1         |
| C22  | 0.8002(6)   | 0.5685(2)   | 0.27201(11) | 0.0448(8)  | 1         |
| C23  | 0.6709(8)   | 0.5058(3)   | 0.24279(13) | 0.0640(11) | 1         |
| C24  | 0.5072(8)   | 0.4553(3)   | 0.26589(15) | 0.0672(12) | 1         |
| C25  | 0.6171(6)   | 0.4051(2)   | 0.30292(12) | 0.0492(9)  | 1         |
| C26  | 0.7813(6)   | 0.4578(2)   | 0.33098(11) | 0.0417(8)  | 1         |
| C27  | 0.8404(5)   | 0.43133(18) | 0.37254(9)  | 0.0404(7)  | 0.5       |
| N27  | 0.8404(5)   | 0.43133(18) | 0.37254(9)  | 0.0404(7)  | 0.5       |
| C28  | 1.0081(6)   | 0.47124(18) | 0.39565(10) | 0.0347(7)  | 1         |
| C29  | 0.8078(7)   | 0.0725(2)   | 0.51705(15) | 0.0657(12) | 1         |
| C30  | 0.4821(7)   | 0.1666(3)   | 0.51310(14) | 0.0660(12) | 1         |
| C31  | 1.1256(7)   | 0.1267(3)   | 0.66111(13) | 0.0644(12) | 1         |
| C32  | 0.7765(7)   | 0.2076(3)   | 0.66089(12) | 0.0589(11) | 1         |
| C33  | 0.9850(7)   | 0.6044(3)   | 0.24761(13) | 0.0650(12) | 1         |
| C34  | 0.6459(7)   | 0.6390(2)   | 0.28378(15) | 0.0665(12) | 1         |
| C35  | 0.7495(7)   | 0.3332(2)   | 0.28583(14) | 0.0601(11) | 1         |
| C36  | 0.4342(8)   | 0.3688(3)   | 0.32869(17) | 0.0783(14) | 1         |
| O1W  | 1.3888(7)   | 0.6241(3)   | 0.53911(14) | 0.0403(11) | 0.5       |

**Table SI-13.** Bond lengths [Å] and angles [°] for mL105c2_0m_sq.

| | | | |
|---|---|---|---|
| Zn1-Zn1#1 | 0.7019(11) | C20-H20 | 0.9400 |
| Zn1-N17#1 | 1.886(3) | C21-C26 | 1.431(5) |
| Zn1-N3#1 | 1.908(3) | C21-C22 | 1.531(5) |
| Zn1-N3 | 2.121(3) | C22-C33 | 1.521(5) |
| Zn1-O1W | 2.138(4) | C22-C23 | 1.528(5) |
| Zn1-N17 | 2.143(3) | C22-C34 | 1.540(5) |
| N1-C2 | 1.330(4) | C23-C24 | 1.514(6) |
| N1-C18#1 | 1.333(4) | C23-H23A | 0.9800 |
| C2-N3 | 1.374(4) | C23-H23B | 0.9800 |
| C2-C14 | 1.464(4) | C24-C25 | 1.510(6) |
| N3-C4 | 1.375(4) | C24-H24A | 0.9800 |
| N3-Zn1#1 | 1.908(3) | C24-H24B | 0.9800 |
| C4-N15 | 1.325(4) | C25-C26 | 1.521(5) |
| C4-C5 | 1.457(4) | C25-C35 | 1.537(5) |
| C5-C6 | 1.348(4) | C25-C36 | 1.540(6) |
| C5-C14 | 1.380(4) | C26-C27 | 1.373(5) |
| C6-C7 | 1.375(4) | C27-C28 | 1.351(4) |
| C6-H6 | 0.9400 | C27-H27 | 0.9400 |
| C7-C12 | 1.419(5) | C29-H29A | 0.9700 |
| C7-C8 | 1.531(5) | C29-H29B | 0.9700 |
| C8-C9 | 1.513(5) | C29-H29C | 0.9700 |
| C8-C29 | 1.531(5) | C30-H30A | 0.9700 |
| C8-C30 | 1.534(5) | C30-H30B | 0.9700 |
| C9-C10 | 1.507(6) | C30-H30C | 0.9700 |
| C9-H9A | 0.9800 | C31-H31A | 0.9700 |
| C9-H9B | 0.9800 | C31-H31B | 0.9700 |
| C10-C11 | 1.514(5) | C31-H31C | 0.9700 |
| C10-H10A | 0.9800 | C32-H32A | 0.9700 |
| C10-H10B | 0.9800 | C32-H32B | 0.9700 |
| C11-C31 | 1.523(5) | C32-H32C | 0.9700 |
| C11-C32 | 1.533(5) | C33-H33A | 0.9700 |
| C11-C12 | 1.533(4) | C33-H33B | 0.9700 |
| C12-C13 | 1.376(4) | C33-H33C | 0.9700 |
| C13-C14 | 1.369(4) | C34-H34A | 0.9700 |
| C13-H13 | 0.9400 | C34-H34B | 0.9700 |
| N15-C16 | 1.343(4) | C34-H34C | 0.9700 |
| C16-N17 | 1.355(4) | C35-H35A | 0.9700 |
| C16-C28 | 1.448(4) | C35-H35B | 0.9700 |
| N17-C18 | 1.373(4) | C35-H35C | 0.9700 |
| N17-Zn1#1 | 1.886(3) | C36-H36A | 0.9700 |
| C18-N1#1 | 1.333(4) | C36-H36B | 0.9700 |
| C18-C19 | 1.465(5) | C36-H36C | 0.9700 |
| C19-C20 | 1.360(4) | O1W-H1WA | 0.8643 |
| C19-C28 | 1.387(4) | O1W-H1WB | 0.8644 |
| C20-C21 | 1.366(5) | | |
| | | | |
| Zn1#1-Zn1-N17#1 | 101.9(2) | N3-Zn1-N17 | 83.97(10) |
| Zn1#1-Zn1-N3#1 | 97.8(2) | O1W-Zn1-N17 | 106.71(14) |
| N17#1-Zn1-N3#1 | 97.51(12) | C2-N1-C18#1 | 123.2(2) |
| Zn1#1-Zn1-N3 | 63.01(17) | N1-C2-N3 | 127.3(3) |
| N17#1-Zn1-N3 | 87.16(11) | N1-C2-C14 | 124.6(3) |
| N3#1-Zn1-N3 | 160.86(4) | N3-C2-C14 | 108.1(3) |
| Zn1#1-Zn1-O1W | 162.0(2) | C2-N3-C4 | 109.4(3) |
| N17#1-Zn1-O1W | 91.56(14) | C2-N3-Zn1#1 | 130.3(2) |
| N3#1-Zn1-O1W | 92.15(14) | C4-N3-Zn1#1 | 117.73(19) |
| N3-Zn1-O1W | 106.33(14) | C2-N3-Zn1 | 123.1(2) |
| Zn1#1-Zn1-N17 | 59.45(17) | C4-N3-Zn1 | 127.40(19) |
| N17#1-Zn1-N17 | 161.30(4) | Zn1#1-N3-Zn1 | 19.14(4) |
| N3#1-Zn1-N17 | 85.97(11) | N15-C4-N3 | 128.2(3) |

S44

| | | | |
|---|---|---|---|
| N15-C4-C5 | 123.8(3) | C19-C20-H20 | 121.2 |
| N3-C4-C5 | 108.0(3) | C21-C20-H20 | 121.2 |
| C6-C5-C14 | 122.9(3) | C20-C21-C26 | 121.3(3) |
| C6-C5-C4 | 129.4(3) | C20-C21-C22 | 118.0(3) |
| C14-C5-C4 | 107.5(3) | C26-C21-C22 | 120.6(3) |
| C5-C6-C7 | 117.4(3) | C33-C22-C23 | 108.5(3) |
| C5-C6-H6 | 121.3 | C33-C22-C21 | 112.1(3) |
| C7-C6-H6 | 121.3 | C23-C22-C21 | 110.4(3) |
| C6-C7-C12 | 120.0(3) | C33-C22-C34 | 108.8(3) |
| C6-C7-C8 | 117.5(3) | C23-C22-C34 | 110.3(3) |
| C12-C7-C8 | 122.3(3) | C21-C22-C34 | 106.6(3) |
| C9-C8-C29 | 111.9(3) | C24-C23-C22 | 113.6(3) |
| C9-C8-C7 | 110.2(3) | C24-C23-H23A | 108.9 |
| C29-C8-C7 | 106.3(3) | C22-C23-H23A | 108.9 |
| C9-C8-C30 | 108.8(3) | C24-C23-H23B | 108.9 |
| C29-C8-C30 | 108.2(3) | C22-C23-H23B | 108.9 |
| C7-C8-C30 | 111.3(3) | H23A-C23-H23B | 107.7 |
| C10-C9-C8 | 112.9(3) | C25-C24-C23 | 113.0(4) |
| C10-C9-H9A | 109.0 | C25-C24-H24A | 109.0 |
| C8-C9-H9A | 109.0 | C23-C24-H24A | 109.0 |
| C10-C9-H9B | 109.0 | C25-C24-H24B | 109.0 |
| C8-C9-H9B | 109.0 | C23-C24-H24B | 109.0 |
| H9A-C9-H9B | 107.8 | H24A-C24-H24B | 107.8 |
| C9-C10-C11 | 113.5(3) | C24-C25-C26 | 110.5(3) |
| C9-C10-H10A | 108.9 | C24-C25-C35 | 110.7(3) |
| C11-C10-H10A | 108.9 | C26-C25-C35 | 106.9(3) |
| C9-C10-H10B | 108.9 | C24-C25-C36 | 108.4(4) |
| C11-C10-H10B | 108.9 | C26-C25-C36 | 112.3(3) |
| H10A-C10-H10B | 107.7 | C35-C25-C36 | 108.1(3) |
| C10-C11-C31 | 109.2(3) | C27-C26-C21 | 119.5(3) |
| C10-C11-C32 | 109.6(3) | C27-C26-C25 | 117.4(3) |
| C31-C11-C32 | 108.7(3) | C21-C26-C25 | 122.8(3) |
| C10-C11-C12 | 111.2(3) | C28-C27-C26 | 117.7(3) |
| C31-C11-C12 | 111.5(3) | C28-C27-H27 | 121.1 |
| C32-C11-C12 | 106.5(3) | C26-C27-H27 | 121.1 |
| C13-C12-C7 | 121.5(3) | C27-C28-C19 | 122.6(3) |
| C13-C12-C11 | 117.4(3) | C27-C28-C16 | 130.4(3) |
| C7-C12-C11 | 121.0(3) | C19-C28-C16 | 106.8(3) |
| C14-C13-C12 | 116.6(3) | C8-C29-H29A | 109.5 |
| C14-C13-H13 | 121.7 | C8-C29-H29B | 109.5 |
| C12-C13-H13 | 121.7 | H29A-C29-H29B | 109.5 |
| C13-C14-C5 | 121.4(3) | C8-C29-H29C | 109.5 |
| C13-C14-C2 | 131.8(3) | H29A-C29-H29C | 109.5 |
| C5-C14-C2 | 106.8(2) | H29B-C29-H29C | 109.5 |
| C4-N15-C16 | 124.6(3) | C8-C30-H30A | 109.5 |
| N15-C16-N17 | 127.9(3) | C8-C30-H30B | 109.5 |
| N15-C16-C28 | 122.6(3) | H30A-C30-H30B | 109.5 |
| N17-C16-C28 | 109.5(3) | C8-C30-H30C | 109.5 |
| C16-N17-C18 | 108.8(3) | H30A-C30-H30C | 109.5 |
| C16-N17-Zn1#1 | 118.9(2) | H30B-C30-H30C | 109.5 |
| C18-N17-Zn1#1 | 129.6(2) | C11-C31-H31A | 109.5 |
| C16-N17-Zn1 | 127.5(2) | C11-C31-H31B | 109.5 |
| C18-N17-Zn1 | 123.7(2) | H31A-C31-H31B | 109.5 |
| Zn1#1-N17-Zn1 | 18.69(4) | C11-C31-H31C | 109.5 |
| N1#1-C18-N17 | 127.2(3) | H31A-C31-H31C | 109.5 |
| N1#1-C18-C19 | 124.2(3) | H31B-C31-H31C | 109.5 |
| N17-C18-C19 | 108.5(3) | C11-C32-H32A | 109.5 |
| C20-C19-C28 | 121.0(3) | C11-C32-H32B | 109.5 |
| C20-C19-C18 | 132.7(3) | H32A-C32-H32B | 109.5 |
| C28-C19-C18 | 106.2(3) | C11-C32-H32C | 109.5 |
| C19-C20-C21 | 117.5(3) | H32A-C32-H32C | 109.5 |
| H32B-C32-H32C | 109.5 | C25-C35-H35B | 109.5 |
| C22-C33-H33A | 109.5 | H35A-C35-H35B | 109.5 |
| C22-C33-H33B | 109.5 | C25-C35-H35C | 109.5 |
| H33A-C33-H33B | 109.5 | H35A-C35-H35C | 109.5 |
| C22-C33-H33C | 109.5 | H35B-C35-H35C | 109.5 |
| H33A-C33-H33C | 109.5 | C25-C36-H36A | 109.5 |
| H33B-C33-H33C | 109.5 | C25-C36-H36B | 109.5 |
| C22-C34-H34A | 109.5 | H36A-C36-H36B | 109.5 |
| C22-C34-H34B | 109.5 | C25-C36-H36C | 109.5 |
| H34A-C34-H34B | 109.5 | H36A-C36-H36C | 109.5 |
| C22-C34-H34C | 109.5 | H36B-C36-H36C | 109.5 |
| H34A-C34-H34C | 109.5 | Zn1-O1W-H1WA | 109.6 |
| H34B-C34-H34C | 109.5 | Zn1-O1W-H1WB | 109.6 |
| C25-C35-H35A | 109.5 | H1WA-O1W-H1WB | 109.2 |

Symmetry transformations used to generate equivalent atoms:
#1 -x+3,-y+1,-z+1

**Table SI-14.** Anisotropic displacement parameters (Å$^2$) for mL105c2_0m_sq.
The anisotropic displacement factor exponent takes the form: $-2\pi^2[ h^2 a^{*2}U^{11} + ... + 2 h k a^* b^* U^{12} ]$

|     | U$^{11}$ | U$^{22}$ | U$^{33}$ | U$^{23}$ | U$^{13}$ | U$^{12}$ |
|-----|----------|----------|----------|----------|----------|----------|
| Zn1 | 0.0235(3) | 0.0261(6) | 0.0219(6) | 0.0052(4) | -0.0039(4) | -0.0076(5) |
| N1  | 0.0602(19) | 0.0240(13) | 0.0228(14) | 0.0029(10) | 0.0085(13) | 0.0018(13) |
| C2  | 0.058(2) | 0.0241(15) | 0.0201(16) | -0.0001(12) | 0.0107(15) | 0.0024(15) |
| N3  | 0.0509(17) | 0.0247(12) | 0.0239(13) | 0.0044(10) | 0.0071(12) | 0.0008(12) |
| C4  | 0.053(2) | 0.0245(15) | 0.0316(18) | 0.0031(13) | 0.0125(15) | 0.0041(15) |
| C5  | 0.048(2) | 0.0214(15) | 0.0295(17) | 0.0023(12) | 0.0100(14) | 0.0044(14) |
| C6  | 0.0522(19) | 0.0354(15) | 0.0234(14) | 0.0011(12) | 0.0041(13) | 0.0099(15) |
| N6  | 0.0522(19) | 0.0354(15) | 0.0234(14) | 0.0011(12) | 0.0041(13) | 0.0099(15) |
| C7  | 0.046(2) | 0.0348(17) | 0.0340(18) | 0.0005(14) | 0.0079(15) | 0.0119(15) |
| C8  | 0.049(2) | 0.049(2) | 0.0392(19) | -0.0031(16) | 0.0024(16) | 0.0106(17) |
| C9  | 0.065(3) | 0.060(2) | 0.053(2) | -0.0006(19) | 0.007(2) | -0.012(2) |
| C10 | 0.079(3) | 0.048(2) | 0.054(2) | 0.0135(18) | 0.000(2) | -0.007(2) |
| C11 | 0.045(2) | 0.052(2) | 0.0362(19) | 0.0154(16) | 0.0047(16) | -0.0002(17) |
| C12 | 0.046(2) | 0.0317(16) | 0.0312(18) | 0.0041(13) | 0.0082(15) | 0.0086(15) |
| C13 | 0.0548(19) | 0.0303(15) | 0.0248(15) | 0.0056(12) | 0.0104(13) | 0.0044(14) |
| N13 | 0.0548(19) | 0.0303(15) | 0.0248(15) | 0.0056(12) | 0.0104(13) | 0.0044(14) |
| C14 | 0.049(2) | 0.0201(14) | 0.0241(16) | 0.0008(12) | 0.0118(14) | 0.0019(13) |
| N15 | 0.0520(18) | 0.0293(13) | 0.0243(14) | 0.0046(11) | 0.0072(12) | 0.0039(12) |
| C16 | 0.055(2) | 0.0250(15) | 0.0249(16) | 0.0013(12) | 0.0090(15) | 0.0038(15) |
| N17 | 0.0569(18) | 0.0264(13) | 0.0212(13) | 0.0039(10) | 0.0037(12) | 0.0007(13) |
| C18 | 0.056(2) | 0.0241(15) | 0.0205(15) | 0.0010(12) | 0.0042(14) | 0.0065(15) |
| C19 | 0.052(2) | 0.0241(15) | 0.0260(16) | -0.0009(12) | 0.0054(14) | 0.0025(14) |
| C20 | 0.0510(19) | 0.0328(15) | 0.0237(15) | 0.0030(12) | 0.0052(13) | 0.0021(14) |
| N20 | 0.0510(19) | 0.0328(15) | 0.0237(15) | 0.0030(12) | 0.0052(13) | 0.0021(14) |
| C21 | 0.048(2) | 0.0360(18) | 0.0354(18) | -0.0011(14) | 0.0137(16) | 0.0086(16) |
| C22 | 0.043(2) | 0.048(2) | 0.042(2) | 0.0088(16) | -0.0002(16) | 0.0076(17) |
| C23 | 0.080(3) | 0.062(3) | 0.047(2) | 0.0016(19) | -0.009(2) | 0.003(2) |
| C24 | 0.063(3) | 0.058(2) | 0.077(3) | 0.000(2) | -0.014(2) | -0.010(2) |
| C25 | 0.047(2) | 0.0400(19) | 0.062(2) | -0.0024(17) | 0.0103(18) | -0.0003(17) |
| C26 | 0.049(2) | 0.0364(18) | 0.042(2) | -0.0015(15) | 0.0149(16) | 0.0072(16) |
| C27 | 0.055(2) | 0.0325(15) | 0.0356(17) | 0.0016(13) | 0.0148(15) | 0.0020(14) |
| N27 | 0.055(2) | 0.0325(15) | 0.0356(17) | 0.0016(13) | 0.0148(15) | 0.0020(14) |
| C28 | 0.051(2) | 0.0276(15) | 0.0260(16) | 0.0002(13) | 0.0055(14) | 0.0036(15) |
| C29 | 0.063(3) | 0.052(2) | 0.083(3) | -0.023(2) | 0.011(2) | 0.011(2) |
| C30 | 0.056(3) | 0.078(3) | 0.063(3) | -0.005(2) | -0.002(2) | 0.015(2) |
| C31 | 0.057(3) | 0.076(3) | 0.061(3) | 0.043(2) | 0.007(2) | 0.000(2) |
| C32 | 0.061(3) | 0.071(3) | 0.047(2) | 0.002(2) | 0.0188(19) | 0.000(2) |
| C33 | 0.058(3) | 0.092(3) | 0.044(2) | 0.029(2) | -0.0033(19) | 0.003(2) |
| C34 | 0.058(3) | 0.051(2) | 0.089(3) | 0.003(2) | -0.005(2) | 0.016(2) |
| C35 | 0.064(3) | 0.046(2) | 0.070(3) | -0.0147(19) | 0.008(2) | 0.003(2) |
| C36 | 0.055(3) | 0.081(3) | 0.102(4) | 0.003(3) | 0.024(3) | -0.007(2) |
| O1W | 0.027(2) | 0.046(3) | 0.047(3) | -0.007(2) | 0.003(2) | -0.001(2) |

**Table SI-15.** Hydrogen coordinates and isotropic displacement parameters (Å$^2$) for mL105c2_0m_sq.

|      | x      | y      | z      | U(eq) | Occupancy |
|------|--------|--------|--------|-------|-----------|
| H6   | 0.7937 | 0.2632 | 0.4967 | 0.044 | 0.5       |
| H9A  | 0.4972 | 0.1366 | 0.5947 | 0.071 | 1         |
| H9B  | 0.5197 | 0.0473 | 0.5750 | 0.071 | 1         |
| H10A | 0.8878 | 0.0393 | 0.6056 | 0.073 | 1         |
| H10B | 0.7234 | 0.0529 | 0.6419 | 0.073 | 1         |
| H13  | 1.2358 | 0.2649 | 0.6298 | 0.043 | 0.5       |
| H20  | 1.1406 | 0.6097 | 0.3269 | 0.043 | 0.5       |
| H23A | 0.5900 | 0.5351 | 0.2186 | 0.077 | 1         |
| H23B | 0.7766 | 0.4684 | 0.2307 | 0.077 | 1         |
| H24A | 0.3981 | 0.4925 | 0.2771 | 0.081 | 1         |
| H24B | 0.4270 | 0.4181 | 0.2450 | 0.081 | 1         |
| H27  | 0.7666 | 0.3870 | 0.3845 | 0.049 | 0.5       |
| H29A | 0.7083 | 0.0278 | 0.5076 | 0.099 | 1         |
| H29B | 0.9329 | 0.0508 | 0.5355 | 0.099 | 1         |
| H29C | 0.8614 | 0.0991 | 0.4920 | 0.099 | 1         |
| H30A | 0.3726 | 0.1229 | 0.5088 | 0.099 | 1         |
| H30B | 0.5300 | 0.1829 | 0.4853 | 0.099 | 1         |
| H30C | 0.4169 | 0.2136 | 0.5265 | 0.099 | 1         |
| H31A | 1.2093 | 0.1748 | 0.6717 | 0.097 | 1         |
| H31B | 1.2170 | 0.0931 | 0.6439 | 0.097 | 1         |
| H31C | 1.0829 | 0.0947 | 0.6854 | 0.097 | 1         |
| H32A | 0.8638 | 0.2541 | 0.6725 | 0.088 | 1         |
| H32B | 0.7291 | 0.1749 | 0.6846 | 0.088 | 1         |
| H32C | 0.6470 | 0.2276 | 0.6431 | 0.088 | 1         |
| H33A | 0.9209 | 0.6314 | 0.2215 | 0.098 | 1         |
| H33B | 1.0830 | 0.5604 | 0.2399 | 0.098 | 1         |
| H33C | 1.0691 | 0.6441 | 0.2658 | 0.098 | 1         |
| H34A | 0.5716 | 0.6623 | 0.2575 | 0.100 | 1         |
| H34B | 0.7330 | 0.6814 | 0.2995 | 0.100 | 1         |
| H34C | 0.5359 | 0.6176 | 0.3018 | 0.100 | 1         |
| H35A | 0.8144 | 0.3006 | 0.3100 | 0.090 | 1         |
| H35B | 0.8668 | 0.3547 | 0.2697 | 0.090 | 1         |
| H35C | 0.6510 | 0.2988 | 0.2670 | 0.090 | 1         |
| H36A | 0.5014 | 0.3338 | 0.3517 | 0.118 | 1         |
| H36B | 0.3323 | 0.3364 | 0.3095 | 0.118 | 1         |
| H36C | 0.3536 | 0.4132 | 0.3412 | 0.118 | 1         |
| H1WA | 1.4889 | 0.6385 | 0.5594 | 0.060 | 0.5       |
| H1WB | 1.2668 | 0.6126 | 0.5504 | 0.060 | 0.5       |

**Table SI-16.** Torsion angles [°] for mL105c2_0m_sq.

___________________________________________________________________

| | | | |
|---|---:|---|---:|
| C18#1-N1-C2-N3 | -1.2(5) | C4-N15-C16-N17 | 0.7(5) |
| C18#1-N1-C2-C14 | 175.7(3) | C4-N15-C16-C28 | 178.3(3) |
| N1-C2-N3-C4 | 175.4(3) | N15-C16-N17-C18 | 178.1(3) |
| C14-C2-N3-C4 | -2.0(3) | C28-C16-N17-C18 | 0.2(3) |
| N1-C2-N3-Zn1#1 | 14.4(5) | N15-C16-N17-Zn1#1 | 15.1(4) |
| C14-C2-N3-Zn1#1 | -163.0(2) | C28-C16-N17-Zn1#1 | -162.8(2) |
| N1-C2-N3-Zn1 | -7.8(4) | N15-C16-N17-Zn1 | -4.8(5) |
| C14-C2-N3-Zn1 | 174.78(18) | C28-C16-N17-Zn1 | 177.3(2) |
| C2-N3-C4-N15 | -177.5(3) | C16-N17-C18-N1#1 | -174.9(3) |
| Zn1#1-N3-C4-N15 | -13.8(4) | Zn1#1-N17-C18-N1#1 | -14.3(4) |
| Zn1-N3-C4-N15 | 5.9(5) | Zn1-N17-C18-N1#1 | 7.9(4) |
| C2-N3-C4-C5 | 0.4(3) | C16-N17-C18-C19 | 1.9(3) |
| Zn1#1-N3-C4-C5 | 164.15(19) | Zn1#1-N17-C18-C19 | 162.5(2) |
| Zn1-N3-C4-C5 | -176.16(19) | Zn1-N17-C18-C19 | -175.29(19) |
| N15-C4-C5-C6 | 4.4(5) | N1#1-C18-C19-C20 | -5.1(5) |
| N3-C4-C5-C6 | -173.6(3) | N17-C18-C19-C20 | 177.9(3) |
| N15-C4-C5-C14 | 179.4(3) | N1#1-C18-C19-C28 | 173.5(3) |
| N3-C4-C5-C14 | 1.4(3) | N17-C18-C19-C28 | -3.4(3) |
| C14-C5-C6-C7 | 1.9(4) | C28-C19-C20-C21 | -2.8(4) |
| C4-C5-C6-C7 | 176.1(3) | C18-C19-C20-C21 | 175.7(3) |
| C5-C6-C7-C12 | 2.9(4) | C19-C20-C21-C26 | -3.0(5) |
| C5-C6-C7-C8 | -171.9(3) | C19-C20-C21-C22 | 172.1(3) |
| C6-C7-C8-C9 | -159.1(3) | C20-C21-C22-C33 | 37.8(4) |
| C12-C7-C8-C9 | 26.2(4) | C26-C21-C22-C33 | -147.2(3) |
| C6-C7-C8-C29 | 79.4(4) | C20-C21-C22-C23 | 159.0(3) |
| C12-C7-C8-C29 | -95.3(4) | C26-C21-C22-C23 | -26.0(4) |
| C6-C7-C8-C30 | -38.3(4) | C20-C21-C22-C34 | -81.2(4) |
| C12-C7-C8-C30 | 147.0(3) | C26-C21-C22-C34 | 93.9(4) |
| C29-C8-C9-C10 | 70.3(4) | C33-C22-C23-C24 | 170.3(3) |
| C7-C8-C9-C10 | -47.8(4) | C21-C22-C23-C24 | 47.0(4) |
| C30-C8-C9-C10 | -170.1(3) | C34-C22-C23-C24 | -70.6(4) |
| C8-C9-C10-C11 | 61.1(5) | C22-C23-C24-C25 | -60.7(5) |
| C9-C10-C11-C31 | -169.2(3) | C23-C24-C25-C26 | 46.8(5) |
| C9-C10-C11-C32 | 71.8(4) | C23-C24-C25-C35 | -71.4(4) |
| C9-C10-C11-C12 | -45.7(5) | C23-C24-C25-C36 | 170.3(4) |
| C6-C7-C12-C13 | -5.0(5) | C20-C21-C26-C27 | 6.1(5) |
| C8-C7-C12-C13 | 169.6(3) | C22-C21-C26-C27 | -168.8(3) |
| C6-C7-C12-C11 | 170.3(3) | C20-C21-C26-C25 | -167.9(3) |
| C8-C7-C12-C11 | -15.1(5) | C22-C21-C26-C25 | 17.2(5) |
| C10-C11-C12-C13 | -160.9(3) | C24-C25-C26-C27 | 159.3(3) |
| C31-C11-C12-C13 | -38.7(4) | C35-C25-C26-C27 | -80.1(4) |
| C32-C11-C12-C13 | 79.8(4) | C36-C25-C26-C27 | 38.2(4) |
| C10-C11-C12-C7 | 23.6(4) | C24-C25-C26-C21 | -26.5(5) |
| C31-C11-C12-C7 | 145.8(3) | C35-C25-C26-C21 | 94.0(4) |
| C32-C11-C12-C7 | -95.7(4) | C36-C25-C26-C21 | -147.7(3) |
| C7-C12-C13-C14 | 2.1(4) | C21-C26-C27-C28 | -3.2(5) |
| C11-C12-C13-C14 | -173.4(3) | C25-C26-C27-C28 | 171.1(3) |
| C12-C13-C14-C5 | 2.7(4) | C26-C27-C28-C19 | -2.5(5) |
| C12-C13-C14-C2 | -174.4(3) | C26-C27-C28-C16 | -178.1(3) |
| C6-C5-C14-C13 | -4.8(5) | C20-C19-C28-C27 | 5.8(5) |
| C4-C5-C14-C13 | 179.8(3) | C18-C19-C28-C27 | -173.1(3) |
| C6-C5-C14-C2 | 172.9(3) | C20-C19-C28-C16 | -177.7(3) |
| C4-C5-C14-C2 | -2.5(3) | C18-C19-C28-C16 | 3.4(3) |
| N1-C2-C14-C13 | 2.8(5) | N15-C16-C28-C27 | -4.3(5) |
| N3-C2-C14-C13 | -179.8(3) | N17-C16-C28-C27 | 173.8(3) |
| N1-C2-C14-C5 | -174.6(3) | N15-C16-C28-C19 | 179.6(3) |
| N3-C2-C14-C5 | 2.8(3) | N17-C16-C28-C19 | -2.4(3) |
| N3-C4-N15-C16 | -1.3(5) | | |
| C5-C4-N15-C16 | -178.9(3) | | |

**Table SI-17.** Hydrogen bonds for mL105c2_0m_sq [Å and °].

________________________________________________________________________________
D-H...A                 d(D-H)      d(H...A)     d(D...A)      <(DHA)
________________________________________________________________________________
O1W-H1WB...N15#2  0.86      1.86         2.583(5)      140.6
O1W-H1WB...N27#2  0.86      2.63         3.307(5)      135.7
________________________________________________________________________________

Symmetry transformations used to generate equivalent atoms:
#1 -x+3,-y+1,-z+1   #2 -x+2,-y+1,-z+1

### 1.7.4 Crystal Structure of [Sppz*BCl]     Crystallographer: Dr. Benjamin Oelkers

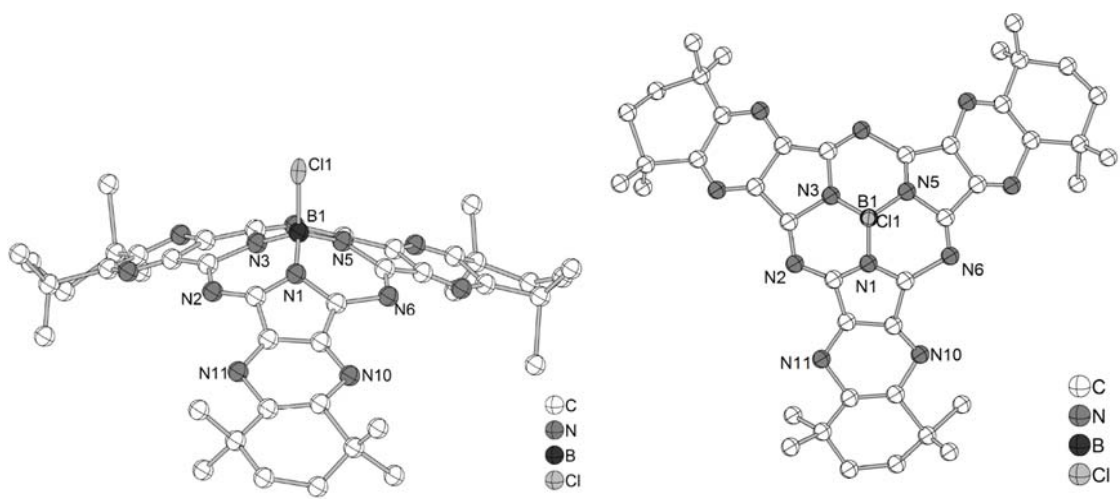

| | |
|---|---|
| Empirical formula | C49 H56 B Cl N12 |
| Formula weight | 859.32 |
| Temperature | 100(2) K |
| Wavelength | 0.71073 Å |
| Crystal system | Monoclinic |
| Space group | P 21/c |
| Unit cell dimensions | a = 7.253(3) Å       α = 90°. |
| | b = 25.008(5) Å     β = 98.83(3)°. |
| | c = 25.698(7) Å     γ = 90°. |
| Volume | 4606(2) Å$^3$ |
| Z | 4 |
| Density (calculated) | 1.239 Mg/m$^3$ |
| Absorption coefficient | 0.132 mm$^{-1}$ |
| F(000) | 1824 |
| Crystal size | 0.3 x 0.06 x 0.03 mm$^3$ |
| Theta range for data collection | 1.14 to 20.00°. |
| Index ranges | -6<=h<=6, -24<=k<=24, -24<=l<=24 |
| Reflections collected | 13040 |
| Independent reflections | 4273 [R(int) = 0.2894] |
| Completeness to theta = 20.00° | 99.8 % |
| Absorption correction | None |
| Refinement method | Full-matrix least-squares on F$^2$ |
| Data / restraints / parameters | 4273 / 0 / 197 |
| Goodness-of-fit on F$^2$ | 0.841 |
| Final R indices [I>2sigma(I)] | R1 = 0.1654, wR2 = 0.3401 |
| R indices (all data) | R1 = 0.4135, wR2 = 0.4094 |
| Largest diff. peak and hole | 0.382 and -0.431 e.Å$^{-3}$ |

Note: Because of the poor diffraction of the crystal, no satisfying refinement could be obtained. To gain a preliminary crystal structure of low resolution only refelexes up to theta = 20° were used. All deflection coefficients C, N, and B were hold constant isotropic and refined with common parameters. The interpretation and orientation of the solvent molecules are, in contrast to the general figure of the complex molecule, not of sufficient quality for a discussion. Because of the poor data, no detailed discussion of bond lengths is possible however a documentation of the unit cell is possible.